\documentclass[journal]{IEEEtran}
\ifCLASSINFOpdf
\else
\fi
\usepackage{amsfonts}
\usepackage{graphicx}
\usepackage{graphics}
\usepackage{subfigure}
\usepackage{caption}
\usepackage{cite}
\usepackage{amsmath} 
\usepackage{bm}
\usepackage{amsthm}
\usepackage{amssymb}
\usepackage{xcolor}
\usepackage{algorithm}
\usepackage{algorithmic}
\usepackage{lettrine}
\usepackage{multirow}
\usepackage{makecell}
\usepackage{lettrine}
\usepackage{booktabs}
\usepackage{amsfonts}
\usepackage{url}
\usepackage{enumitem}
\usepackage{booktabs}
\usepackage{array}
\usepackage{bbding}

\begin{document}

\title{Resource Scheduling in Edge Computing: A Survey}
\author{Quyuan~Luo, Shihong~Hu,
        Changle~Li, ~\IEEEmembership{Senior Member,~IEEE},
        Guanghui~Li,
        and Weisong~Shi,~\IEEEmembership{Fellow,~IEEE}%
\thanks{This work was supported in part by the National Natural Science Foundation of China (No. 62101463, No. U1801266, No. 62072216 and No. 61731017), in part by the Fundamental Research Funds for the Central Universities (
No. 2682021CX044), in part by the 111 project (No. 111-2-14), in part by Jiangsu Agriculture Science and Technology Innovation Fund (No. CX (19)3087), in part by Wuxi International Science and Technology Research and Development Cooperative Project (No. CZE02H1706), in part by the scholarship from China Scholarship Council. Quyuan Luo and Shihong Hu contributed equally to this work. (\textit{Corresponding author: Shihong Hu.})}
\IEEEcompsocitemizethanks{\IEEEcompsocthanksitem Q. Luo is with the School of Information Science and Technology, Southwest Jiaotong University, Chengdu 611756, China, the State Key Laboratory of Integrated Services Networks, Xidian University, Xi'an 710071, China, and the Department of Computer Science, Wayne State University, Detroit, MI 48202, USA (e-mail: qyluo@swjtu.edu.cn).
\IEEEcompsocthanksitem S. Hu is with the school of Artificial Intelligence and Computer, Jiangnan University, Wuxi, Jiangsu, 214122, China, and is also with the Department of Computer Science, Wayne State University, Detroit, MI 48202 (e-mail: Shihong@wayne.edu).
\IEEEcompsocthanksitem C. Li is with the State Key Laboratory of Integrated Services Networks, Xidian University, Xi'an 710071, China 
(e-mail: clli@mail.xidian.edu.cn)
\IEEEcompsocthanksitem G. Li are with the school of Artificial Intelligence and Computer, Jiangnan University, Wuxi, Jiangsu, 214122, China, and is also with theResearch Center for IoT Technology Application Engineering (MOE), Wuxi, Jiangsu, 214122 China (e-mail: ghli@jiangnan.edu.cn).
\IEEEcompsocthanksitem W. Shi is with the Department of Computer Science, Wayne State University, Detroit, MI 48202 (e-mail: weisong@wayne.edu).}

}

\maketitle
\begin{abstract}
With the proliferation of the Internet of Things (IoT) and the wide penetration of wireless networks, the surging demand for data communications and computing calls for the emerging edge computing paradigm. By moving the services and functions located in the cloud  to the proximity of users,  edge computing can provide powerful communication, storage, networking, and communication capacity.
The resource scheduling in edge computing, which is the key to the success of edge computing systems, has attracted increasing research interests. In this paper, we survey the state-of-the-art research findings to know the research progress in this field.
Specifically, we present the architecture of edge computing, under which different collaborative manners for resource scheduling are discussed. Particularly, we introduce a unified model before summarizing the current works on resource scheduling from three research issues, including computation offloading, resource allocation, and resource provisioning. Based on two modes of operation, i.e., centralized and distributed modes, different techniques for resource scheduling are discussed and compared. Also, we summarize the main performance indicators based on the surveyed literature. To shed light on the significance of resource scheduling in real-world scenarios, we discuss several typical application scenarios involved in the research of resource scheduling in edge computing. 
Finally, we highlight some open research challenges yet to be addressed and outline several open issues as the future research direction.
\end{abstract}

\section{Introduction}
\label{introduction}
\subsection{From Cloud Computing to Edge Computing}

With the rapid development of the mobile Internet, smart devices have become an indispensable part of people's life. Increasingly complex applications such as mobile payment, smart healthcare, mobile games, and virtual reality (VR) put higher requirements on the resource capacity of smart devices. 
Since Google put forward the concept of cloud computing in 2008~\cite{hayes2008cloud}, cloud computing was gradually accepted and introduced into the mobile environment, which breaks through the resource limitations of smart devices and provides highly demanding applications for users. Cloud computing is a cost-effective model that provides abundant applications and services while making information technology (IT) management more accessible and responding to users' demands faster~\cite{velte2009cloud}. The services (computing, communication, storage, and all necessary services) are delivered and implemented in a simplified way: on-demand, regardless of the users' location and the type of smart devices.

\begin{figure}[h]
\centering
\includegraphics[width=0.45\textwidth]{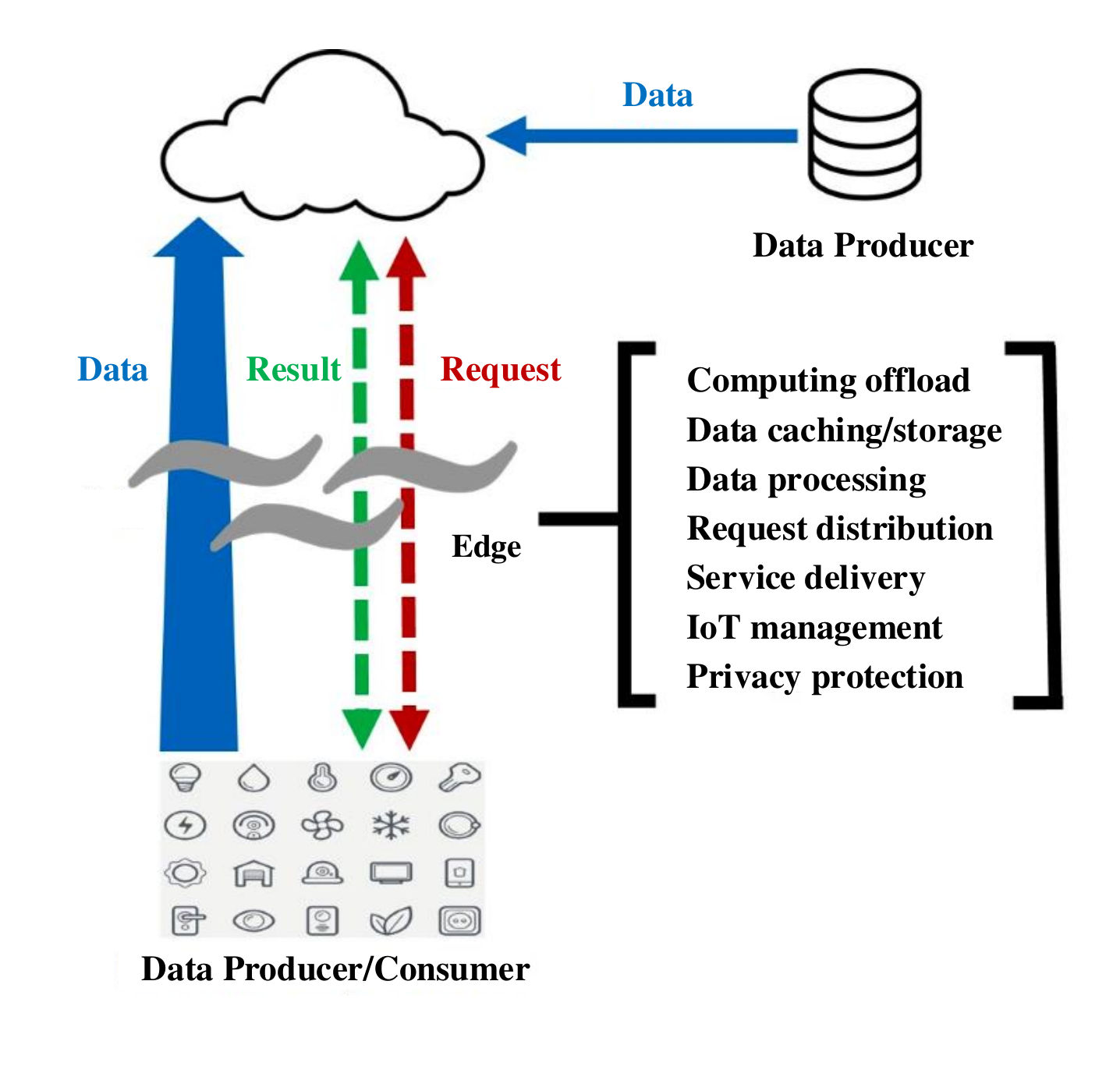}
\caption{ Edge computing paradigm \cite{shi2016edge}. The things not only are data consumers but also play as data producers. At the edge, the things can not only request service and content from the cloud but also perform the computing tasks from the cloud. Edge can perform computing offloading, data storage, caching and processing, as well as distribute request and delivery service from cloud to user.}
\label{fig edge computing paradigm}
\end{figure}

Thanks to rapid advances in underlying technologies, the Internet of Things (IoT) is opening tremendous opportunities for a large number of novel applications that promise to improve the quality of our lives \cite{xia2012internet}. Technically, all applications we discussed in this survey belong to the category of IoT. Applications such as unmanned aerial vehicle (UAV), connected and autonomous vehicle (CAV), video service, smart city, smart health, smart manufactory, and smart home are all committed to improving the quality of our lives through various technologies of IoT.
However, in recent years, the IoT era has brought higher requirements for transmission bandwidth, latency, energy consumption, application performance, and reliability. 
In this context, due to the limited bandwidth, high latency, and high energy consumption in the centralized processing model of cloud computing, it is hard to meet the high-performance requirements of users. Fortunately, it can be estimated that tens of billions of edge nodes (ENs) will be deployed in the near future \cite{CISCO2018}. By integrating these large amounts of idle resources distributed at the edge of the network to seamlessly provide services for users, a new computing paradigm - edge computing is proposed, which is regarded as the key technology and architectural concept of the transition to 5G~\cite{hu2015mobile}.  Fig.~\ref{fig edge computing paradigm} illustrates the edge computing paradigm. Edge computing refers to the enabling technologies allowing computation to be performed at the edge of the network, on downstream data on behalf of cloud services and upstream data on behalf of IoT services. Edge computing moves the services and functions originally located in the cloud to the proximity of users, which integrates the cloud computing platform and the network to provide powerful computing, storage, networking, and communication capacity at the edge of the network. Edge computing is interchangeable with fog computing, but edge computing focuses more on the things side, while fog computing focuses more on the infrastructure side~\cite{shi2016edge}. Since the services and functions are closer to users in edge computing, a better quality-of-experience (QoE) and quality-of-service (QoS) can be obtained by users. 
Let's take the edge computing in mobile communication/5G communication as an example. With the development of mobile communication, especially the 5G communication, the demand for high-quality wireless services shows a trend of exponential growth. In the age of 5G, in addition to mobile phones, tablets, a lot of new business scenarios in mobile network service emerges, such as autonomous driving, VR, and augmented reality (AR), and more close to the life business scenarios, such as smart grid, smart agriculture, smart city, and environmental monitoring. The emergence of these new service scenarios has higher requirements for 5G key technical indicators such as time delay, energy efficiency and reliability. In this context, due to the limited bandwidth, high latency, and high energy consumption in the centralized processing model of cloud computing, it is hard to meet the high-performance requirements of users. To cope with the issue in mobile communication, a new emerging concept, known as mobile edge computing (MEC), has been introduced. The MEC brings computation and storage resources to the edge of the mobile network enabling it to run the highly demanding applications at the user equipment while meeting strict performance requirements \cite{mach2017mobile}.

\subsection{Resource Scheduling in Edge Computing}
In recent years, resource scheduling in edge computing has attracted widespread interest from industry and academia.
Before introducing resource scheduling in edge computing, two questions should be answered firstly:

\textit{ 1) What is resource scheduling in edge computing?}

Generally, resource scheduling refers to the set of \textit{\textbf{actions}} and \textit{\textbf{methodology}} that \textit{\textbf{participants}} used to efficiently assign\textit{ \textbf{resources}} to the\textit{ \textbf{tasks}} that need to complete, and achieve the \textit{\textbf{objectives} }of participants based on resource availability. Specifically, according to edge computing characteristics, the key terms of resource scheduling in edge computing can be detailed as follows. 
\begin{itemize}
\item \textit{Resources}: Various resources existing in the edge network, by which the powerful serviceability is provided and the tasks can be completed. The resource in edge network can be categorized into three types, i.e., communication resources, storage resources (also as caching resources), and computing resources \cite{tan2017virtual,wang2018joint}.
\item \textit{Tasks}: Tasks generally  refer to data generated from users. The task types may vary based on different application scenarios for different objectives. For example, the  data from LiDAR and high-definition camera on CAVs is for safety purpose \cite{lin2019computation,li2018end}; the  data from body area networks (BAN) is for  health monitoring; and the  data from surveillance cameras is for  security \cite{wang2019edge}.    
\item \textit{Participants}: To complete tasks, there are different collaborative processing modes  that involves different participants. For ``things-edge collaboration", users (referred as ``things") and edge are the participants \cite{chen2018efficient}. For ``things-edge-cloud collaboration", users, edge, and  cloud are the participants \cite{dinh2020online}. For ``edge-cloud collaboration", edge and cloud center are the participants \cite{yoon2016low}.
\item \textit{Objectives}: Different users pursue different objectives during task processing. For example, CAVs aim to obtain low latency for traffic safety \cite{xu2019edge}. UAVs and smart health devices aim to reduce energy consumption for long battery life \cite{zhou2018uav}. The objectives can also be referred to as performance indicators.
\item \textit{Actions}:  The ways to achieve the objectives of participants are referred to as actions. In edge computing, there are mainly three actions: 1) computation offloading, which decides whether a task is offloaded to the edge or the  cloud to process \cite{li2018computing}; 2) resource allocation, which means allocating the communication, storage resources, and computing resources for tasks \cite{zhang2018jointresource}; 3) resource provisioning, which decides the user-resource pair association from the perspective of users, or actively conducts resource placement from the perspective of service providers (SPs) \cite{dai2018joint,wang2017online}.
\item \textit{Methodology}: Methodology refers to the methods, techniques, and algorithms to better take  the above actions for the objectives of participants. Basically, the methodology can be mainly categorized into centralized and distributed manners. The centralized methodology needs a control center to collect global information while the distributed methodology does not \cite{guo2018computation,meskar2018fair}. 
\end{itemize}

\begin{table*}[!htbp]
\centering
\caption{A summary of surveys on edge computing.}
\label{surveys}%
\begin{tabular}{lll}
\toprule
\textbf{Paper} &\textbf{Year} &  \textbf{Topic}\\
\midrule
Mao \textit{et al}.~\cite{mao2017survey} & 2017 &Joint  radio-and-computational resource management in edge computing.\\
Wang \textit{et al}.~\cite{wang2017survey} & 2017 & Issues on computing, caching and communication techniques in edge computing. \\
Mach \textit{et al}.~\cite{mach2017mobile} & 2017& User-oriented use case of computation offloading in edge computing. \\
Abbas \textit{et al}.~\cite{abbas2017mobile}& 2017 & Relevant research and technological developments in edge computing. \\
Peng \textit{et al}.~\cite{peng2018survey}& 2018 & Service adoption and provision in edge computing. \\
Tocze \textit{et al}.~\cite{tocze2018taxonomy}& 2018 & Resource management and optimization of multiple resources in edge computing. \\
Lin \textit{et al}.~\cite{lin2019computation}& 2019 & Research on computation offloading in edge computing. \\
Duc \textit{et al}.~\cite{duc2019machine}& 2019 & Resource provisioning in Edge-Cloud computing from a machine learning perspective. \\
Hong \textit{et al}.~\cite{hong2019resource} & 2019& Resource management from the architecture, infrastructure and algorithms in edge computing. \\
Ghobaei \textit{et al}.~\cite{ghobaei2019resource}& 2019 & Resource management approaches in edge computing. \\
Santos \textit{et al}.~\cite{santos2019resource}& 2019 & Resource provisioning from theory to practice in edge computing\\
Ren \textit{et al}.~\cite{ren2019survey} & 2019&Issues on different computing paradigms in edge computing. \\
Varghese \textit{et al}.~\cite{varghese2020survey}& 2020 & Different dimensions of research works in edge benchmarking. \\
\bottomrule
\end{tabular}
\end{table*}

\textit{ 2) Why do we need resource scheduling in edge computing?}

While edge computing greatly strengthens the serviceability of edge network by providing powerful computing, storage, and communication capacities,  it also requires appropriate resource scheduling strategies from three perspectives. 
\begin{itemize}
    \item 

\textbf{User}. Tens of billions of heterogeneous end-devices are geographically deployed in a distributed manner, the data volume generated from those end-devices and their corresponding applications are also heterogeneous. Orchestrating the limited  edge resources to better process those data requires appropriate resource scheduling strategies.  In the edge computing network, there are not only static  end-devices (e.g., sensors in smart homes, video cameras in public 
places), but also dynamic ones such as UAVs and vehicles, making the resource management even more challenging. Appropriate resource scheduling can alleviate this situation. Besides, the data from different application scenarios may have different service requirements. For example, the CAVs in intelligent transportation systems (ITS) need to process data within several milliseconds for traffic safety;  thus low latency is their main objective. The UAV-assisted edge computing usually focuses more on long battery life;  thus the objective of low energy consumption is expected during data processing. Also, some mobile devices (MDs) and IoT devices aim to achieve low data processing cost. Therefore, it needs proper resource scheduling strategies to meeting those service requirements.
 \item

\textbf{Service provider}. In addition to users, the edge computing ecosystem incorporates multiple actors,  such as edge infrastructure SPs, edge computing service providers, application service providers, and mobile network operators. Although these SPs and operators are resource-rich and have powerful serviceability, they are all commercial entities aiming at earning revenue by providing services \cite{samanta2019adaptive}.  In this context, designing an appropriate resource scheduling strategy can help them get a maximal revenue during service providing competition at a minimal cost. 
    
\item    
\textbf{Edge network}. Edge resources are distributed and scattered in the edge network. It is a waste of resources if scattered ones can not be efficiently utilized by resource scheduling. For example, the parked vehicles (PVs) account for a large portion of the global vehicles and have idle time to perform computational workloads \cite{huang2018parked,hou2016vehicular}. If an efficient resource strategy is applied, they can be combined to establish an available and cost-effective computing resource pool \cite{abdelhamid2015vehicle}, which helps to alleviate workloads of edge computing servers and promote the distributed computing environment. Besides, since both users and SPs try to  earn their benefits from edge computing, it is more like a game between buyers and sellers in terms of resources and services. An effective  resource strategy can jointly consider their interests and improve the edge system utility \cite{liu2019decentralized}.
\end{itemize}

\subsection{Related Surveys}
In recent years, many surveys on edge computing from various perspectives have been published, as shown in Table~\ref{surveys}. Mao \textit{et al}.~\cite{mao2017survey} presented a survey with the focus of joint radio-and-computational resource management in edge computing. Likely, a more recent survey~\cite{hong2019resource} also focused on resource management in edge computing. The difference is that this survey is from the viewpoint of architecture, infrastructure, and the underlying algorithms about resource management. Furthermore, both~\cite{tocze2018taxonomy} and~\cite{ghobaei2019resource} presented a comprehensive survey of resource management in edge computing, the work in~\cite{tocze2018taxonomy} surveyed related literature in terms of resource type, objective, resource location, and resource while Ghobaei \textit{et al}.~\cite{ghobaei2019resource} provided a systematic review from application placement, resource scheduling, task offloading, load balancing, resource allocation and provisioning six fields in resource management.
Wang \textit{et al}.~\cite{wang2017survey} summarized the related works on computing, caching, and communication techniques in the area of edge computing. Mach \textit{et al}.~\cite{mach2017mobile} surveyed the research on computation offloading in the area of edge computing. Later, Lin \textit{et al}.~\cite{lin2019computation} presented a more comprehensive survey on computation offloading. The review angle of the survey~\cite{abbas2017mobile} is more macro. It comprehensively elaborated on the definition, architecture, application areas, and advantages of edge computing. Besides, Varghese \textit{et al}.~\cite{varghese2020survey} presented a systematic survey on edge benchmarking, which summarized the research from the system under test, techniques, quality metrics, and benchmark runtime in the edge computing. Some surveys focus on one topic, like service adoption and provision~\cite{peng2018survey}, resource provision from a machine learning perspective~\cite{duc2019machine} or computing paradigms~\cite{ren2019survey} in edge computing. 

 It can be concluded that some existing surveys summarized the research in edge computing only from a single angle in the resource scheduling field, like computation offloading or resource provisioning. Some surveys in previous years mostly discussed topics in edge computing from a high level and failed to comprehensively address these topics at the depth. With the increasing enthusiasm of the academic community for edge computing research in recent years, a large number of new research results have emerged, among which the research on resource scheduling is particularly prominent. Although the existing surveys listed in Table \ref{surveys} have reviewed edge computing from various perspectives, none of them focus on the resource scheduling issue in a comprehensive way. This motivates us to present a systematic survey on resource scheduling, so we review from multiple perspectives, including architecture, research issue, techniques, indicators, and applications to provide a comprehensive, informative and up-to-date viewpoint for researchers.

\subsection{Contribution and Organization}
\begin{figure}[h]
\centering
\includegraphics[width=0.48\textwidth]{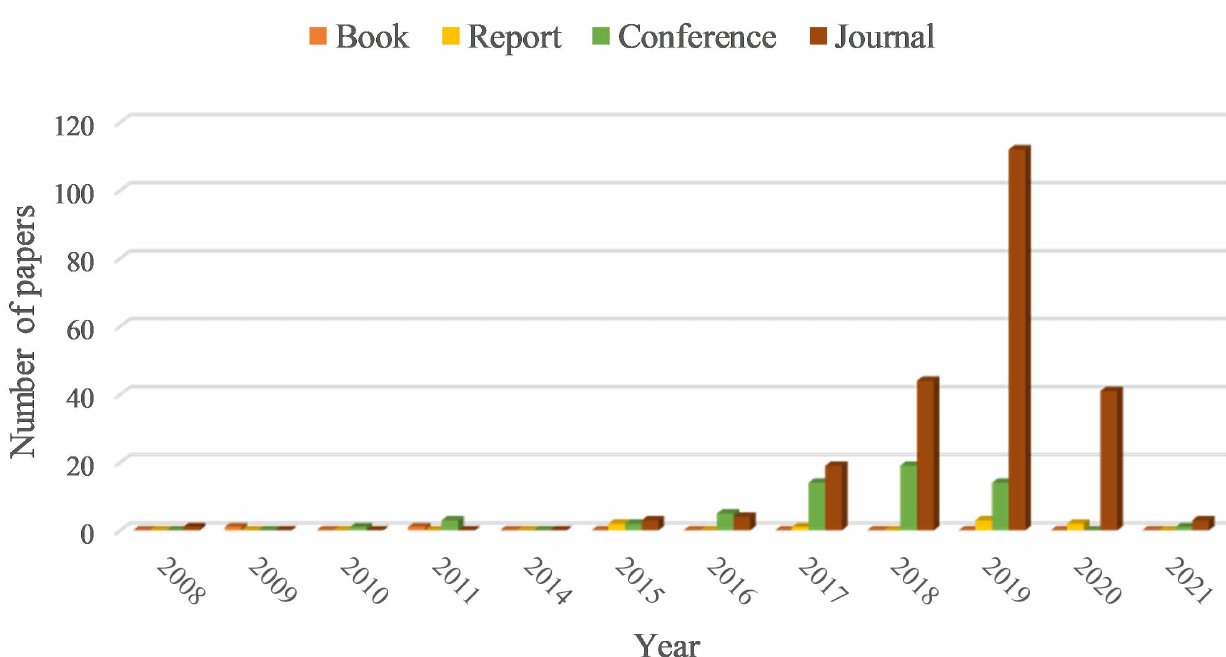}
\caption{ The distribution of papers surveyed by year and source. \textit{Book} includes books and book chapters; \textit{Report} includes arXiv articles, website articles and white papers; \textit{Conference} includes conference and symposium papers; \textit{Journal} includes journal and magazine articles.}
\label{fig histogram of publications}
\end{figure}
This article provides a comprehensive survey of the state-of-the-art research with a focus on resource scheduling in edge computing. Fig.~\ref{fig histogram of publications} shows the distribution of papers surveyed by year and source. Specifically, the focus of this article  is five-fold. 

\begin{table*}[]
\caption{ Summary of Acronyms Frequently Used in the Paper.}
\label{table list of acronym}
\centering
\begin{tabular}{ll|ll}
\toprule
\hline
\textbf{Acronym} & \textbf{Definition} & \textbf{Acronym} & \textbf{Definition} \\ \hline
    ADMM     &    Alternating Direction Method of Multipliers   & MD   &     Mobile Device      \\ \hline        
    AI      &      Artificial Intelligence      &  MDC   &     Micro Data Center \\ \hline 
    AR      &     Augmented Reality       & MDP   &    Markov Decision Process            \\ \hline 
    BAN      &      Body Area Network      &  MEC   &     Mobile Edge Computing \\ \hline 
    BS      &     Base Station       &  MILP   &  Mixed Integer Linear Programming    \\ \hline 
    CAV      &     Connected and Autonomous Vehicle       &  MU   &      Mobile User     \\ \hline 
    CC     &    Computing  and  Communication        & NFV   &     Network Function Virtualization   \\ \hline 
    CCS     &    Computing, Communication, and Storage        &  NSGA   &    Non-dominated Sorting Genetic Algorithm  \\ \hline 
    CN     &    Core Network        &  NOMA   &    Non-orthogonal Multiple Access   \\ \hline      
    DQN     &    Deep Q-network        &  PVEC   &    Parked Vehicle Edge Computing   \\ \hline 
    DRL     &    Deep Reinforcement Learning        &   PSO  &     Particle Swarm Optimization     \\ \hline 
    DSRC     &    Dedicated Short-Range Communications        &   PV  &    Parked Vehicle     \\ \hline 
     EC   &    Edge Cloud        &   QoE  &    Quality of Experience \\ \hline 
    EG    &    Edge Gateway        &  QoS  &    Quality of Service   \\ \hline 
    EN    &    Edge Node        &   RSU  &    Road Side Unit    \\ \hline 
    ES    &     Edge Server        &   SP  &     Service Provider  \\ \hline 
    FiWi     &    Fiber-Wireless       &   SCA  &  Successive Convex Approximation   \\ \hline 
    FL     &     Federated Learning         &   SDN  &   Soft-defined Network  \\ \hline 
     GA   &   Genetic Algorithm       &  TDMA  &  Time Division Multiple Access   \\ \hline 
    IIoT   &   Industrial Internet of Things       &  UAV  &    Unmanned  Aerial Vehicle  \\ \hline 
    IoT     &     Internet of Thing        &     UE  &  User Equipment  \\ \hline 
    IT     &     Information Technology  &    VEC  &   Vehicle Edge Computing  \\ \hline 
    ITS    &    Intelligent Transportation Systems   
       &    VM  &   Virtual Machine  \\ \hline 
    LSTM   &   Long Short-Term Memory &    WAN  &  Wireless Access Network  \\ \hline 
\end{tabular}
\end{table*}

\begin{figure*}[!htb]
\centering
\includegraphics[width=0.75\textwidth]{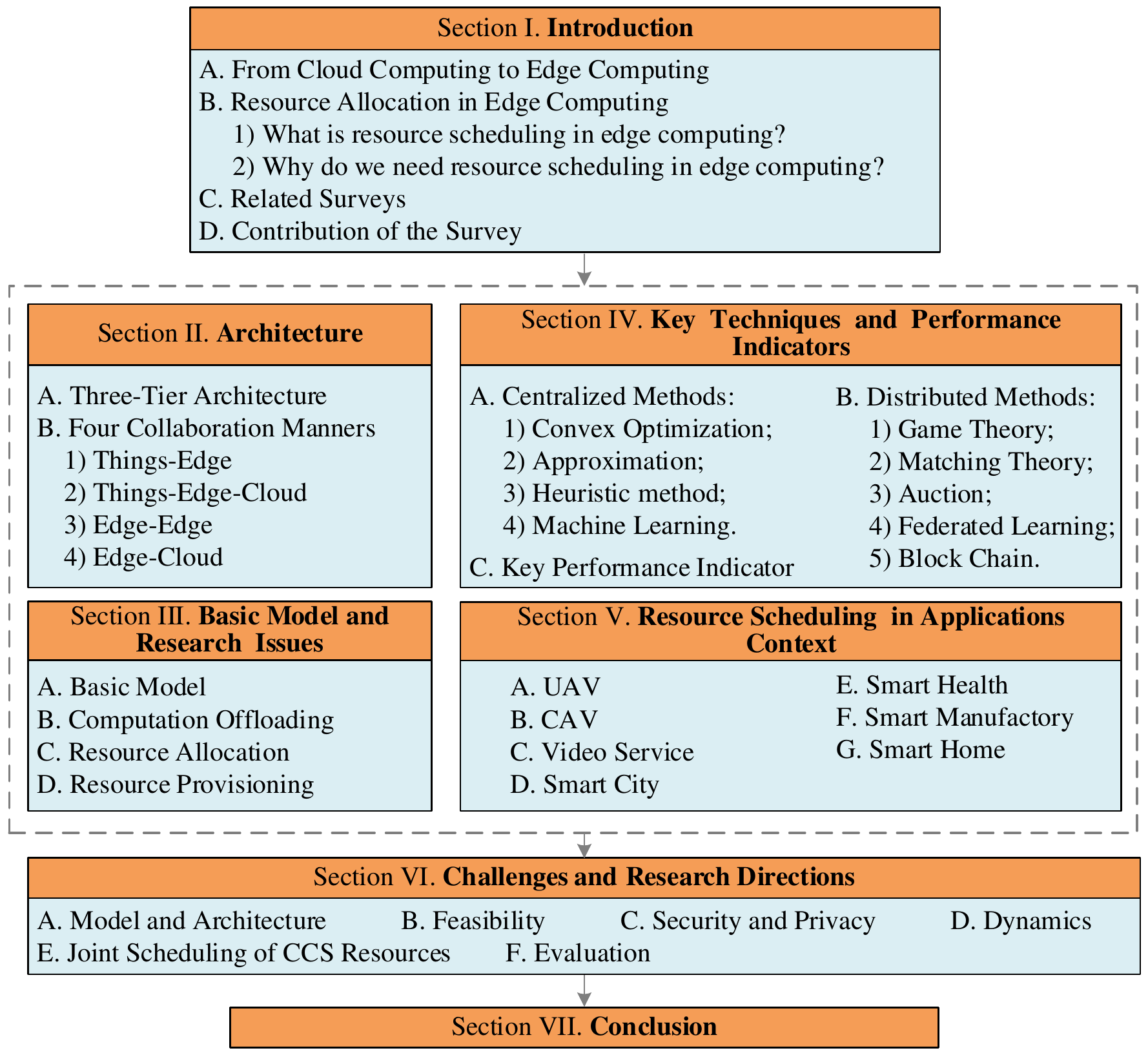}
\caption{Road map of the survey.}
\label{fig organization of the survey}
\end{figure*}

\begin{itemize}
\item \textit{Architecture  (Section \ref{sec architecture}):} A three-tier edge computing architecture including the \textit{thing layer}, the \textit{edge layer}, and the \textit{cloud layer} is first introduced. Then we elaborate on four different collaborations for resource scheduling under the three-tier architecture, i.e., \textit{things-edge}, \textit{things-edge-cloud}, \textit{edge-edge}, and \textit{edge-cloud}.
\item  \textit{ Basic Model and Research issue  (Section \ref{sec research issues}):} To achieve the different requirements of both  end-devices and the system for QoS and QoE, several basic models are first introduced. Based on those models, we then present three aspects involved in resource scheduling, which forms the three key research issues, i.e., \textit{computation offloading}, \textit{resource allocation}, and \textit{resource provisioning}. 
\item \textit{Technique and indicator  (Section \ref{sec techniques}):} We  summarize the main performance indicators such as \textit{latency}, \textit{energy consumption}, \textit{cost}, \textit{utility}, \textit{profit}, and \textit{resource utilization} in existing works. To achieve those objectives, we also elaborate on the resource scheduling techniques both in \textit{centralized} and \textit{distributed} ways.
\item \textit{Application  (Section \ref{sec application}):} We  summarize several typical application scenarios involved in the research on resource scheduling in edge computing, mainly including \textit{UAV}, \textit{CAV}, \textit{video service}, \textit{smart city}, \textit{smart health}, \textit{smart manufacturing}, and \textit{smart home}.
\item \textit{Challenge and open issue  (Section \ref{sec challenges}):} The lessons learned in the area of resource scheduling in edge computing are highlighted and several challenges yet to be addressed are presented for future research directions.
\end{itemize}

 To help the readers have a comprehensive picture of the structure of this survey, Fig.~\ref{fig organization of the survey} outlines the organization of the survey, and Table~\ref{table list of acronym} lists the acronyms that will be frequently used in the survey.

\section{Architecture}
\label{sec architecture}
\begin{figure*}[!htb]
\centering
\includegraphics[width=0.9\textwidth]{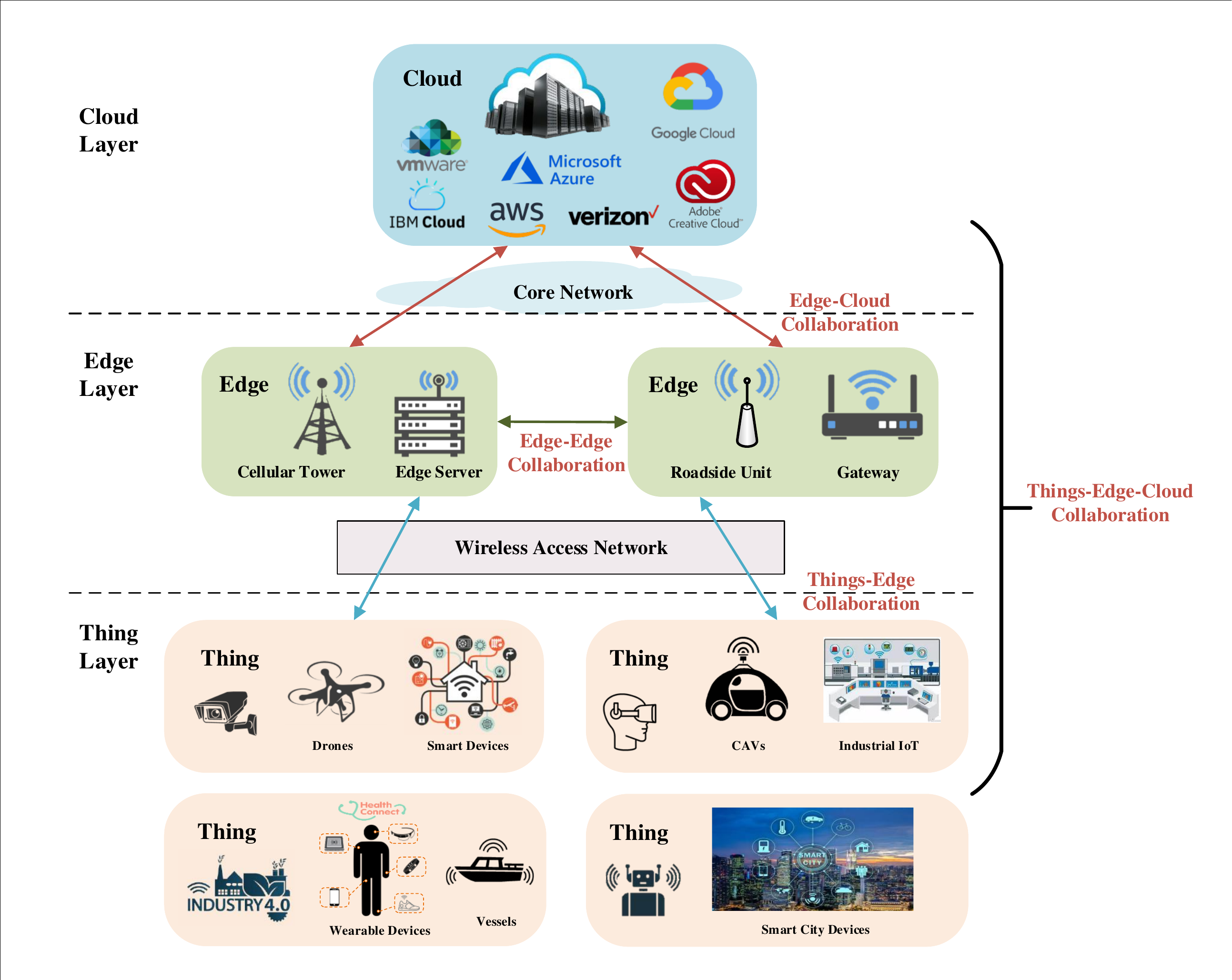}
\caption{Architecture for Resource Scheduling in Edge Computing.}
\label{fig architecture}
\end{figure*}

This section introduces the edge computing architecture for resource scheduling. We overview the composition of the architecture and introduce a three-tier heterogeneous edge computing network, where the first tier is the \textit{thing layer}, the second tier is the \textit{edge layer}, and the third one is the \textit{cloud layer}. Based on the three-tier architecture, we then present different collaborative manners for resource scheduling in edge computing.
\subsection{Overview of the Architecture for Resource Scheduling in Edge Computing}

\begin{figure}[!htb]
\centering
\includegraphics[width=0.48\textwidth]{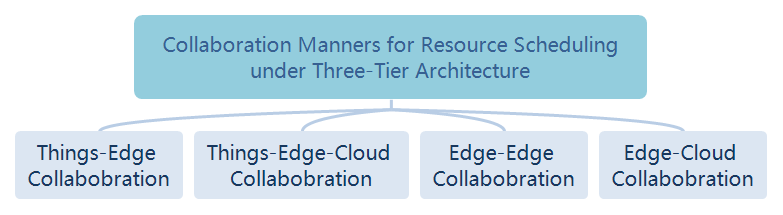}
\caption{Four different collaboration manners for resource scheduling under three-tier architecture.}
\label{fig collaboration manners}
\end{figure}

Traditional  cloud computing has difficulty to meet the high requirements of users in real-time response and low energy consumption due to bandwidth congestion and heavy load on the core network (CN). Nevertheless, the edge computing paradigm itself cannot be a substitute for cloud computing because it does not have as powerful resource capacity as cloud computing. In some cases, however, the advantages of edge computing can be leveraged to offload computing services from the cloud to the edge to improve users' QoE. Accordingly, cloud computing and edge computing are complementary and mutually reinforcing. Thus, the resource scheduling in edge computing is not only operated among users and the edge, but also among users, the edge, and the cloud. 
The three-tier heterogeneous architecture for resource scheduling in edge computing is presented, as shown in Fig.~\ref{fig architecture}, including the thing layer (a.k.a, the user layer), the edge layer, and the cloud layer. 
The three-tier architecture is a widely popular and accepted
paradigm by many existing works \cite{lin2019computation,hong2019resource,mao2017survey,abbas2017mobile,mach2017mobile}. The function of this kind of architecture is to illustrate the relationship among components that make up the edge computing system.
In the following, we first give a brief introduction on the three layers. Then, we elaborate on four different collaborations for resource scheduling under the three-tier architecture, i.e., \textit{things-edge collaboration}, \textit{things-edge-cloud collaboration}, \textit{edge-edge collaboration}, and \textit{edge-cloud collaboration}, as shown in Fig~\ref{fig collaboration manners}.

\subsubsection{Thing Layer}
The thing layer, also known as the user layer, is composed of various  end-devices (a.k.a., things), such as UAVs \cite{li2020energy}, CAVs \cite{xu2019edge}, AR equipment \cite{ren2019edge}, surveillance cameras for  smart city \cite{taleb2017mobile}, sensors for smart health \cite{lin2018task}, IoT devices for smart manufacturing \cite{liao2019learning,sun2018double}, smart devices for smart home \cite{deng2019parallel}. In different works,  end-devices are also called MDs or mobile users (MUs).
Various things can perceive and have certain storage and computing capability. Things continuously generate and collect multiple types of data. Based on the QoE and QoS requirements of things, the data can be processed locally, or be offloaded to the edge and  the cloud. 
In the edge computing network, there are not only static end-devices (e.g.,
sensors in smart homes, video cameras in public places) but also dynamic ones such as
UAVs and vehicles, making the resource management even more
challenging. Therefore, different solutions are proposed to address this issue, which are discussed
in Section~\ref{sec techniques}.
\subsubsection{Edge Layer}
The edge layer, as the core of the three-tier architecture, is an intermediate layer between the thing layer and the cloud layer.
From the perspective of hardware composition, the edge layer consists of various networking and computing equipment, such as cellular tower, edge server (ES), roadside unit (RSU), gateway, edge controller, etc. The edge layer provides wireless access to smart devices through the radio access technology, such as Long Term Evolution (LTE), Wireless Fidelity (WiFi), and Dedicated Short-Range Communications (DSRC). Basically, the edge layer can provide more powerful storage and computing capabilities than the thing layer. 
From the perspective of software composition, the edge layer has edge management capabilities that offer service orchestration and invocation and schedule the ESs to complete tasks.
The edge layer can receive, process, and forward data streams from the thing layer, and achieve  intelligent sensing, privacy protection, data analysis, intelligent computing, process optimization, and real-time control. Besides, since the edge and the cloud are complementary and mutually reinforcing, services in the cloud can be offloaded to the edge layer for load balancing and better QoE.  
With the objective of reducing bandwidth usage and energy conption of the CN as well as reducing the communication overhead between the edge and the cloud, the edge layer is expected to schedule edge resources to enable rapid service response.

\subsubsection{Cloud Layer}
The cloud layer consists of the existing cloud computing infrastructures, such as computing units, storage units, and micro data centers (MDCs), connected with the edge layer through the CN (a.k.a, backbone network). Among the three layers, the cloud layer is undoubtedly the most powerful data processing and storage center. 
While ESs in the edge layer can process large amounts of data to reduce latency and energy consumption, the edge computing paradigm still requires the computing power and high-capacity storage infrastructure of the cloud to handle some tough tasks and global information.
For example, the cloud layer can receive data streams from the edge layer, and send control information to the edge layer, and then from the edge layer to the thing layer, thereby optimizing the resource scheduling and field production process from a global perspective.
Besides, based on the network resource distribution,  the cloud layer can also dynamically adjust the deployment strategies and algorithms.
Furthermore, it also provides decision-support systems, intelligent production, networking collaboration, service extension, personalized and customized service, and other domain-specific application services.

\subsection{Things-Edge Collaboration}
\label{subsec things-edge collaboration}
The resource scheduling in a things-edge collaboration manner involves the things layer and the edge layer.  The task generated from smart devices can be processed locally or offloaded to ESs. Whether to offload these data depends on the things-edge collaboration strategy and the QoS and QoE requirements of smart devices. For example, Ali \textit{et al.} in \cite{ali2019deep} proposed to select an optimal set of computation components to offload to ESs, aiming at minimizing the energy consumption of MDs.
In addition to offloading  task to the ES in a local region, Wang \textit{et al.} in \cite{wang2018energy} proposed that the  task can also be offloaded to the ES in a nearby region to reduce overall system costs and guarantee users' QoE. Since the service requests of MUs and location may be dynamically changing, the static ES deployment may cause a “service hole”. To compensate for this issue  and to improve the resource utilization as well as the system utility, Liu \textit{et al.} in \cite{liu2019deep} explored a vehicle edge computing (VEC) network architecture and regarded the moving vehicles as vehicular ESs to assist the fixed ES to process the  task from MUs. Besides, regarding UAVs as ESs is also a research treading.
Yang \textit{et al.} in \cite{yang2019energy} considered a UAV-enabled mobile edge computing (MEC) network, where the computation tasks from MUs can be processed by UAVs aiming at minimizing the  power consumption of all MUs and UAVs.
Unlike previous studies in which users first offload  task to ES and results are then fed back, Chen \textit{et al.} in \cite{chen2019efficient} investigated the relay-assisted computation offloading (RACO). In the considered RACO scenario, a mobile-edge relay server (MERS) is utilized to assist the results of computational tasks among users by allocating computing and communication resources.

\begin{table*}[!htb]
\newcommand{\tabincell}[7]{\begin{tabular}{@{}#1@{}}#2\end{tabular}}
\centering
\caption{Comparison of Papers Focusing on Different Collaboration Manner for Resource Scheduling.  Acronyms used in this Table: user equipment (UE), edge server (ES), mobile device (MD), vehicular edge server(VES), fixed edge server (FES), mobile edge relay server (MERS), base station (BS), unmanned arerial vehicle (UAV), edge gateway (EG), parked vehicle (PV), mobile user (MU), micro data center (MDC).}
\begin{tabular}{|c|m{1.5cm}<{\centering}|m{1cm}<{\centering}|m{1.5cm}<{\centering}|m{2.5cm}<{\centering}|m{5.3cm}|m{2.4cm}<{\centering}|}
\hline
\makecell*[c]{\textbf{Paper}} &\textbf{Collaboration Manner}
&\makecell*[c]{\textbf{Things}}
&\makecell*[c]{\textbf{Edge}}
&\makecell*[c]{\textbf{Research Issue}}
&\makecell*[c]{\textbf{  Characteristics}}
&\makecell[c]{\textbf{ Methodology}}
\\
\hline
\cite{ali2019deep}
& Things-edge & UE & ES & Offloading  strategy & Minimize the energy consumption  of  MDs by selecting  an  optimal  set  of  computation components to offload to ESs. 
&Deep learning\\
\hline
\cite{wang2018energy}
& Things-edge & UE & ES in local and nearby region & Offloading  strategy & Formulate the computation offloading problem as a potential game
& Game theory, Jacobi algorithm\\
\hline
\cite{liu2019deep}
& Things-edge & UE & VES and FES & Offloading  strategy; resource allocation & Consider the stochastic vehicle traffic, dynamic computation requests and time-varying communication conditions
& Reinforcement learning\\
\hline
\cite{yang2019energy}
& Things-edge & UE & UAV & Resource allocation & Jointly optimize user association, power control, computation capacity allocation and location planning
& Compressive sensing, search
method\\
\hline
\cite{chen2019efficient}
& Things-edge & User & MERS & Computation offloading; resource allocation
& Jointly optimize transmit powers, processor speeds, bandwidth, and offloading ratio& Iterative algorithm\\
\hline

\cite{guo2018collaborative}
& Things-edge-cloud & MD & ES & Offloading  strategy
& Minimize all MDs' energy consumption while satisfying the MDs’ computation execution time constraint & Game theory\\
\hline
\cite{hong2019multi}
& Things-edge-cloud & IIoT devices & BS enabled ES & offloading  strategy
& Minimize energy consumption and computing time of task processing & Game theory\\
\hline
\cite{dinh2020online}
& Things-edge-cloud& User & ES & Resource allocation
& Consider the edge’s local processing cost and
capacity, the cloud’s multiple rental options& Offline and online algorithms\\
\hline
\cite{wang2019hetmec}
& Things-edge-cloud& Smart device & ES & Resource allocation
& The communication and computing resources, the task assignment among multiple layers are jointly coordinated & Latency minimization algorithm\\
\hline

\cite{na2018frequency}
& Things-edge; edge-edge & IoT devices & EG; ES& Resource allocation
& Consider computing capacities of ES and EGs, and interference among EGs & Lagrangian and KKT condition.\\
\hline
\cite{huang2018parked}
& Things-edge; edge-edge & Mobile vehicles & PVs; VES & Resource allocation
& Fully utilize the idle resource of parked vehicles & Stackelberg game, iterative algorithm\\
\hline
 \cite{alameddine2019dynamic}
& Things-edge; edge-edge & UE & eNB enabled ES & Computation offloading; resource allocation
& The tasks from UEs is scheduled among different ESs & benders decomposition technique\\
\hline
\cite{miao2020intelligent}
& Things-edge-cloud; edge-edge & MD & ES & Computation offloading
& Integrate artificial intelligence (AI), local computing, edge computing, and cloud computing & Deep learning, LSTM\\
\hline
\cite{thai2019workload}
& Things-edge-cloud; edge-edge & Smart device & ES & Computation offloading
& Vertical and horizontal offloading; workload and capacity optimization problem & branch-and-bound method\\
\hline
\cite{yoon2016low}
& Things-edge-cloud; edge-cloud & MU & ES & Resource placement
& Place the video  transcoding function at edge layer; provide higher video bit-rates without causing video stall or rebuffering & Video transcoding at edge\\
\hline
\cite{xu2017zenith}
& Things-edge-cloud; edge-cloud & Smart device & MDC & Resource allocation; resource provisioning
& SPs put resource in the edge layer; a latency-aware task scheduling mechanism & Auction-based contracts\\
\hline
\cite{zhang2019dmra}
& Things-edge-cloud; edge-cloud & UE & ES & Resource allocation
& SPs at the edge layer assign the tasks from UEs to be processed in base staion or cloud center & Decentralized multi-SP resource allocation\\
\hline
\end{tabular}
\label{table architecture}
\end{table*}

\subsection{Things-Edge-Cloud Collaboration}
\label{subsec things-edge-cloud collaboration}

Although the things-edge collaboration manner has a relatively powerful capacity, it ignores the huge computing resources in the cloud computing center. With the ever-increasing smart devices and their resource-hungry applications, it will become increasingly difficult to rely on the resources in the edge layer alone to meet the service requirements of smart devices. Therefore, it is particularly important and necessary to take full advantage of both edge computing and cloud computing and make them complementary to design a collaborative paradigm, the things-edge-cloud collaboration manner. 
Guo \textit{et al.} in 
\cite{guo2018collaborative} introduced the concept of a hybrid fiber-wireless (FiWi) network, in which the multi-access edge computing and the centralized cloud computing  cooperated to provide better offloading performance and good scalability as computation tasks increase. The combination of edge computing and cloud computing FiWi takes the complementary advantages of good scalability, high mobility, and supports diverse wireless access technologies in edge computing, large capacity, high reliability, and low-latency in fiber-enabled cloud computing. For the resource-intensive applications, such as big-data analytics, AI processing, and 3D sensing from industrial Internet of things
(IIoT) devices, Hong \textit{et al.} in \cite{hong2019multi}
proposed a multi-hop IIoT-edge-cloud collaborative computation offloading paradigm, aiming at minimizing energy consumption and computing time of task processing.
Wang \textit{et al.} in \cite{wang2019hetmec} proposed the concept of "HetMEC", which refers to heterogeneous multi-layer MEC. In HetMEC, if the  task offloaded from smart devices cannot be processed on time by the ES, it can be offloaded to the cloud center, aiming at minimizing transmission and computing time.  
Different from previous studies, Dinh \textit{et al.} in \cite{dinh2020online} considered renting computing resources termed virtual machines (VMs) from the cloud layer to scale up the capacity of the edge layer, with the goal of minimizing the total cost, including the processing cost at the edge, the remote on-demand VMs cost, the reserving and using remote reserved VMs cost.

\subsection{Edge-Edge Collaboration}
\label{subsec edge-edge collaboration}
Generally, the edge-edge collaboration manner for resource scheduling in edge computing does not arise in isolation. Instead, it usually comes along with the things-edge collaboration  manner or the things-edge-cloud collaboration manner. Through an edge-edge collaboration manner, there is one more option for task processing. Many studies have investigated this collaboration manner. 
Huang \textit{et al.} in \cite{huang2018parked} proposed a parked vehicle edge computing (PVEC) architecture, where idle resources of PVs can be fully utilized. In PVEC architecture, VEC servers explore opportunistic resources from PVs to allocate workloads, and provide rewards to PVs for their assistance. When necessary, VEC servers can also undertake the residual workloads. As a result, VEC servers and PVs cooperate to process  task in an edge-edge collaboration manner. To alleviate the workload on ESs, Na \textit{et al.} in \cite{na2018frequency} proposed to utilize edge gateways (EGs)  at the edge layer to assist task processing. A resource orchestration scheme among EGs and/or between ES and EGs is also proposed, aiming to maximize the efficiency of IoT systems.
Alameddine \textit{et al.} in \cite{alameddine2019dynamic} studied the dynamic
task offloading and scheduling problem (DTOS) in multi-access edge computing, where application's task assignment and the order of execution are jointly considered. The tasks that cannot be processed by its corresponding eNB-enabled ES can be offloaded to another ES in an edge-edge collaboration manner to meet UE's QoE requirement.
Miao \textit{et al.} in \cite{miao2020intelligent} proposed an intelligent offloading strategy based on the mobile-edge cloud computing architecture, where tasks are scheduled among MDs, ESs, and the cloud based on task prediction, aiming at reducing the total task delay. Besides, the ES in this strategy can decide whether to migrate its overload to other ES in an edge-edge collaboration manner. Differently, Thai \textit{et al.} in \cite{thai2019workload} proposed a cloud-edge computing architecture to provide horizontal and vertical collaborations, aim to minimize the total cost. Horizontal collaboration means that offloading operations can be conducted among the nodes in the same tier, while vertical collaboration means that offloading operations can be conducted among the cross-tier nodes.

\subsection{Edge-Cloud Collaboration}
\label{subsec edge-cloud collaboration}
If most computing tasks are performed in the cloud computing center in the considered three-tier architecture, long latency will be produced, which can not satisfy users' QoE. The long latency problem can be improved by offloading some or all of the tasks in the cloud center to the edge in an edge-cloud collaboration manner, such as the edge accelerated web platform (EAWP) by Nippon Telegraph and Telephone Corporation \cite{NTT2020eawp}. The edge-cloud collaboration manner can be used in many applications. For example, mobile client shopping has become popular where customers frequently operate the shopping cart. The change of the shopping cart status is first completed in the cloud center, and then the product view is updated on the MD, which results in long latency. If shopping cart data can be cached and relevant actions can be performed on the edge, the new product view will be pushed to the MD once the customer's request reaches the edge, thus greatly improving the customer's QoE.
Another example is the video transcoding application. Online video traffic on MDs is growing exponentially in network traffic \cite{forecast2019cisco, erman2011over}, and MUs have high QoE requirements for streaming video.
The video transcoding has become an optimized technique for video data transmission. However, since video transcoding consumes a great  quantity of computing and storage resources, it is typically executed in the offline media server (located in the cloud layer). Unfortunately, this approach may increase the latency when the video stream is redirected from the media server and the real-time streaming service cannot be provided. 
To this end, Yoon \textit{et al.} in \cite{yoon2016low} proposed to run the video transcoding on ENs such as home WiFi access point. The experimental results show that their solution is low-cost, transparent, and scalable.
Besides, Xu \textit{et al.} in \cite{xu2017zenith} proposed to regard the edge layer as MDCs to provide edge computing services. A model, named Zenith, was also proposed, where SPs can establish resource sharing contracts with edge infrastructure providers, aiming to increase resource utilization and minimize job execution latency.
Similarly, Zhang \textit{et al.} in \cite{zhang2019dmra} proposed
to deploy SPs in the edge layer to manage the task processing for MUs. The SPs can schedule the task to the edge or the cloud in an edge-cloud collaboration manner, aiming at providing high-quality services and maximizing the total profit of all SPs.

For simplicity, a comparison of papers focusing on different collaboration manner for resource scheduling are summarized in Table~\ref{table architecture}.
\section{Basic Model and Research Issues}
\label{sec research issues}

In this section, we first present the \textit{basic model} for resource scheduling in edge computing, which guides users to decide whether to take offloading action based on the current communication and computing resource state as well as their QoE requirements. Then, we elaborate on the state-of-the-art research on resource scheduling in edge computing from \textit{three aspects}: computation offloading, resource allocation,  and resource provisioning.

\subsection{Basic Model}
\label{labc}
In a typical edge computing scenario, various tasks would be generated from user devices. Generally, an arbitrary task $T$ can be described by five items, i.e., $T = \{D, c, \alpha, \gamma, \tau \}$, where $D$ is the data size of $T$, $c$ represents the processing density (in CPU cycles/bit) of $T$, $\alpha$ ($0\leq \alpha \leq 1$) stands for the parallelizable fraction of $T$, $\gamma$ denotes the ratio of the data size of processing result to the data size of $T$, and $\tau$ represents the delay constraint of $T$ \cite{lin2019computation}.
The  end-devices, CAVs and UAVs, can be connected to the edge through various communication channels (such as 4G/5G, WiFi, LTE/DSRC, etc.).
We denote the wireless bandwidth assigned to the  end-devices for task $T$ as $B$.
The generated task $T$ can be processed locally or offloaded to the edge or the cloud to be processed. The offloading action is taken based on different requirements for energy consumption, latency, cost, and computing acceleration. 
Let $\lambda$ ($0 \leq \lambda \leq 1$) denote the offloading decision variable, which represents the ratio of the offloaded data size to the total data size of task $T$. If $\lambda=0$, task $T$ will be processed locally; if $\lambda=1$, task $T$ will be fully offloaded; otherwise, the data with size $\lambda D$ will be offloaded, the data with size $(1-\lambda)D$ will be processed locally.
In the following, we will demonstrate the local processing part and offloading part, respectively.

\subsubsection{Task $T$ processed locally}
The number of cores of the users is denoted as $n_1$, and the processing capability (i.e., the amount of CPU frequency in cycles/s) of each core assigned for local computing as $f^{l}$, then the power consumption of each core for a user to process data locally is expressed as $p^l=\kappa_1(f^{l})^3$, where $\kappa_1$ is a coefficient reflecting the relationship between processing capability and power consumption at the  end-device side \cite{miettinen2010energy}.

\textbf{Local computing time:} Based on the Amdahl's law \cite{amdahl1967validity}, the local computing time for $(1-\lambda) D$ bits data of the task, which consists of the computing time of the serialized part $t_s^l=c(1-\alpha)(1-\lambda) D/f^l$ and the computing time of the parallelizable part $t_{p}^l=c \alpha (1-\lambda) D/f^l n_1$, can be calculated as
\begin{equation}\label{eq local execution time}
t^l=t_{s}^l+t_{p}^l=\frac{c(1-\lambda)D}{f^{l}} (1-\alpha+\frac{\alpha}{n_1}).
\end{equation}

\textbf{Local energy consumption:} The energy consumption for local computing is formulated as
\begin{equation}\label{eq energy consumption for local processing}
E^l=p^l t_{s}^l+n_1p^l t_{p}^l =\kappa_1 c D (1-\lambda) (f^{l})^2.
\end{equation}

\subsubsection{Task $T$ offloaded to the edge}
The data of task $T$ can be offloaded to the edge through wireless communication links. For the data transmission rate, we use $r$ to denote it. The data transmission rate can be characterized by various wireless transmission models based on Shannon's formula.  For example, Wang \textit{et al.} in \cite{wang2016mobile} model the path loss as $d^{-\vartheta}$, where $d$ denotes the distance from the  end-device to the edge, and $\vartheta$ denotes the path loss exponent. Based on Shannon's formula, when data is offloaded from the  end-device to the edge over the assigned wireless bandwidth $B$, the transmission rate can be expressed as
$r_1 = B{\log _2}(1 + \frac{P_{1} |h|^2 }{\omega_0 d^{\vartheta}})$,
where $P_{1}$ is the transmission power of the  end-device, $h$ is the channel fading coefficient, and $\omega_0$ denotes the white Gaussian noise power.

\textbf{Transmission delay for offloading:} Based on the analysis above, the transmission delay for offloading $\lambda D$ bits of data to the edge can be obtained by
\begin{equation}\label{eq uplink delay}
t^{up}=\frac{\lambda D}{r_1}
\end{equation}

\textbf{Transmission energy consumption for offloading:} Accordingly, the energy consumption of the  end-device for transmitting the offloaded $\lambda D$ bits of data is expressed as
\begin{equation}\label{eq energy consumption of a vehicle for transmitting data}
E^{up}=P_{1}t^{up}=\frac{ \lambda D P_{1}}{r_1}.
\end{equation}

\textbf{Computing time at the edge:} After the $\lambda D$ bits of data is offloaded to the edge, the edge would process the data. Let $n_2$ denote the number of cores assigned for task processing of the edge, $f^e$ denote the processing capability (i.e., the amount of CPU frequency in cycles/s) of each core ($f^e \gg f^{l}$). The power consumption of each core of the edge to process data can be expressed as $p^e=\kappa_2(f^e)^3$, where $\kappa_2$ is a coefficient reflecting the relationship between processing capability and power consumption at the edge  side \cite{miettinen2010energy}. And the computing time for the offloaded $\lambda D$ bits of data, which consists of the computing time of the serialized part $t_s^e=c\lambda(1-\alpha)D/f^e$ and the computing time of the parallelizable part $t_p^e= c\lambda \alpha D /n_2 f^e$, can be formulated as
\begin{equation}\label{eq computing time of RSU for the offloaded data}
t^e=t_s^e+t_p^e=\frac{c\lambda D}{f^e} (1-\alpha+\frac{\alpha}{n_2}).
\end{equation}

\textbf{Energy consumption at the edge:} The energy consumption of the edge for computing the $\lambda D$ bits of data is formulated as
\begin{equation}\label{eq energy consumption of RSU for computing data}
E^e=p^e t_{s}^e+n_2 p^e t_{p}^e =\kappa_2 c D(f^e)^2.
\end{equation}

\subsubsection{Result return}
After the task $T$ has been processed, the result will be returned to the  end-device. Generally, the return process has been neglected in many works since the processing result is usually very tiny \cite{wang2017computation,mao2017stochastic,du2018computation}. As a general model, we still consider the result return process. Let $r_2$ denote the data transmission rate in the result return process, then similar to the offloading data rate, $r_2$ can be formulated as $r_2 = B{\log _2}(1 + \frac{P_{2} |h|^2 }{\omega_0 d^{\vartheta}})$,
where $P_{2}$ is the transmission power of the EN. 

\textbf{Transmission delay for result return:} Based on the analysis above, the transmission delay for $\gamma D$ bits result return can be obtained by
\begin{equation}\label{eq downlink delay}
t^{down}=\frac{\gamma D}{r_2}.
\end{equation}

\textbf{Transmission energy consumption for result return:}  Accordingly, the energy consumption of the EN for transmitting the $\gamma D$-bits of processing result to the  end-device is expressed as
\begin{equation}\label{eq energy consumption of downlink}
E^{down}=P_{2}t^{down}=\frac{\gamma D P_2}{r_2}.
\end{equation}

\subsubsection{Total delay}
Based on the analysis above, the total delay of processing task $T$ is a combination of local computing time, transmission delay for offloading, computing time at the edge, and transmission delay for result return, which is formulated as
\begin{equation}
\label{eq total delay}
t=\min \ \{t^l,t^{up}+t^{e}+t^{down}\}.
\end{equation}
\subsubsection{Total cost}
The total cost of processing task $T$ comes from three aspects, including energy consumption, use of bandwidth resources, and use of computing resources.
For the energy consumption, let $\varrho$ denote the weight coefficient that indicates the energy consumption cost of one unit energy during task computing and transmitting \cite{kim2017dual}, then the energy consumption cost can be formulated as
\begin{equation}
\label{eq total energy cost}
C^{energy}=\varrho (E^l+E^{tr}+E^e+E^{down}).
\end{equation}
For the bandwidth cost, let $p_1$ denote the cost of using per unit of bandwidth per unit of time, the the bandwidth cost can be formulated as
\begin{equation}
\label{eq total bandwidth cost}
C^{comm}=p_1B (t^{up}+t^{down}).
\end{equation}
For the computing cost, let $p_2$ denote the cost of using per unit of processing capability per unit of time, then the computing cost can be formulated as
\begin{equation}
\label{eq total computing cost}
C^{comp}=p_2n_2f^e t^e.
\end{equation}
Therefore, the total cost for processing task $T$ can be expressed as
\begin{equation}
\label{eq total cost}
C=C^{energy} + C^{comm} + C^{comp}.
\end{equation}

\subsubsection{Computing acceleration}
Before the task offloading decision is made, some other QoE requirement such as computing acceleration is also a key consideration. The computing acceleration refers to the speedup of processing a task at the edge when compared with computing it locally.
According to Amdahl's law, the speedup can be obtained if the $(1-\lambda) D$ bits of task data is computed locally as follows,
$S_1=\frac{1}{(1-\alpha)+\frac{\alpha}{n_1}}$.
Similarly, the speedup can be obtained if the $\lambda D$ bits of task data is computed at the edge by the following formula,
$S_2=\frac{1}{(1-\alpha)+\frac{\alpha}{n_2}}$.
However, when task data is offloaded to the edge for processing, the actual latency comes from computing delay and transmission delay. In this circumstance, the actual computing acceleration is expressed as,
\begin{equation}\label{eq actual computing acceleration}
A=\frac{t^l}{t^{up}+t^{e}+t^{down}}.
\end{equation}

\begin{figure}[!htb]
\centering
\includegraphics[width=0.45\textwidth]{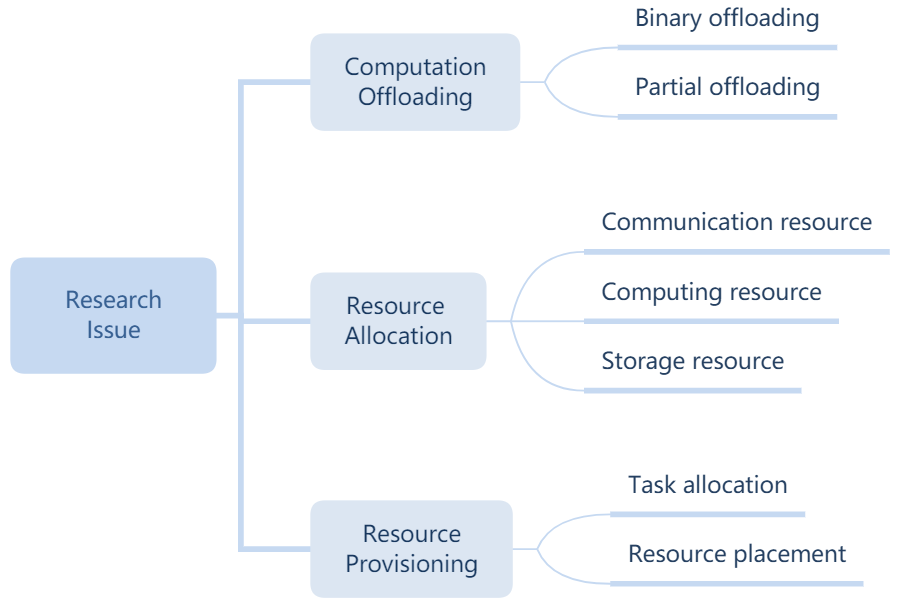}
\caption{ Research issues of resource scheduling in edge computing.}
\label{fig research issues}
\end{figure}

According to the above basic model, many aspects should be considered to achieve the different requirements of both  end-devices and the system for energy consumption, latency, cost, and computing acceleration. The first aspect is to decide the offloading variable $\lambda$, i.e., an efficient \textbf{\textit{computation offloading}}. The second aspect is to decide the variables $B$, $n_1$, $n_2$, $f^l$, $f^e$, i.e., \textbf{\textit{resource allocation}} of the communication and computing resources. The third aspect is to decide the association between tasks and ENs and the placement of computing resources, i.e., \textbf{\textit{resource provisioning}}. 
 The outline of the three research issues is shown in Fig.~\ref{fig research issues} and is described in detail below. 

\subsection{Computation offloading}
\begin{figure}[!htb]
\centering
\includegraphics[width=0.48\textwidth]{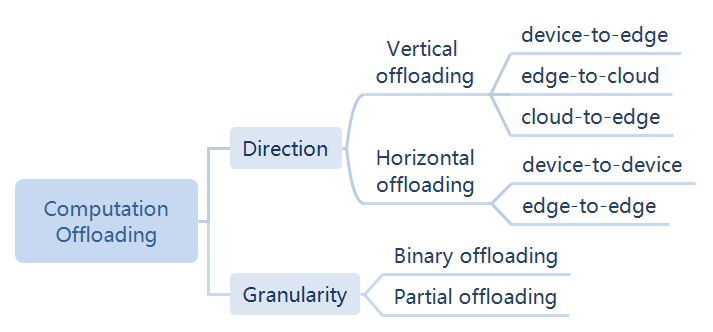}
\caption{A classification of computation offloading for resource scheduling in edge computing.}
\label{fig five offloading directions}
\end{figure}

The computation offloading is a very important research issue for resource scheduling in edge computing, which brings services to the proximity of data source \cite{varghese2020survey}. This subsection reviews the research on this issue. As shown in Fig.~\ref{fig five offloading directions}, the computation offloading can be broadly classified on the base of: \textit{a)} the direction of offloading, namely from device to edge, from edge to cloud, from cloud to edge, from device to device, and from edge to edge, and \textit{b)} the granularity of offloading, namely binary offloading and partial offloading.\\

\noindent \textbf{\textit{B.1. Direction}}

Since  end-devices in the thing layer are mostly resource-constrained, resource-intensive tasks need to be fully or partially offloaded to ENs with powerful computing resources. The computation offloading from  end-devices to ENs compensates for the deficiency of  end-devices in computing performance, storage, and energy efficiency. Also, the computation offloading from  end-devices to ENs can alleviate the overload of the cloud computing center and reduce the delay caused by wireless transmission. For example, video data from surveillance cameras can be offloaded to the EN for low-delay and privacy-protecting analysis and process, compared with being offloaded to the cloud computing center. 
In addition, the upward offloading has also promoted the development of the super low-delay applications such as video services and CAVs. The application data of real-time perception need to be offloaded to ENs for rapid processing, which guides vehicles to take right driving actions. Similarly, if ENs are unable to process the task data offloaded from  end-devices in a timely manner, it can be offloaded to the cloud center. The computation offloading ways both from  end-devices to ENs and from ENs to the cloud center can be referred to as \textit{upward offloading}.

The computation offloading also concentrates on \textit{downward offloading}, which means the offloading from the cloud center to the edge. In the edge-cloud collaboration manner discussed in the last section, this kind of offloading is adopted.
Both upward offloading and downward offloading are regarded as \textit{vertical offloading}.
In addition to vertical offloading, the computation offloading manner also includes \textit{horizontal offloading}. There are two research issues in horizontal offloading. The first one is that  end-devices can offload their resource-intensive tasks to other  end-devices with idle computing resources. The second is that one EN can also migrate their task data to other ENs for processing. Thus, there are in total five different offloading directions in the vertical offloading and horizontal offloading, which will be discussed in the following.

\begin{table*}[]
\centering
\caption{Comparison of Papers Focusing on Computation Offloading.  Acronyms used in this Table: virtual machine (VM). }
\begin{tabular}{|m{1cm}<{\centering}|m{1cm}<{\centering}|m{3.3cm}|m{8.5cm}|}
\hline
\makecell*[c]{\textbf{Gran.}}
&  \makecell*[c]{\textbf{Paper}}
&  \makecell*[c]{\textbf{ Objective}} 
&  \makecell*[c]{\textbf{Research Content}} \\ \hline
\multirow{13}{*}{\rotatebox{90}{Binary Offloading}}  
&    \cite{ding2019code}
&  Delay, energy consumption
&  a) Offloading decision; b) transmission power allocation; c) CPU frequency allocation; 
\\ \cline{2-4}   
&    \cite{feng2019computation}
&  Utility
&  a) Offloading proportion determining; b) power allocation; c) energy harvesting; 
\\ \cline{2-4}                                     &
  \cite{gu2018context}
&  Energy consumption
&  a) Task-destination association; b) offloading decision; 
\\ \cline{2-4}                        &
  \cite{liu2019energy}
&  Energy consumption
&  a) Task-destination association; b) offloading decision; c) task ready time determining;
\\
\cline{2-4}                              
&     \cite{liu2020distributed}
& Utility
&  a) Task-destination association; b) offloading decision; 
\\ 
\cline{2-4}                              
&     \cite{nguyen2019computation}
& Energy consumption
&  a) Transmission power allocation; b) offloading decision;  c) CPU clock allocation; 
\\ 
\cline{2-4}                              
&     \cite{yang2018distributed}
& Latency, energy consumption
&  a) Task-destination association; b) wireless channel allocation;  c) computation capability allocation; 
\\
\cline{2-4}                     
&     \cite{yang2018mobile}
& Energy consumption
&  a) Task-destination association; b) computing  capability allocation; 
\\
\hline

\multirow{11}{*}{\rotatebox{90}{Partial Offloading}} 
&
  \cite{chen2018multi}
&  Revenue 
&  a) Task-destination association; b) offloading workload amount determining; c) energy harvesting; 
\\ \cline{2-4}                                     &
  \cite{kiran2019joint}
&  Delay, energy consumption
&  a) Computing resource allocation; b) offloading ratio determining; 
\\ \cline{2-4}                    &
  \cite{liu2018offloading}
&  Latency
&  a) Task-destination association; b) offloading ratio determining; 
\\ \cline{2-4}         
&
  \cite{ning2018cooperative}
&  Delay
&  a) Task-destination association; b) offloading decision;
\\ \cline{2-4}  
&
  \cite{pan2018energy}
&  Energy consumption
&  a) Offloading data amount determining; b) transmission power allocation;  c) transmission time allocation;
\\
\cline{2-4}                   
&              
  \cite{saleem2018performance}
&  Latency
&  a) Subcarrier assignment; b) offloading ratio determining; c) transmission power allocation;
\\ 
\cline{2-4}                   
&              
  \cite{shu2019multi}
&  Execution time 
&  a) Subtask placement; b) topology/schedules of
the IoT tasks; 
\\
\cline{2-4}                   
&              
  \cite{xu2019joint}
&  Latency, resource utilization
&  a) Task placement; b) VM instance provisioning; 
\\
\hline
\end{tabular}
\label{table offloading strategy}
\end{table*}

\textit{1) Device-to-Edge:} For applications that require powerful capacity or edge data aggregation, various  end-devices will offload their tasks to ENs. This offloading direction is the focus of computation offloading, and it is operated under the things-edge collaboration manner as discussed in Section~\ref{subsec things-edge collaboration}. The offloading from  end-devices to ENs can achieve different QoS and QoE requirements for  end-devices. For example, for reducing the task processing latency, Chen \textit{et al.} in \cite{chen2018computation} considered to offload the computation tasks from MDs to small-cell base stations (BSs) with cloud-like computing and storage capabilities, with the aim of minimizing the long-term system delay. 
For reducing energy consumption, Guo \textit{et al.} in \cite{guo2018anefficient} proposed to offload the computation tasks from MDs to small BSs, and an efficient computation offloading scheme by jointly considering offloading decision-making and resource allocation was proposed, aiming at reducing the energy consumption of MDs. Also, Guo \textit{et al.} in \cite{guo2018anefficient} considered an ultra-dense edge computing network, where MDs' energy consumption is minimized by offloading their tasks to ENs. 
Besides, Jošilo \textit{et al.} in \cite{jovsilo2020computation} proposed a computation offloading scheduling scheme to determine whether to offload the tasks of  end-devices to ENs, aiming to minimize the cost that is a combination of delay and energy consumption.

\textit{2) Edge-to-Cloud:} Generally, the tasks offloaded from  end-devices are processed by computing nodes in the edge layer. The computing nodes, including cloudlets, ENs, BSs, mini data centers, etc., can provide different capacities. If the task data in the edge layer cannot be processed by the computing node in time, they can be further offloaded to the cloud center to achieve a balanced overload. This kind of offloading direction, from the edge to the cloud, is actually operated under the edge-cloud collaboration manner, as discussed in Section~\ref{subsec edge-cloud collaboration}. For example, in the area of CAVs, Zhang \textit{et al.} in \cite{zhang2017optimal} proposed to improve the system utility by utilizing a multi-level offloading scheme among ENs and cloud servers. Also, Zhao \textit{et al.} in \cite{zhao2019computation} considered to jointly optimize the offloading decision and resource allocation by an edge-cloud collaborative offloading approach.

\textit{3) Cloud-to-Edge:}
This kind of offloading direction is also operated under the edge-cloud collaboration manner as discussed in Section~\ref{subsec edge-cloud collaboration}, which brings computation tasks from the distant cloud to the edge to achieve lower data transmission latency, thereby shortening the application response time. The typical issues of the cloud-to-edge offloading mainly include: (i) \textit{video transcoding} on ENs \cite{yoon2016low}; (ii) \textit{application cloning} from cloud to edge to provide users with better QoE \cite{chen2015early}; (iii) \textit{data replication} on the edge \cite{xu2017zenith,lin2007enhancing,gao2003application,luo2020edgevcd}; (iv) \textit{edge discovery and management}, where workloads are offloaded from the cloud to the chosen ENs and the orchestration across multiple ENs is evaluated\cite{amento2016focusstack,liu2016paradrop}.

\textit{4) Edge-to-Edge:}
The edge-to-edge offloading is actually operated under the edge-edge collaboration manner, as discussed in Section~\ref{subsec edge-edge collaboration}, which can alleviate the workload of some overloaded EN by offloading (or migrating) some workloads to a peer. The typical issues of the edge-to-edge offloading mainly include: (i) \textit{task scheduling}, which can  orchestrate the task processing among different ENs \cite{na2018frequency,alameddine2019dynamic,miao2020intelligent,thai2019workload};
(ii) \textit{service migration}, by which services are dynamically migrated across multiple heterogeneous ENs \cite{chen2019dynamic,ma2017efficient};
(iii) \textit{offload forwarding}, in which an EN is regarded as a relay to forward workloads to neighboring ENs \cite{xiao2017qoe}.

\textit{5) Device-to-Device:}
The device-to-device offloading can be operated under both the things-edge collaboration manner and the things-edge-cloud 
collaboration manner, as discussed in Section~\ref{subsec things-edge collaboration} and Section~\ref{subsec things-edge-cloud collaboration}, which offloads the workloads from one  end-device to a peer by making full use of idle resources. For example, Luo \textit{et al.} in \cite{luo2020collaborative} proposed a collaborative task data scheduling scheme in VEC, where the computation tasks of vehicles can be not only processed locally, i.e., offloaded to RSUs, but also can be migrated to other vehicles with idle computing resources.\\

\noindent \textbf{\textit{B.2. Granularity}}

As one of the important research issues in computation offloading, the offloading decision-making problem focuses on whether and how much to offload. Depending on whether the computation task is dividable or not,
the granularity of offloading can be classified into two categories: \textit{a)} binary offloading, and \textit{b)} partial offloading, which will be presented in the following.

\textit{1) Binary Offloading:}
Binary offloading, also known as ``0-1 offloading", means the whole computation task is either processed locally or offloaded to elsewhere. ``0" and ``1" are the indicators of whether the task is offloaded or not. Generally, ``0" means the whole task is processed locally, and ``1" means it is offloaded to elsewhere \cite{hong2019multi,liu2020distributed}.
When the whole task is processed locally, the computing time, energy consumption, and the cost of processing task are determined by the local capacity. When the whole task is offloaded to other nodes to process, the computing time mainly includes task transmission time and task processing time. Similarly, energy consumption mainly includes transmission energy consumption and processing energy consumption. The cost mainly includes transmission cost and processing cost. From this point of view, the factors that affect the offloading performance include wireless channel conditions, wireless bandwidth, and processing capability of the destination node (i.e., the node to which the task is offloaded).
The research on binary offloading involves in the association between tasks and destination nodes \cite{hu2019learning,li2019energy,liu2019dynamic,mazouzi2019dm2,yang2018multivessel}, which refers to the determination of the offloading of a specific task to a destination node, among various tasks and destination nodes.

\textit{2) Partial Offloading:}
Partial offloading allows flexible components/data partitioning, which means that a task can be divided into separated parts \cite{zhou2018uav,yu2020joint,xiao2019task,luo2021minimizing}.
The research on partial offloading is to determine how much and in what way of the whole task can be offloaded to the destination node. Generally, a ratio known as ``offloading ratio" is set to indicate the proportion of offloading part of the task. Partial offloading involves two parts of task processing, the local processing part and the offloading part. Accordingly, the task processing performance is jointly determined by the computing time, energy consumption, and the cost of processing task locally and at the destination side. Actually, in addition to deciding and optimizing the offloading ratio to achieve various QoS requirements, the study of partial offloading also involves in the association between the offloading part of the task and the destination node \cite{liu2019reliability}.

In most existing works, neither binary offloading or partial offloading issues can be addressed alone, and other issues such as\textbf{ \textit{resource allocation}} 
\cite{bi2018computation,chen2018task,elgendy2019resource,gu2018optimal} 
and\textbf{ \textit{resource provisioning} } \cite{bahreini2019energy,kiani2019hierarchical} 
are jointly studied with computation offloading, which will be presented in later sections. To enable readers to grasp basic ideas of computation offloading on both binary offloading and partial offloading, a comparison of papers focusing on this research issue is presented in Table~\ref{table offloading strategy}.

\subsection{Resource Allocation}
As another important research issue in resource scheduling, resource allocation studies how to reasonably and effectively allocate resources in the edge computing system to complete offloading and task processing. 
Generally, the main resources involved in the current research on resource allocation are computing, communication, and storage resources. Computing resources typically refer to CPU cycles and resource blocks (VMs/containers). Communication resources refer to wireless resources including bandwidth, spectrum, power, and link used for data transmission during computation offloading. Storage resources are used to cache computation tasks and popular content (e.g., on-demand video, AR/VR, road surveillance, etc.) to the edge of the network, reducing the service response time and the burden on the network. Some research on resource allocation only focuses on allocating one kind of resource while most research considering the joint resource allocation, which will be elaborated on in the following.

\subsubsection{Single resource}
The existing works involved in the single-resource allocation mainly focus on the allocation of computing  or communication resources. In the computation offloading decision-making problem, many works consider the allocation of communication resources. Like the works in \cite{khalili2019joint} and \cite{kuang2019partial}, both focused on communication resources and studied how to allocate the transmission power during the offloading process, with the goal of minimizing the system's energy consumption. Differently, Li \textit{et al.} in \cite{li2018energyaware} 
studied the channel selection for task offloading. The effect of multi-channel interference on the energy efficiency of task offloading was taken into account. Obviously, the most important thing in the offloading process is the allocation of computing resources. The work in \cite{lyu2018selective} designed the selective offloading scheme for IoT devices, and it studied how to allocate the best EN for offloading tasks to minimize energy consumption. Similarly, Xu \textit{et al.} in \cite{xu2019computation} studied the computation offloading problem for IoT-enabled cloud-edge computing, and they focused on how to allocate the computing resource for tasks to minimize the execution time and energy consumption for MDs.  Also, 
some studies only consider storage resources in terms of caching data \cite{yu2018computation} and caching service \cite{xu2018joint,nikoloudakis2018edge}. Yu \textit{et al.} in \cite{yu2018computation} proposed a collaborative offloading with data caching enhancement strategy to minimize the total delay. Caching services such as databases or libraries on ENs for task execution can effectively reduce the total delay. The study in \cite{xu2018joint} focused on dynamic service caching and task offloading, and proposed an online algorithm based on Lyapunov optimization and Gibbs sampling.

\subsubsection{Computing and communication (CC)}
The offloading process often involves the joint allocation of communication and computing resources. Many existing works have studied this topic \cite{guo2018joint,guo2019adaptive,qian2019noma,wang2018dynamic,wang2019effective,xing2019joint,yang2019efficientresource,zhao2019context,zhao2020mobile,elgendy2020efficient}. Guo \textit{et al.} in \cite{guo2019adaptive} proposed an adaptive resource allocation framework for MEC, which applied the idea of blockchain into the framework design. They formulated an optimization problem for spectrum and block allocation. The study in \cite{qian2019noma} formulated the problem of optimizing the joint allocation of computing resources on ENs and radio resources under the non-orthogonal multiple access (NOMA) protocol and used an efficient layer algorithm to solve it. Likely, to maximize the total revenue, Wang \textit{et al.} in \cite{wang2019effective} studied the optimization problem for bandwidth and computation allocation with the QoS-guaranteed constraint, and they proposed an algorithm based on alternating direction method of multipliers (ADMM) to solve it. Under the transmission protocol of time division multiple access (TDMA), the authors in \cite{xing2019joint} studied how to assign the time and rate of local users for task offloading and how to allocate computation frequency for task execution, aiming to minimize the computation latency. Similarly, the work in \cite{yang2019efficientresource} also adopted TDMA transmission protocol. Millimeter-wave (mmWave) communication as one of the promising transmission protocols was applied in the work \cite{zhao2019context}. This paper formulated the joint beamforming vectors at the users and computation ratios at ENs allocation problem to minimize the system delay, and proposed a penalty dual decomposition technique to solve this optimization problem. 

\subsubsection{Computing, communication, and storage (CCS)}
Many works have considered communication, computing, and storage resources simultaneously in the resource allocation problem \cite{wang2019edgeai,liang2017energy,tan2017virtual,zhou2017resource,wang2017computation}. In recent years, the prevalence of edge intelligence has attracted widespread attention from academia and industry. In the work \cite{wang2019edgeai}, the authors designed an In-Edge AI framework for optimizing computing, communication, and caching allocation. They utilized both deep reinforcement learning and federated learning (FL) techniques to optimize the edge system's performance. Liang \textit{et al.} in \cite{liang2017energy} studied the bandwidth provisioning and content source selection problem by introducing caching and computing functions in MEC. They proposed a decentralized approach based on ADMM to solve it. Likely, the work in \cite{wang2017computation} addressed the optimization problem for joint computation offloading, resource allocation, and content caching, in which computing, spectrum, and caching resources were considered simultaneously.
Particularly, all resources in the study \cite{tan2017virtual} were in the form of virtual resources. The authors formulated a joint virtual resource (including spectrum, caching, and computing) allocation problem, intending to maximize the system's utility. Similarly, the authors in \cite{zhou2017resource} also studied the virtual resource allocation problem in which the communication, computation, and caching resources can be shared among all users. Besides, they presented a distributed algorithm based on ADMM to address the formulated problem.
Moreover, a few research focus on joint communication and storage resource allocation problems \cite{cui2017energy,hao2018energy}. 

A comparison of papers focusing on resource allocation is presented in Table \ref{tabl works on resource allocation}. It can be observed that communication, computing, and storage resources are rarely allocated individually in resource scheduling. Many works combine two or three of them to model and jointly optimize the allocation simultaneously. 

\begin{table*}[!htb]
\centering
\caption{Comparison of Papers Focusing on Resource Allocation.  Acronyms used in this Table: non-dominated sorting genetic algorithm (NSGA), Deep Q-network (DQN), alternating direction method of multipliers (ADMM), federated learning (FL).}
\begin{tabular}{|c|c|c|c|l|l|}
\hline
\textbf{Paper}
&\textbf{ Computing}
&\textbf{ Communication}
&\textbf{ Storage}
&\textbf{\makebox[5cm][c]{Algorithm}}
&\makecell*[c]{\textbf{ Objective}}

\\
\hline

\cite{khalili2019joint}
&\XSolid
&\Checkmark
&\XSolid
&Majorization minimization method &
Energy consumption \\
\hline
\cite{kuang2019partial}
&\XSolid
&\Checkmark
&\XSolid&
Genetic algorithm&
Energy consumption\\
\hline
\cite{li2018energyaware}
&\XSolid
&\Checkmark
&\XSolid&
Auction-based approach&
Energy consumption\\
\hline
\cite{xu2019computation}
&\Checkmark
&\XSolid 
&\XSolid 
&NSGA-III algorithm&
Delay, energy consumption\\
\hline
\cite{yu2018computation}
&\Checkmark
&\XSolid
&\XSolid
&Game-based&
Delay\\
\hline
\cite{xu2018joint}
&\Checkmark
&\XSolid
&\XSolid
&Lyapunov optimization&
Delay\\
\hline
\cite{guo2019adaptive}
&\Checkmark
&\Checkmark
&\XSolid
&DQN
&Performance\\
\hline
\cite{qian2019noma}
&\Checkmark
&\Checkmark
&\XSolid
&Many-to-one matching algorithm&
Cost\\
\hline
\cite{wang2019effective}
&\Checkmark
&\Checkmark
&\XSolid
&ADMM&
Revenue
\\
\hline
\cite{xing2019joint}
&\Checkmark
&\Checkmark
&\XSolid&
Heuristic-based algorithm&
Latency
\\
\hline
\cite{zhao2019context}
&\Checkmark
&\Checkmark
&\XSolid&
Penalty dual decomposition technique&Delay
\\
\hline

\cite{wang2019edgeai}
&\Checkmark
&\Checkmark
&\Checkmark&
DQN, FL&
Performance
\\
\hline

\cite{liang2017energy}&\Checkmark
&\Checkmark
&\Checkmark&
ADMM&
Energy consumption
\\
\hline

\cite{tan2017virtual}&\Checkmark
&\Checkmark
&\Checkmark&
ADMM&
Utility
\\
\hline

\cite{zhou2017resource}&\Checkmark
&\Checkmark
&\Checkmark&
ADMM&
Utility
\\
\hline

\end{tabular}
\label{tabl works on resource allocation}
\end{table*}

\subsection{Resource Provisioning}
Since loads of users' requests vary over time, edge computing systems experience constant fluctuations in workload. These fluctuated workloads may cause problems such as over-provisioning or under-provisioning of edge resources. In the case of over-provisioning, where the resources allocated to some users are greater than the actual load demanded by users, the edge system may be unnecessarily costly. Besides, in under-provisioning, the resources allocated to users for the service are less than the actual load demanded by users, resulting in a poor QoS or even the inability to complete users' tasks. Therefore, allocating the appropriate amount of edge resources to users dynamically to minimize the system cost and meet users' QoS requirement is an important issue. Based on the analysis and summary of current research, the studies on resource provisioning in edge computing can be divided into two categories: \textit{a)} \textit{task allocation}, 
which is a passive resource provisioning from users' perspective. The task allocation problem in edge computing refers to the optimal placement and matching plan between users' tasks and edge resources; \textit{b)} \textit{resource placement}, which is an active resource provisioning from resource providers' perspective. The resource placement mainly includes cloud service decentralization to the edge, optimized deployment of ESs, quantity allocation of edge resources, and virtual edge resource placement issues. In the following, we will elaborate on the two aspects.

\subsubsection{Task allocation}
Yang \textit{et al.} in~\cite{yang2019cloudlet} studied the cloudlet placement and task allocation problem. Then, they formed a mixed integer linear programming (MILP) problem and used the benders decomposition-based approach to solve it. Before task allocation, the authors investigated the resource placement, aiming to calculate the task delay and energy consumption of different ENs. It provides systematic conditions for task allocation. The work in~\cite{breitbach2019context} focused on data management in edge computing, and it presented a multi-layer scheduler considered the various context dimensions of data. In the multi-layer scheduler design, the tasks generated by data are allocated based on the current context and the system state during runtime. Fan \textit{et al.} in~\cite{fan2017deadline} proposed a deadline-oriented task allocation mechanism and formed a task scheduling problem as a multi-dimensional 0-1 knapsack problem. They adopted an efficient task allocation algorithm based on ant colony optimization to increase the system's total profit while satisfying the deadline and resource constraints of the task. 
There are some works on application placement, which focus on assigning tasks from users' applications to the appropriate edge resources for processing~\cite{cao2018performance,mahmud2019quality,mahmud2020profit}. It is essentially a task allocation problem. In~\cite{cao2018performance}, the authors designed a third-party platform responsible for allocating MUs' application tasks to edge resource providers. MUs subscribe to the platform that collects the information of ENs to place tasks on ENs optimally. A programming algorithm was proposed to select the best task placement server from the users' perspective to avoid task migration, thus minimizing the time cost. From the platform's point, the efficient heuristic algorithm is presented to schedule tasks to minimize the total cost. Likely, Mahmud \textit{et al.} in~\cite{mahmud2019quality} proposed a QoE-aware scheme for application placement. The proposed scheme prioritized different tasks of applications and updated the capabilities of ENs according to their current status, thus facilitating optimal task allocation decisions. Later, for the edge-cloud environment, they proposed another application placement policy~\cite{mahmud2020profit}, aiming to maximize the edge system's profit and ensure the user's QoE.

\begin{table*}[!htb]
\centering
\caption{Comparison of Papers Focusing on Resource Provisioning.  Acronyms used in this Table: quality of experience (QoE), quality of service (QoS), mixed integer linear programming (MILP), edge cloud (EC), network function virtualization (NFV).}
\begin{tabular}{|c|l|l|l|l|}
\hline
\makecell*[c]{\textbf{Paper}} 
&\makecell*[c]{\textbf{Research Content}}
&\makecell*[c]{\textbf{ Solution}}
&\makecell*[c]{\textbf{ Objective}}
&\makecell*[c]{\textbf{What's to be scheduled}}

\\
\hline
\cite{yang2019cloudlet}& 
Cloudlet placement and task allocation
&Benders decomposition-based algorithm&
Energy consumption&
Task from users
\\
\hline
\cite{breitbach2019context}&
Data placement and task allocation&Multi-level scheduler&
Latency, overhead&
Data\\
\hline
\cite{fan2017deadline}&
Task allocation&Ant colony optimization&
Profit&
Users' tasks
\\
\hline
\cite{cao2018performance}& 
Application placement&Game model
Cost&
Uses' applications\\
\hline
\cite{mahmud2019quality}& 
Application placement&Separate Fuzzy logic based approaches&
QoE&
Uses' applications\\
\hline
\cite{fan2019cost}& 
Cloudlet placement&Lagrangian heuristic algorithm &
Delay&
Cloudlet
\\
\hline
\cite{santoyo2020network}&
EC placement&MILP mathematical model&
Cost&
EC
\\
\hline
\cite{jin2018cooperative}& 
Data placement&Graph-based iterative algorithm&
Cache hit rate&
Data
\\
\hline
\cite{chiti2019virtual}& 
NFV placement&Matching game&
Delay&
NFV\\
\hline
\cite{choi2019scalable}&
Service placement&Logical fog network&
Resource utilization&
Service
\\
\hline
\cite{maia2019optimized}&
Service placement&Genetic-based algorithm&
QoS&
Service
\\
\hline
\cite{suresh2019fnsched}&
Resource provisioning&Serverless scheduler&
Cost&
CPU cycles
\\
\hline
\cite{aske2018supporting}&
Service provisioning&Adaptive scheduling&
QoS&
Service
\\
\hline
\end{tabular}
\label{tabl works on resource provisioning}
\end{table*}

\subsubsection{Resource placement}
In terms of resource placement, a portion of works focus on how to place ENs~\cite{li2018energy,fan2019cost,meng2019joint,xiao2018heuristic,santoyo2020network}. The location and number of edge services have a crucial impact on both the cost of the edge computing network and users' average latency. The study in \cite{fan2019cost} presented a cost-aware cloudlet placement scheme for MEC, considering the cost of cloudlet deployment and the average latency of users. A Lagrange-based heuristic algorithm was used to achieve sub-optimal solutions, and a workload allocation scheme was designed to minimize the delay between users and cloudlet considering the mobility of users. The edge server placement has raised concerns on the expenditure of deployment and operation, the current backhaul network capacity, and non-technical placement constraints. In \cite{santoyo2020network}, the authors proposed a new framework for edge server placement aiming to reduce the overall costs of deploying and operating edge computing networks. The framework addressed the server placement problem by implementing service placement and optimization strategies.

Notably, there are lots of current research focusing on service placement. On the one hand, some research study decentralized cloud services to the edge \cite{li2019joint,lin2019time,tang2019new,jin2018cooperative,chen2019effective,huang2019latency}. Nowadays, many data-intensive tasks are computed at the edge. If the data required for the task is not stored at the edge, it needs to be downloaded from the cloud, which may cause additional delay. Therefore, it is valuable to study how to decentralize cloud data to the edge. Jin \textit{et al.} in~\cite{jin2018cooperative} proposed an efficient graph-based algorithm for the data placement problem, aiming to maximize the cache hit rate to reduce the task delay. Combining edge computing and cloud computing to place data for scientific workflows to minimize the transmission time across different data centers, the authors in~\cite{lin2019time} proposed a self-adaptive discrete particle swarm optimization (PSO) algorithm for the data placement problem. The proposed algorithm considered the bandwidth, the number of the edge, and the storage capacity of the edge that affect transmission delay. Similarly, Chen \textit{et al.} in~\cite{chen2019effective} also explored the data placement problem for scientific workflows, and they proposed the model based on GA and PSO to solve the problem. 
On the other hand, more works have studied the service or application placement at the edge based on users' requirements~\cite{choi2019scalable,maia2019optimized,gao2019winning,ouyang2018follow,ouyang2019adaptive,poularakis2019joint,yousefpour2019fogplan,chen2019collaborative,goudarzi2020application}. The objective functions and constraints in those works are determined by considering various aspects of the edge computing environment, such as the application (or service) architecture, the edge architecture or the edge-cloud architecture, the network condition, and the network topology. In \cite{choi2019scalable}, the authors proposed a service placement mechanism based on a logical edge network to meet users' needs and the resource constraints of ENs. The proposed service placement mechanism aimed to minimize the number of services placed on ENs to optimize the resource utilization of ENs. The work in \cite{maia2019optimized} studied the load distribution and layout of scalable IoT services, including vertical and horizontal, to minimize the possibility of QoS violations due to edge computing resource constraints. Similarly, the study \cite{yousefpour2019fogplan} introduced the problem of dynamic edge computing service 
placement, which was designed to dynamically deploy IoT services on edge resources to meet QoS requirements such as service delay and bandwidth usage.
At present, the difficulty and trend of this subject are how to place tasks with data dependencies when the service or application is composed of multiple dependent tasks. Usually, in the dependent category, related works modeled their service or application by Directed Acyclic Graph (DAG) \cite{mutichiro2019usage,wu2019efficient,neto2018uloof,wang2020joint,zhao2021offloading}. The placement purpose of their research is to find a group of tasks for scheduling, by which the execution time of service or application and energy consumption of MD become reduced.

Although built on less powerful hardware, edge computing faces similar challenges as cloud computing in effectively managing the hardware resources. Therefore, edge computing also employs virtualization as one of its fundamental technologies. The virtualization technology, no matter in the form of VMs or containers, provides flexible and reliable services for edge computing at a high level. VM placement is a popular research in resource provisioning at the edge, which can be regarded as a process to find the optimal network path to allocate VM. Therefore, the task can be quickly executed, and energy usage can be reduced. Li and Wang \cite{li2018energyawareedge} proposed the method to find out a VM placement scheme that can reduce the total energy consumption and keep the access delay in a reasonable range. In \cite{plachy2016dynamic}, the authors exploited the prediction of users' movement. The prediction is used for dynamic VM placement and to find the most suitable communication path according to expected users' movement. To date, there are several pioneer projects proposed by the industry that aims at building general-purpose edge computing frameworks, including OpenStack \cite{OpenStack2019}, Kubernetes \cite{Kubernetes2020}, and OpenEdge \cite{OpenEdge2019}. Applying container techniques to the edge environment is a natural trend because of the facts of rapid construction, instantiation, and initialization of virtualized instances \cite{tao2019survey}. Morabito \cite{morabito2017virtualization} evaluated the performance of container-based virtualization on IoT devices on the edge. They conducted more practical
experiments on Advanced RISC Machine (ARM)-based IoT end-devices (Raspberry Pi). Performance evaluation on the CPU, memory, disk I/O, and network shows that container-based virtualization can represent an efficient and promising way to enhance the features of edge architectures.
In \cite{zhang2021joint}, the authors found that inter-container communications, and container management consume significant CPU resources by experiments. Then, a joint task scheduling and containerizing scheme are introduced to tackle this problem.
In the past two years, research on resource provisioning based on serverless computing architecture has attracted much attention~\cite{suresh2019fnsched,aske2018supporting,aslanpour2021serverless}. Serverless computing is an emerging paradigm for running user-specified functions on resource providers with infinite scalability. Suresh \textit{et al.} in \cite{suresh2019fnsched} proposed Fnsched,  a novel resource provisioning framework that aims to meet users' performance requirements while minimizing the cost of SPs. Fnsched implemented the autoscale ability by carefully regulating resource usage on each resource scheduler. Besides, the authors in \cite{aske2018supporting} proposed an MPSC framework for serverless computing that supports multiple edge resource providers. MPSC monitored the performance of serverless providers in real-time and dispatched users' application tasks to appropriate resources. 

A comparison of papers focusing on resource provisioning is presented in Table~\ref{tabl works on resource provisioning}. 
 Since the virtualization technology brings high flexibility and resource isolation to the edge, it can be predicted that more research will be devoted to resource provisioning based on container-based or serverless-based edge computing architecture in the future.
\section{Key Techniques and Performance Indicators}
\label{sec techniques}
\begin{figure}[!htb]
\centering
\includegraphics[width=0.48\textwidth]{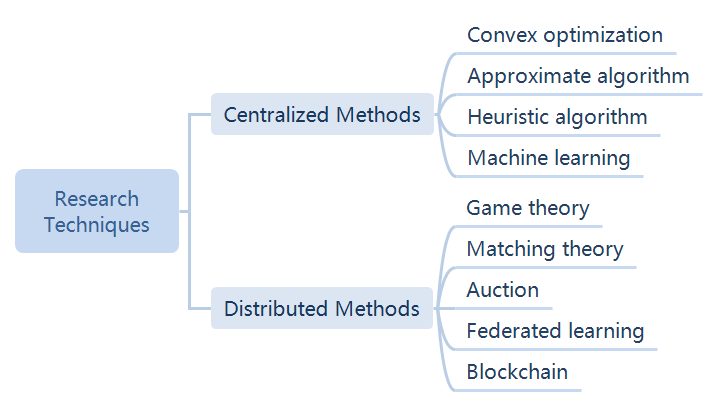}
\caption{Research techniques of resource scheduling in edge computing.}
\label{fig research techniques}
\end{figure}

Advanced scheduling strategies and techniques are indispensable for realizing optimal scheduling of edge computing resources and thus meeting the QoS and QoE requirements of both end-devices and the system. In recent years, many state-of-the-art resource scheduling techniques have emerged. Based on whether a control center is needed to collect global information, resource scheduling can be operated in centralized manner or distributed manner. Generally, centralized methods mainly include convex optimization, approximate algorithm, heuristic algorithm, and machine learning; distributed methods mainly include game theory, matching theory, auction, federated learning (FL), and blockchain, 
as shown in 
Fig.~\ref{fig research techniques}.
In the following, we elaborate on the centralized and distributed resource scheduling methods before summarizing six performance indicators, i.e., latency, energy consumption, cost, utility, profit, and resource utilization.

\subsection{Centralized Methods}
\subsubsection{Convex optimization}
The optimization models developed in the issues of computation offloading, resource allocation, and resource provisioning are typically non-convex or NP-hard problems. A significant portion of studies transform the non-convex problem into a near-convex or convex optimization problem, thus adopting a feasible convex optimization method.
Deng \textit{et al.} in \cite{deng2019parallel} studied the offloading problem under the green and sustainable MEC framework for the IoT system. To minimize the response time, they proposed a DPCOEM algorithm based on the Lyapunov technique and achieve approximately optimal performance. Similarly, some research \cite{chen2018computation,he2019peace,lyu2017optimal,mao2016dynamic,zhang2020dynamic,li2019dynamic} also used Lyapunov technique to solve the optimization problem. Lyapunov optimization, as a stochastic optimization approach, can enable online decision-making while preserving sub-optimal performance. The work \cite{kherraf2019optimized} modeled the problem of resource allocation in MEC as a mixed-integer program. Due to the NP-hardness nature of the formulated problem, the authors proposed a decomposition method to solve it. They decomposed the original problem into two sub-problems, one is the workload assignment and another is the edge node dimensioning. Also, the studies in \cite{saleem2020latency,wang2020optimal} employed the decomposition method to solve the complicated optimization problem. 
The authors in \cite{li2020energy} investigated the computation offloading problem in the UAV scenario, and the formulated non-convex optimization problem was solved using the Dinkelbath algorithm and successive convex approximation (SCA) technique. Similarly, Liu \textit{et al.} \cite{liu2019uav} also used the SCA technique to solve a non-convex optimization problem. The idea of SCA is to iteratively solve a series of convex optimization problems similar to the original non-convex problem, to find a local optimal solution of the original problem. 
Yang \textit{et al.} in \cite{yang2019joint} formulated a non-convex problem for computation offloading and data caching. To solve the problem, they transformed it into a near-convex problem and then designed an algorithm based on ADMM. ADMM is a simple method for solving decomposable convex optimization problems. Using the ADMM algorithm, the original problem can be equivalently decomposed into some solvable sub-problems, which can be solved in parallel. Finally, the solutions of the sub-problems were coordinated to obtain the global solution of the original problem. Besides, the ADMM technique was also utilized in \cite{do2015proximal, zhou2017virtual}.

\textit{Summary}: The main techniques of convex optimization include the Lyapunov technique, decomposition technique, SCA technique, and ADMM technique. In general, techniques based on convex optimization have the following advantages: \textit{a)} mature, and widely used; and \textit{b)} sub-optimal optimization results can be easily obtained. However, the calculations of methods based on these techniques are often complex and challenging to implement in real systems.

\subsubsection{Approximate algorithm}
In addition to the transformation to traditional convex optimization methods, a large number of studies adopt various approximation algorithms to solve the non-convex and NP-hard problems in resource scheduling. For MEC systems, Badri \textit{et al.} in \cite{badri2019energy} built the application placement problem as a multi-stage stochastic programming problem. They adopted a parallel sample averaging approximation (SAA) algorithm to solve this problem and obtained an effective solution. In \cite{meng2019closed}, the computation problem was modeled as an infinite horizon average cost Markov decision process (MDP) process and was approximated to a virtual continuous-time system before a multi-level offloading policy was proposed. The work in \cite{guo2020user} studied the edge-cloud placement problem and described it as a multi-objective optimization problem, which was solved by an approximate method using k-means and hybrid quadratic programming. Lu \textit{et al.} in \cite{lu2019cost} modeled a multi-user resource allocation problem in edge computing and utilized an approximation algorithm for local search to solve the NP-hard problem. The work in \cite{pasteris2019service} studied the problem of maximizing revenue by placing multiple services in an edge system. The authors first proved that the formulated problem is NP-hard and then proposed a deterministic approximation algorithm to solve it.

\textit{Summary}: The basic idea of the approximate algorithm is utilizing the existing approximate methods, such as relaxation, bounded, local search, and dynamic planning techniques, to solve the established NP-hard problems. In general, the approximate algorithm has the following advantages: \textit{a)} simple, flexible, and easy to implement; and \textit{b)} not difficult to design a local search algorithm for most difficult NP-hard problems.
However, the approximation algorithm has some disadvantages: \textit{a)} easy to fall into a local optimum; and \textit{b)} the performance of the solution can not be guaranteed due to randomness.

\subsubsection{Heuristic algorithm}
Nowadays, one of the most popular ways to solve NP-hard problems is utilizing heuristic algorithms including simple heuristics and meta-heuristics. Using principles similar to bionics, heuristic algorithms abstract some phenomena in nature and animals into algorithms to deal with corresponding problems \cite{luo2018optimal}. 
In resource scheduling research, most of the current works utilize greedy algorithms while some works utilize local search algorithms. Huang \textit{et al.} in \cite{huang2019latency} modeled a multi-replica data placement problem for MEC. They analyzed the complexity of the formulated problem and designed a greedy strategy to solve the problem. Similarly, the works in \cite{kiani2019hierarchical,li2019virtual} also employed the greedy idea to solve the NP-hard problem. The study in \cite{meng2019joint} jointly studied the problem of edge server placement and application allocation, and they proposed a heuristic algorithm based on local search to effectively solve the problem. Likely, the local search heuristic algorithm was also used in \cite{zhang2020energy}. Meta-heuristics in heuristics is widely used in various fields, including genetic algorithm, ant colony algorithm, PSO, simulated annealing, and tabu search.
Canali \textit{et al.} in \cite{canali2019gasp} designed a heuristic algorithm based on a genetic algorithm for the service placement problem. There are also some works \cite{peng2019energy,xu2020dynamic,hu2019dynamic,xu2019computation,xu2019energy,guo2018efficient2} utilizing the non-dominated sorting genetic algorithm (NSGA) to solve the formulated multi-objective optimization problem.
Hu \textit{et al.} in \cite{hu2019dynamic} formulated the request scheduling problem as a mixed-integer nonlinear program. The problem was analyzed as a double decision-making problem, and the authors presented an optimization approach based on NSGA to address the problem. Besides, the authors in \cite{mseddi2019joint} proposed a PSO-based heuristic strategy to solve the joint problem of service placement and task provisioning. The study in \cite{wu2020efficient} designed a heuristic algorithm based on tabu search for task scheduling in IoVs. In \cite{huang2019bilevel}, the authors studied the problem of computation offloading and resource allocation and solved the upper-level optimization problem with an ant colony based heuristic algorithm.

\textit{Summary}: The research that utilizes heuristic algorithms to solve NP-hard problems in resource scheduling tends to employ greedy-based and genetic-based algorithms. The simple heuristic algorithm is efficient, but easy to fall into a local optimal solution. The meta-heuristic algorithm has too many parameters, which makes it difficult to reuse the calculation results. Also, it is impossible to adjust those parameters quickly and effectively.

\begin{table*}[]
\caption{Comparison of Papers Using Centralized Methods.  Acronyms used in this Table: markov decision process (MDP), successive convex approximation (SCA), alternating direction method of multipliers (ADMM),non-dominated sorting genetic algorithm (NSGA), Deep Q-network (DQN),Quality of service (QoS), quality of experience (QoE), long short-term memory (LSTM), federated learning (FL).}
\begin{tabular}{|m{0.6cm}<{\centering}|m{0.7cm}<{\centering}|m{2cm}|m{0.8cm}<{\centering}|m{6cm}|m{2.5cm}|m{2.5cm}|}
\hline
\makecell*[c]{\textbf{Tech.}}  
& \makecell*[c]{{\textbf{Paper}}}
& \makecell*[c]{\textbf{ Objective}} 
& \makecell*[c]{\textbf{ Online}}
& \makecell*[c]{\textbf{Method}}
& \makecell*[c]{\textbf{Advantages}}   
& \makecell*[c]{\textbf{Disadvantages}}\\
\hline
\multirow{11}{*}{\rotatebox{90}{Convex optimization}}
&\cite{deng2019parallel}
& Response time 
&\XSolid
& a) Use Lyapunov technique to decompose the formulated problem to be a convex optimization; b) Proposed a DPCOEM algorithm to solve the problem. 
& \multirow{11}{*}{\makecell*[c]{a) Mature and widely \\used;\\ b) Near-optimal \\results can be \\easily obtained.}}
& \multirow{11}{*}{\makecell*[c]{a) High complexity; \\b) Poor practicality.}} \\
\cline{2-5} 
&\cite{kherraf2019optimized}
&Cost 
&\XSolid
&a) Divide the formulated problem into two sub-problems; b) Propose a trade-off approach to solve it. 
&
&\\
\cline{2-5} 
&\cite{li2020energy}
&Energy efficiency 
&\XSolid
&a) Decompose the problem into sub-problems; b) Use the Dinkelbath algorithm and SCA technique to solve it
&
&\\
\cline{2-5} 
&\cite{yang2019joint}
&Execution delay 
&\XSolid
&a) Use McCormick envelopes to transformed the problem into a near-convex one; b) Designed an algorithm based on ADMM to achieve near optimal results.
&
&\\
\hline

\multirow{11}{*}{\rotatebox{90}{Approximate algorithm}}
&\cite{badri2019energy}          
&QoS    
&\XSolid
& a) Use a sample averaging approximation algorithm to solve muti-stage stochastic programs; b) Design a fast parallel greedy algorithm to solve application placement. 
& \multirow{11}{*}{\makecell*[c]{a) Simple, flexible \\and easy to implement;\\ b) Easy to design a \\local search algorithm.}} 
& \multirow{11}{*}{\makecell*[c]{a) Easy to fall into \\a local optimum;\\ b) The performance \\of the solution \\can not be \\guaranteed.}} \\
\cline{2-5} 
&\cite{guo2020user}
&Service delay 
&\XSolid
&a) Prove the formulated problem is NP-hard; b) Propose an approximate approach with k-means and hybrid quadratic programming. 
&
&\\
\cline{2-5} 
&\cite{lu2019cost}
&Cost 
&\Checkmark
&a) From a simple case to a complicated case; b) Prove the formulated problem is NP-hard; c) Propose an approximation algorithm for local search.
&
&\\
\cline{2-5} 
&\cite{pasteris2019service}
&Revenue 
&\Checkmark
&a) Prove the formulated problem is NP-hard; Propose a deterministic approximation algorithm to solve it.
&
&\\
\hline

\multirow{10}{*}{\rotatebox{90}{Heuristic algorithm}}
&\cite{huang2019latency}
&Latency 
&\XSolid
& a) Prove the problem is NP-hard; b) design a greedy-based heuristic algorithm to address it.  
& \multirow{10}{*}{\makecell*[c]{a) Efficient; \\b) Obtain the optimal\\ solution quickly}} 
& \multirow{10}{*}{\makecell*[c]{a) Easy to fall into \\the local optimal\\ solution; \\b) Too many \\parameters.}} \\
\cline{2-5} 
&\cite{meng2019joint}
&Service cost
&\XSolid
&a) Prove the formulated problem is NP-hard; b) Propose SPAC based on local research. 
&
&\\
\cline{2-5}
&\cite{canali2019gasp}
&Latency 
&\XSolid
&a) Prove the formulated problem is NP-hard; Propose a scalable heuristic approach based on genetic algorithm.
&
&\\
\cline{2-5} 
&\cite{hu2019dynamic}
&Latency 
&\XSolid
&a) Analyze the problem as a double decision-making problem; b) Propose an heuristic approach based on NSGA.
&
&\\
\hline

\multirow{10}{*}{\rotatebox{90}{Machine learning}}
&\cite{qiu2019online}
&Performance  
&\Checkmark
& a) Formulate the offloading problem as an MDP; b) design a DQN-based offloading policy.  
& \multirow{10}{*}{\makecell*[c]{a) Strong parallel \\ processing capability; \\b) Strong distributed \\storage and learning \\capabilities;\\ 3) Approximate the \\ complex nonlinear \\relationship.}} 
& \multirow{10}{*}{\makecell*[c]{a) Require a large \\number of \\parameters; \\b) A black-box \\process; \\c) Long learning \\time.}} \\
\cline{2-5} 
&\cite{ning2019deep}
&QoE
&\Checkmark
&a) Divide the original problem into two sub-problems; b) Develop a two-side matching scheme and a DQN approach to schedule requests. 
&
&\\
\cline{2-5}
&\cite{lu2020optimization}
&Performance
&\Checkmark
& a) Propose a DQN algorithm to solve the offloading problem; b) use LSTM network layer and candidate network to improve DQN algorithm.
&
&\\

\cline{2-5} 
&\cite{shen2019computation}
&Utility 
&\Checkmark 
&a) Prove the formulated problem is NP-hard; b) Design an offloading method based on DQN and FL.
&
&\\
\hline

\end{tabular}
\label{table centralized methods}
\end{table*}

\subsubsection{Machine learning}
In recent years, advanced AI techniques have been applied in various fields due to the development of machine learning, such as deep learning and reinforcement learning techniques. In the research on resource scheduling for edge computing, traditional methods (e.g., convex optimization and approximation algorithms) are usually static solutions to complex optimization problems. They cannot achieve optimal decisions based on dynamic environments. Generally, the interaction with the edge environment during resource scheduling can be modeled as an MDP problem, which can be effectively solved by the reinforcement learning technique.
Therefore, many studies utilize reinforcement and deep learning methods for resource scheduling problem in edge computing. 
In \cite{qiu2019online}, the authors modeled the online offloading problem as an MDP and proposed a deep Q-network (DQN) technique to accommodate dynamic environments and solve the problem.
Ning \textit{et al.} in \cite{ning2019deep} utilized the DQN technique to design an intelligent scheduling approach for VEC. Similarly, the works in \cite{liu2020resource,wang2019computation,zhang2019deep}, \cite{xiong2020resource} and \cite{zhai2020toward} respectively studied the computation offloading, resource allocation, and request scheduling problems of IoT users, and all utilized the DQN technique to learn the optimal strategy. Lu \textit{et al.} in \cite{lu2020optimization} utilized the LSTM network layer and candidate network combined with the actual edge computing environment to improve the DQN algorithm and achieve better performance. The work in \cite{shen2019computation} studied the computation offloading optimization problem and proved it is NP-hard before proposing an offloading algorithm based on DQN and FL. 
Besides, the work in \cite{yu2017computation} described the offloading decision problem as a multi-label classification problem and utilized a deep supervised learning technique.
Chen \textit{et al.} in \cite{chen2020resource} proposed a novel prediction-enabled feedback control with reinforcement learning based resource allocation method, which effectively obtain adaptive and efficient resource allocation for cloud-based software services.

\textit{Summary}: 
Generally, the machine learning technique used for resource scheduling in edge computing has the following advantages: \textit{a)} strong parallel processing capability; \textit{b)} strong distributed storage and learning capability; and \textit{c)} has the function of associative memory and can fully approximate the complex nonlinear relationship. However, it also has the following disadvantages:  \textit{a)} require a large number of parameters; \textit{b)} a black-box process, and the learning process cannot be observed, and the output results are difficult to interpret, which will affect the credibility and acceptability of the results; and \textit{c)} long learning time, and may fall into a local optimal solution or may not even achieve the learning purpose.

\subsection{Distributed Methods}

\subsubsection{Game Theory}
Game theory is a powerful framework to analyze the interactions among entities that act for their self-interests with low complexity \cite{lasaulce2011game}. In a game, all players are rational and aware that their interests are affected by others and also affect others. All players can change their actions in response to others' actions to maximize their own interests. Li \textit{et al.} \cite{li2019cooperative} proposed a game-theoretic scheme to optimize the offloading strategy considering computing resource and bandwidth to minimize the system cost. Liu \textit{et al.} \cite{liu2017price} formulated a Stackelberg game to model the interactions between ENs and users, where the EN determines the price at which services are provided to maximize its revenue, and users make offloading decisions based on the price to minimize their own costs. 
Also, Ranadheera \textit{et al.} \cite{ranadheera2018computation}
developed a distributed mechanism for computation offloading by utilizing a minority game-based method, aiming to guarantee users' QoE requirement for latency and energy-efficient activation of servers.
Similarly, some research \cite{zhang2018joint,wang2018energy,asheralieva2019hierarchical,bai2020risk} also utilized game theory to analyze and solve the resource scheduling problem in edge computing. Besides, some solutions combine game theory with other techniques. For example, Meng \textit{et al.} \cite{meng2019fault} proposed a game-theoretic based resource allocation mechanism to optimally allocate resources for each component task of a mobile application. They combined the mechanism with a reverse-auction based allocation mechanism and a  Partial Critical Path (PCP) strategy.
Zhan \textit{et al.} in \cite{zhan2020deep} proposed a computation offloading game framework that does not need information of network bandwidth and preference. To obtain the optimal offloading decision for a maximal utility in terms of processing time and energy consumption, an MDP and a policy gradient based deep reinforcement learning (DRL) are utilized to solve the problem. 
Zhang \textit{et al.} \cite{zhang2017data} proposed a coalitional game-based method to analyze the data offloading from MDs to MEC servers, aiming to improve bandwidth efficiency and user latency, and gain the payoff of MEC servers. To stimulate the offloading, the authors utilized a pricing mechanism to combined with the coalitional game-based method.

\textit{Summary}: The basic idea of a game theory-based distributed method is to regard each user in the game as a player. The best response decision is made through a collaborative or non-collaborative manner among players to gain their best interests. All those game theory-based methods need to prove the existence of Nash Equilibrium, where a mutually satisfactory solution among users is obtained, and no user is willing to change its decision unilaterally.
Generally, the game theory-based method has the following advantages: \textit{a)} simple, flexible, and easy to implement; and \textit{b)} practical and rational for the participants. However, it also has the following disadvantages: \textit{a)} the mutually satisfactory solution may not be the global optimal solution; and \textit{b)} continuous iteration to achieve the Nash Equilibrium.

\begin{table*}[]
\centering
\caption{Comparison of Papers Using Distributed Methods.  Acronyms used in this Table: markov decision process (MDP), deep reinforcement learning (DRL), non-dominated sorting genetic algorithm (NSGA), vehicular edge computing (VEC), mobile device (MD), edge node (EN), federated learning (FL). }
\begin{tabular}{|m{0.6cm}<{\centering}|m{0.7cm}<{\centering}|m{2cm}|m{0.8cm}<{\centering}|m{6cm}|m{2.5cm}|m{2.5cm}|}
\hline
\makecell*[c]{\textbf{Tech.}}  
& \makecell*[c]{\textbf{Paper}} 
& \makecell*[c]{\textbf{ Objective}}
& \makecell*[c]{\textbf{ Online}}
& \makecell*[c]{\textbf{Method}} 
& \makecell*[c]{\textbf{Advantages}}        
& \makecell*[c]{\textbf{Disadvantages}}\\
\hline
\multirow{10}{*}{\rotatebox{90}{{Game Theory}}} 
&\cite{li2019cooperative}           
&Cost
&\XSolid
& a) The formulated problem is decoupled into resource allocation and offloading decision-making problems;
b) The offloading decisions are obtained via potential game;
c) The resource allocation is achieved by using the Lagrange multiplier.
& \multirow{10}{*}{\rotatebox{0}{\makecell*[c]{a) Simple, flexible\\ and easy to \\implement;\\
b) Practical and \\rational strategy \\for the \\ participants.} }}
& \multirow{10}{*}{\rotatebox{0}{\makecell*[c]{
a) The mutually  \\satisfactory \\solution may \\not the  global \\optimal  solution;\\
 b) Continuous \\iteration to \\achieve the \\Nash  Equilibrium.}
 }} \\
\cline{2-5} 
&\cite{liu2017price} 
& Revenue, cost 
&\XSolid
& Depending on the edge node’s knowledge of the network information, developed the uniform and differentiated pricing algorithms. 
&
&\\
\cline{2-5} 
&\cite{ranadheera2018computation}
& Energy efficiency
&\Checkmark
&A distributed learning algorithm to solve server mode selection problem
&
&\\
\cline{2-5} 
&\cite{zhan2020deep}
&Utility 
&\Checkmark
& a) Formulate the problem as a partially observable MDP; b) Solve it by a policy gradient DRL based approach.
&
&\\
\hline

\multirow{11}{*}{\rotatebox{90}{Matching Theory}}
&\cite{pham2018decentralized}          
&Overhead
&\XSolid
& a) Users make the offloading decisions;
b) Approximate the inter-cell interference and find the transmit power of offloading users using a bisection method.  
& \multirow{11}{*}{\rotatebox{0}{\makecell*[c]{
a) Effective in  \\high dynamic \\networks;\\ 
b) Extendable, \\decentralized, and  \\practical solutions 
\\for some  complex \\networks.}}
}
& \multirow{11}{*}{\rotatebox{0}{\makecell*[c]{
a) Generally used \\to solve binary \\ offloading\\  problem; \\ b) Ineffective in \\solving partial\\ offloading \\problem. }
}} \\
\cline{2-5} 
&\cite{gu2019task}
&Delay 
&\XSolid
&a) Formulate the task assignment problem in VEC as a matching game;
b) Propose two methods, one is one-to-many matching method and another is a heuristic swap-matching method. 
&
&\\
\cline{2-5} 
&\cite{liao2019learning}
&Throughput
&\Checkmark
& Propose a learning-based channel selection framework by leveraging the combined power of machine
learning, Lyapunov optimization, and matching theory.
&
&\\
\hline

\multirow{11}{*}{\rotatebox{90}{Auction}} 
&\cite{he2019truthful}           
&Welfare  
&\Checkmark    
& a) Propose a VCG-based offline optimal auction Mechanism;
b) Propose a Myerson Theorem-based allocation
rule of online truthful auction.  
& \multirow{11}{*}{\rotatebox{0}{\makecell*[c]{
a) Economic \\efficiency to \\achieve a trade-off\\ between requests \\and services;\\ b) Practical in \\real scenarios.}
}} 
& \multirow{11}{*}{\rotatebox{0}{\makecell*[c]{
a) The solution \\may not be the \\global optimal\\ solution;\\ b) Extra overhead \\will be induced \\ since a third \\trusted party \\is needed.}
}} \\
\cline{2-5} 
&\cite{sun2018double} 
&Successful trades
&\XSolid
&a) Propose a breakeven-based double auction (BDA);
 b) Propose a more efficient dynamic pricing 
based double auction (DPDA). 
&
&\\
\cline{2-5}
&\cite{li2019online}
&Welfare
&\Checkmark
& a) Proposed a primal-dual framework based online auction. b) Schedule transmission and computing times, and optimally allocate communication and computing resources;
&
&\\

\cline{2-5} 
& \cite{jiao2018social}
&Energy consumption
&\XSolid
&a) Determine the MD user classification and priority;
b) Proposed a reverse auction-based offloading algorithm.
&
&\\
\hline

\multirow{10}{*}{\rotatebox{90}{Federated Learning}} 
&\cite{ren2019federated}          
&Utility   
&\XSolid
& a) Multiple DRL agents are deployed on multiple ENs to indicate the decisions of the IoT devices; b) FL is used to train DRL agents in a distributed fashion.  
& \multirow{10}{*}{\rotatebox{0}{\makecell*[c]{
a) Privacy-\\protected;\\
b) Reduce the \\burden of \\wireless channel;\\
c) Low overhead \\of learning.}
}} 
& \multirow{10}{*}{\rotatebox{0}{\makecell*[c]{
a) Involve in\\ multiple devices;\\
b) Vulnerable to \\malicious attacks.}
}} \\
\cline{2-5} 
&\cite{wang2019edgeai}
&Utility
&\Checkmark
&a) Integrate the DRL and FL methods with edge computing system;
b) Exchange the training model parameters among  end-devices and servers in a collaborative way. 
&
&\\

\cline{2-5} 
& \cite{qian2019privacy}
&Privacy, service demands
&\XSolid
&a) Model the problem of whether service is placed on edge node or not as a 0-1 problem;
b)Propose a hybrid algorithm combining a distributed FL method and a centralized greedy algorithm.
&
&\\
\hline

\multirow{6}{*}{\rotatebox{90}{Blockchain} }
& \cite{xiong2018mobile}           
& Profit   
&\XSolid
& a) A prototype of an edge computing system for mobile blockchain; b) A pricing schemes.  
& \multirow{6}{*}{\rotatebox{0}{\makecell*[c]{
a) Maintain data \\security;\\ b) Maintain data \\ integrity}
} }
& \multirow{6}{*}{\rotatebox{0}{\makecell*[c]{
a) Relatively high \\latency; \\b) Involve in \\ multiple devices.}
}} \\
\cline{2-5} 
& \cite{xu2019blockchain} 
& Latency
&\XSolid
&a) blockchain-based framework is designed degrade the
data loss possibility; 
b) NSGA-III is leveraged to acquire
the balanced offloading strategies;
&
&\\

\cline{2-5} 
&\cite{xiao2020edgeabc}
&Profit
&\XSolid
&a) subtask-virtual machine mapping strategy;
b) stack cache supplement mechanism;
&
&\\
\hline

\end{tabular}
\label{table distributed methods}
\end{table*}

\subsubsection{Matching theory}
The matching theory is a sub-field of economics, which is a promising concept in distributed resource management and scheduling. Besides, the matching theory provides distributed self-organizing solutions to resource scheduling problems with low complexity. 
In matching theory-based resource scheduling, each agent (such as an EN, a radio resource, or a transmitter node) sorts the others and allocates resources using a preference relation. Generally, a match is defined as: for a given graph $G=(V,E)$, a match of the graph $M$ is a sub-graph of $G$ that consists of a portion of vertexes and edges of the original graph $G$. And there are no common vertex and no adjacent edge in the sub-graph. 
A vertex has at most one edge in a matching graph, and if a vertex has one edge, this vertex is called a matched vertex.
Gu \textit{et al.} \cite{gu2018context}
studied the problem of how to efficiently assign computing tasks to reduce energy consumption in the edge computing system under the constraints of the computing capacity of both MDs and ENs, wireless channel conditions, and delay. In this regard, this paper utilized a one-to-many matching theory for modeling and analysis, and proposed a heuristic swap-matching based algorithm to solve the task assignment problem.
Pham \textit{et al.} \cite{pham2018decentralized} proposed two matching algorithms to solve the computation offloading decision problem and joint resource allocation problem, aiming to minimize the system-wide computation overhead.
Similarly, the study in 
\cite{xiao2019task,gu2019task,liao2019learning,chiti2019virtual} also utilized matching theory-based methods to solve resource scheduling problems in edge computing.

\textit{Summary}: Matching theory is a strong tool for analyzing the mutually and dynamic beneficial relations between users and SPs \cite{gu2015matching,zhou2018mobile}.  
Generally, the matching theory-based method has the following advantages: \textit{a)} effective in high dynamic networks; and \textit{b)} extendable, decentralized, and practical for some complex networks. However, since it is generally used to solve binary offloading problems, it is not very appropriate in solving partial offloading problems.

\subsubsection{Auction}
Auction is inherited from economics and is widely used for resource management and scheduling problems. In an auction mechanism framework for resource scheduling, the entities with tasks to be processed act as bidders, and the entities providing task processing service act as sellers. A trusted entity acts as a third auctioneer to administrate trading and makes online decisions. To understand the auction concept easily, we take the work in \cite{zhang2019near} as an example. IoT devices first published their computation tasks and the corresponding rewards to the edge computing system. Then, the MDs providing computing services analyzed the rewards they can obtain through computing tasks and submitted their bids to the system. Finally, the system assigns the task to the MD who submitted the highest bids.  
The auction-based resource scheduling technique can provide a polynomial complexity solution, which has been verified to achieve near-optimal performance.
He \textit{et al.} in \cite{he2019truthful} considered regarding the resourceful MDs as collaborative nodes to process tasks offloaded from  end-devices. And an online auction-based incentive mechanism is proposed to maximize the long-term system welfare.  
Sun \textit{et al.} in \cite{sun2018double} investigated joint resource allocation and network economics in edge computing. They proposed two double auction schemes with dynamic pricing in MEC to maximize the number of successful trades, one is called breakeven-based double auction (BDA), and another is called dynamic pricing based double auction (DPDA). 
Li \textit{et al.} in \cite{li2019online} integrated time scheduling, resource allocation, and task executor selection for collaborative task offloading, and proposed an online auction mechanism based on primal-dual optimization
framework to maximize the social welfare.
Also, the work in \cite{jiao2018social} proposed a reverse auction theory-based method to solve the 0-1 nonlinear integer programming optimization problem to decide the offloading target channel.
Similarly, the research in \cite {li2018energyaware,jin2015auction} also utilized the auction-based method to solve resource scheduling problem in edge computing.

\textit{Summary}: Like the game theory-based method, in an auction-based resource scheduling framework, both SPs and users try to maximize their own welfare.
Generally, the matching theory-based method has the following advantages: \textit{a)} economic efficiency to achieve a trade-off between requests and services; and \textit{b)} practical in real scenarios. However, it also has the following drawbacks: \textit{a)} the solution may not be the global optimal solution; and \textit{b)} extra third trusted party for auction management may induce extra overhead.

\subsubsection{Federated learning}
FL, also known as collaborative learning, is a machine learning technique that can train resource scheduling algorithm on multiple distributed edge devices or servers that do not exchange local data samples \cite{konevcny2015Federated}. FL is a distributed machine learning algorithm, which not only takes the advantages of machine learning in solving dynamic resource scheduling problems, but also develops and improves it. 
In this regard, Ren \textit{et al.} in \cite{ren2019federated}
studied the computation offloading problem for IoT devices in an energy harvesting scenario. To jointly allocate communication and computing resources during the offloading process, DRL agents are deployed in IoT devices to guide them to make offloading decisions. Meanwhile, to make the DRL-based algorithm feasible and reduce the transmission overhead between IoT devices and servers, the FL method is adopted to train DRL agents in a distributed manner.
Also, to jointly allocate communication, computing, and storage resources in edge computing,  the authors in \cite{wang2019edgeai} integrated the DRL method and FL method in edge computing and proposed an In-Edge-AI framework, where the parameters of the training model are exchanged between  end-devices and edge node to better optimize the resource scheduling model.
Besides, Qian \textit{et al.} in \cite{qian2019privacy} combined the FL method with a centralized greedy algorithm to address the problem of service placement with privacy-awareness in the edge computing system. 

\textit{Summary}: Compared with the traditional centralized machine learning algorithm, FL has the following advantages: \textit{a)} since the training process is carried out on distributed devices, there is no need to upload local data to the dedicated server for centralized training, which can protect the user privacy and reduce the data transmission burden of wireless channels; \textit{b)} users only upload the parameters of their own training models, and the synthesized parameters from multiple devices are fed back to users, which can effectively reduce the individual training time. However, it also has the following disadvantages: \textit{a)} involves in multiple devices; and \textit{b)} is vulnerable to malicious attacks.
The FL method for resource scheduling in edge computing is a new method, and we look forward to more works in the future.

\subsubsection{Blockchain}
Blockchain technology, as an emerging decentralized security system, has attracted more and more attention due to its unique functions such as decentralization, non-tampering, irreversible and traceable, and has been applied in many applications, such as bitcoin, smart grid, and IoT \cite{luong2018optimal,jiao2018social}. 
The introduction of blockchain technology into edge computing can ensure the integrity of resource transaction data and the SP's profits.
There are several works considering integrating the blockchain technology into edge computing \cite{xiong2018mobile,xu2019blockchain,xiao2020edgeabc,huang2021resource}.
To manage edge computing resources effectively, the work in \cite{xiong2018mobile} introduced a novel concept of edge computing for mobile blockchain and presented a prototype for IoT blockchain mining tasks offloading.
Xu \textit{et al.} in \cite{xu2019blockchain} proposed BCD, a blockchain-based computation offloading method in edge computing. The proposed method can address the unequal resource distribution problem and ensure QoS requirements of users with an offloading strategy that preserves data integrity and balance.
Also, to ensure the integrity of resource transaction data and SPs' profits, Xiao \textit{et al.} in \cite{xiao2020edgeabc} proposed an emerging IoT architecture, name EdgeABC, where the computation offloading algorithm is implemented on the blockchain in the form of smart contracts.

\textit{Summary}: The blockchain-based method has the following advantages: \textit{a)} can maintain data security; and \textit{b)} can maintain data integrity. However, it also has the following disadvantages: \textit{a)} has relatively high latency; and \textit{b)} involves in multiple devices. 
The blockchain-based resource scheduling method in edge computing is also a new method, we expect more future works dedicated to this direction.

From the above analysis, since centralized methods need to collect global information from users, it can obtain a better optimal solution and incur more overhead than distributed methods. Differently, distributed methods are more simple, flexible, easy-implement, and adaptive to a dynamic environment than centralized methods.
We summarize centralized and distributed methods in  Tables \ref{table centralized methods} and \ref{table distributed methods}, respectively.

\subsection{Performance Indicators}
\subsubsection{Latency}
From the objectives designed in current research (Table \ref{table offloading strategy}-Table \ref{table distributed methods}), we find that latency is a key performance indicator that affects users' QoE. For delay-sensitive applications, designing a resource scheduling algorithm to reduce latency is one of the main focuses. 
Since the computing, communication, and storage resources in the edge system are limited, if multiple delay-sensitive task requests are sent to the edge simultaneously, not only the latency requirements should be considered but also the constraints of resource capacity and energy consumption should be weighed, which would form a complex optimization problem. Generally, the latency of a task in resource scheduling consists of: \textit{a)} local computing time; \textit{b)} transmission time for task offloading; \textit{c)} processing time at the edge or cloud; and \textit{d)} transmission time for result return. The idea of current research is generally establishing a delay model for specific application scenarios, and formulating an optimization problem by considering various constraints to reduce latency, before solving it by different algorithms.

\subsubsection{Energy Consumption}
Energy consumption is an important performance indicator for users' QoE in edge computing system, especially for small smart devices. The energy consumption in the research of resource scheduling in edge computing mainly consists of: \textit{a)} the energy consumption for local computing; \textit{b)} the energy consumption for offloading; \textit{c)} the energy consumption for processing tasks at the edge or cloud; and \textit{d)} the energy consumption for transmitting result back. Many works just aim to reduce energy consumption \cite{bai2019energy,dai2018jointoffloading,zhang2020energy} while some works aim to reduce latency and energy consumption simultaneously \cite{chen2019efficient,yang2019efficientresource,chen2018task,wang2020agent,yu2020joint}. 
Besides, there are also some works considering  end-devices have the function of energy harvesting and wireless charging during the energy consumption minimization \cite{bi2018computation,chen2018multi,feng2019computation,li2019dynamic}.
\subsubsection{Cost}
Research on minimizing the cost of the edge computing system as a performance indicator is generally a comprehensive performance indicator established under satisfying user service quality. As described in Section \ref{labc}, when the task is offloaded, its costs include the energy cost (for transmission and processing tasks), the cost for using communication channels for transmission, and the cost for processing tasks at the edge. The current research generally seeks the best solution by establishing different cost models with the objective of minimizing the cost \cite{alam2019edge,liu2020resource,ding2019code}.

\subsubsection{Utility}
The concept of utility in edge computing refers to the satisfaction users obtain 
under a certain resource scheduling scheme. And the utility is generally represented by the utility function. According to different objectives, the utility function is represented and mathematically transformed by different service quality parameters, such as data transmission rate, delay, energy consumption, and cost. The mathematical transformation mainly includes reciprocal, logarithm, and weighted summation.
Finally, effective optimization algorithms are designed to maximize the utility \cite{dai2018jointoffloading,dinh2018learning,xiao2019task,li2020joint,zhao2019computation}.
\subsubsection{Profit}
The profit is generally measured from the perspective of edge SPs when deploying, allocating, and scheduling edge resources for users. The obtained profit is calculated by subtracting the SPs' operating costs from users' payment. Under the condition of satisfying the users' QoS, a profit maximization problem is generally developed before some marvelous solutions (such as game theory, matching theory, and auction) is proposed \cite{mahmud2020profit,chen2020stackelberg}. 
Similarly to profit maximization problem, some works also aim to maximize the welfare of society in edge computing system \cite{li2019online,he2019truthful,jiao2018social}.
\subsubsection{Resource Utilization}
Resource utilization is also measured by edge resource providers. Since the resources in edge are limited compared to that in cloud, the utilization of edge resources becomes particularly important with the increasing users. A proper resource scheduling strategy can make full advantage of edge resources and meet users' requirements simultaneously. Existing works typically aim to maximize resource utilization, which is defined as the ratio of the resource usage volume to the total resource volume \cite{shi2019share,xiong2020resource,ullah2020task,zheng2020task}.

\begin{figure*}[!htb]
\centering
\includegraphics[width=0.55\textwidth]{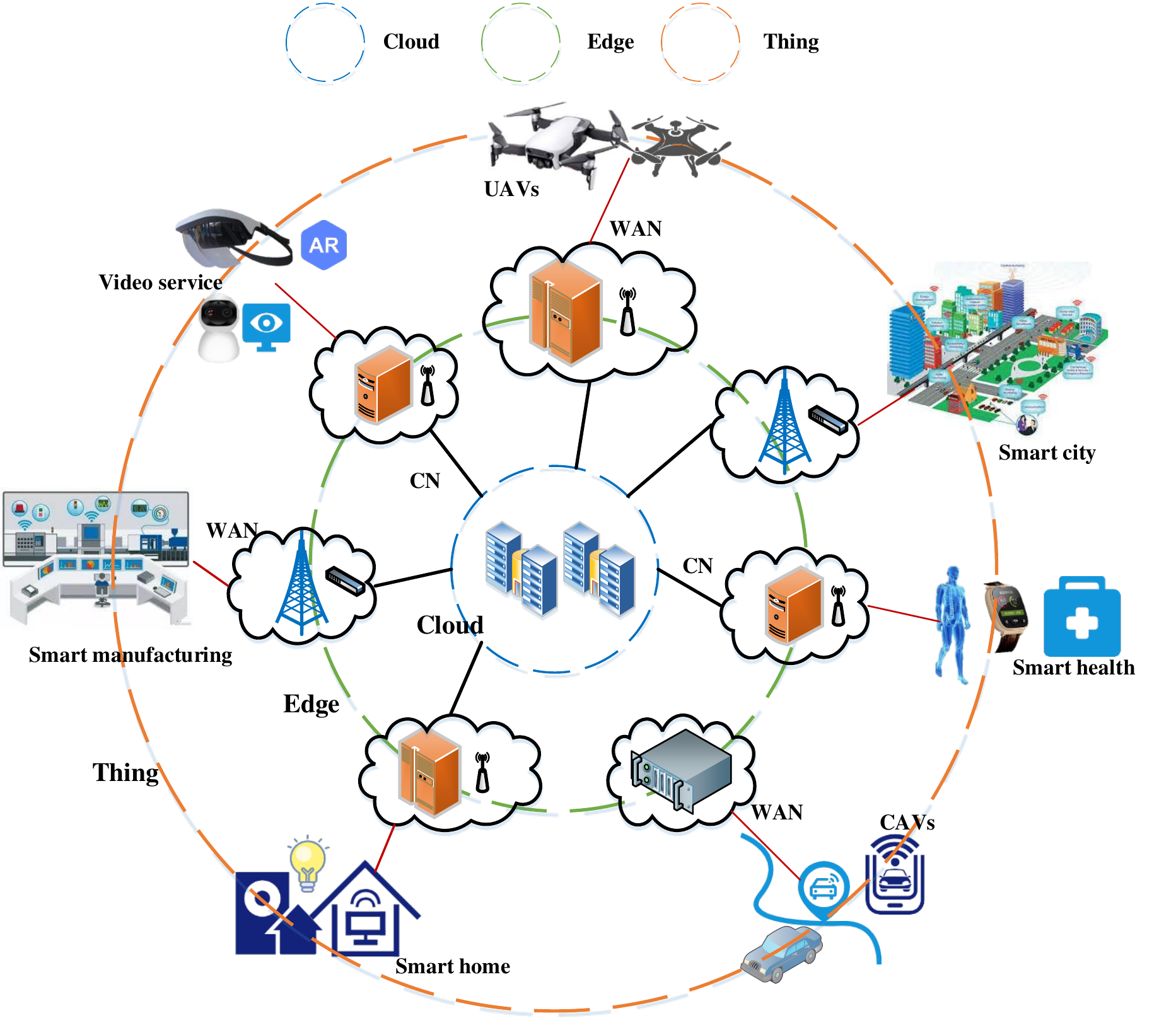}
\caption{Various application scenarios under edge computing architecture.}
\label{fig various application scenarios}
\end{figure*}
\section{ Resource Scheduling  in Applications Context}
\label{sec application}
New applications are the main driving force for edge computing. Edge computing involves optimal resource scheduling in many application scenarios due to users' stringent requirements for latency, energy consumption, cost, privacy, etc.  
In this section, we introduce several typical application scenarios involved in the research on resource scheduling in edge computing. When we were analyzing references, we recorded the applications involved in each paper. Through statistics, we have summarized several more researched and more common applications, which serve as the typical applications of this survey, including UAV, CAV, video service, smart city, smart health, smart manufacturing, and smart home, as shown in Fig.~\ref{fig various application scenarios}.

\subsection{UAV}

UAVs, especially low-cost quad-rotor aircraft, are experiencing explosive growth and have been widely used in civil and military fields, such as traffic monitoring, public safety, disaster detection, search, and rescue. And the research on resource scheduling in the field of UAVs can be divided into two directions:

\subsubsection{UAVs as users}
In some computing-intensive applications, the UAVs are unable to meet the task requirements due to the limited resources. 
In this case, the resources at the edge of the wireless network, such as cellular BSs, can provide cloud-like computing services to assist UAVs to complete the task processing~\cite{cao2018mobile,liu2019minimization}. Cao \textit{et al.} in~\cite{cao2018mobile} studied how to offload the latency-sensitive tasks of UAVs to the ground BSs, subject to the speed constraint of UAVs. Similarly, the authors in~\cite{liu2019minimization} studied the offloading problem based on two-tier UAVs, aiming to minimize the latency of tasks and the system cost.

\subsubsection{UAVs as edge resources}
Due to the convenient mobility, UAVs can be regarded as mobile edge resources or cooperate with traditional edge servers on the ground to improve their connectivity, which can provide high-quality services for users~\cite{wang2019joint,zhang2019computation,hu2019uav,wang2020agent,yu2020joint,peng2021multi}.
In~\cite{wang2019joint}, multiple UAVs are regarded as flying edge nodes for MUs. The authors presented ToDeTaS, a two-layer optimization method, to jointly solve the deployment and task scheduling problem, aiming to minimize the system energy consumption. 
Likely, Zhang \textit{et al.} in~\cite{zhang2019computation} formulated a computation efficiency maximization problem in a UAV-assisted MEC system. 
Yu \textit{et al.} in~\cite{yu2020joint} proposed a UAV-enabled MEC system to provide the computing service to the IoT devices, which cannot access any service due to the  sparse distribution of the existing ENs. They studied the resource allocation problem to minimize the service delay of IoT devices. Similarly, in~\cite{wang2020agent}, under the UAV-aided MEC architecture, the authors studied the task offloading problem and adopted the agent to conduct an offloading plan based on the perceived information of users, UAV, and edge nodes. 

We summarize the studies on UAVs mentoined above in Table~\ref{tabl works on UAVs}.

\subsection{CAV}
With the development of AI, computer vision, depth perception and sensing technologies, vehicles have gradually evolved from traditional travel tools into CAVs with intelligent and interconnected computing systems.
According to Intel, 4TB of raw data would be generated from a CAV in one day, which poses a great challenge on processing capacity of CAVs to support various low-latency and computation-intensive applications. Therefore, the research on computation offloading from vehicles to edge or cloud has attracted much attention. Also, considering the enhancement of the computing, communication, and storage capabilities of vehicles and the widespread distribution, vehicles can also be regarded as edge resources to provide users with flexible computing services. Accordingly,
the research on resource scheduling in edge computing under the CAV environment includes two directions: 

\subsubsection{Vehicle as users}
In this case, the focus is to schedule the tasks generated by vehicles to the edge (e.g., RSU)~\cite{zhang2017optimal,li2020joint,zhang2019task,zhou2019energy,dai2018jointoffloading,wang2019computationvehicular,xu2019multi,wu2020efficient,luo2021selflearning}. Li \textit{et al.} in~\cite{li2020joint} considered the vehicular edge computing framework where the computation tasks of autonomous vehicles can be scheduled to RSUs. They investigated the task offloading problem based on the time-varying channel characteristics to maximize the system utility. Likely, by offloading vehicles' tasks to RSUs, the work in~\cite{zhang2019task} took load balancing into account and used FiWi technology to manage network due to the dynamic vehicular network. Then, the authors proposed a soft-defined network (SDN) based offloading scheme aiming to minimize the task delay. Zhou \textit{et al.} in~\cite{zhou2019energy} studied the energy-efficient offloading problem and presented a distribution method based on consensus ADMM. The work in \cite{xu2019multi} developed a multi-objective optimization problem for computation offloading in an IoV edge system to reduce energy consumption and delay simultaneously. And the authors adopted a non-dominated sorting genetic algorithm to solve the problem. Moreover, the work in \cite{luo2021selflearning} formulated a computation offloading problem as a distributed offloading decision-making game, in which each vehicle as a player makes its best response decision to minimize its joint cost (including latency and offloading cost).

\subsubsection{Vehicle as SPs}
In this case, vehicles can be the supplement to the edge, providing computing services for MUs~\cite{huang2019social,huang2018parked,zhang2019optimalPV,feng2017ave}. Utilizing the idle resources of parked vehicles (PVs), the authors in~\cite{huang2019social} studied how to schedule the tasks generated by MUs that can be partitioned into sub-tasks to PVs, aiming to maximize the social welfare. Besides, Huang \textit{et al.} in~\cite{huang2018parked} regarded PVs as available edge resources that can collaborate with the existing edge servers to provide computing services for MUs. They proposed an interactive protocol for service provisioning considered the security and privacy requirements of users. Similarly, in~\cite{zhang2019optimalPV}, collaborated with edge servers, PVs are employed to execute tasks of MUs with delay constraints. The authors proposed a distributed approach based on the Stackelberg game to solve the task assignment problem. Particularly, AVE was presented in~\cite{feng2017ave} as a job scheduling framework, where autonomous vehicles collaborate to provide computation services for each other.

\begin{table*}[!htb]
\newcommand{\tabincell}[5]{\begin{tabular}{@{}#1@{}}#2\end{tabular}}
\centering
\caption{Comparison of Papers Focusing on UAVs.  Acronyms used in this Table: edge server (ES), base station (BS), unmanned arerial vehicle (UAV), mobile user (MU).}
\begin{tabular}{|m{1cm}<{\centering}|m{2.7cm}|m{2cm}<{\centering}|m{3cm}<{\centering}|m{5cm}|}
\hline
\makecell*[c]{\textbf{Paper}} &\makecell*[c]{\textbf{Research issue}}
&\makecell*[c]{\textbf{Edge}}
&\makecell*[c]{\textbf{What's to be scheduled}}
&\makecell*[c]{\textbf{Key points}}
\\
\hline
\cite{cao2018mobile}& Computation offloading&
BSs&
Tasks from UAVs &
Minimize the response time; Optimizing the trajectory of UAVs; the constraints:  the speed of UAVs and the computation capacity of BSs\\
\hline
\cite{liu2019minimization}& Computation offloading&
BSs&
Tasks from MUs &
Minimize latency and cost; Stackelberg game\\
\hline
\cite{wang2019joint}& Joint deployment and task scheduling&
UAVs&
Tasks from MUs &
Minimize system energy consumption; a two-layer optimization method\\
\hline
\cite{zhang2019computation}& Joint Computation offloading and trajectory scheduling&
UAVs&
Tasks from MUs &
Maximize computation efficiency; the constraints: user association, computing and spectrum resources; non-convex problem\\
\hline
\cite{yu2020joint}& Joint task offloading and resource placement&
UAVs and ESs&
Tasks from MUs &
Maximize service delay; maximize the energy efficiency; non-convex problem\\
\hline
\cite{wang2020agent}& Joint UAV deployment and computation offloading&
UAVs and ESs&
Tasks from MUs &
Maximize task delay and energy consumption\\
\hline
\end{tabular}
\label{tabl works on UAVs}
\end{table*}
\begin{table*}[!htb]
\newcommand{\tabincell}[5]{\begin{tabular}{@{}#1@{}}#2\end{tabular}}
\centering
\caption{Comparison of Papers Focusing on CAVs.  Acronyms used in this Table: road side unit (RSU), edge server (ES), unmanned arerial vehicle (UAV), parked vehicle (PV), mobile user (MU), soft-defined network (SDN), alternating direction method of multipliers (ADMM).}
\begin{tabular}{|m{1cm}<{\centering}|m{2.7cm}|m{2cm}<{\centering}|m{3cm}<{\centering}|m{5cm}|}
\hline
\makecell*[c]{\textbf{Paper}} &\makecell*[c]{\textbf{Research issue}}
&\makecell*[c]{\textbf{Edge}}
&\makecell*[c]{\textbf{What's to be scheduled}}
&\makecell*[c]{\textbf{Key points}}
\\
\hline
\cite{li2020joint}& Computation offloading&
RSUs&
Tasks from vehicles &
Maximize the system utility; time-varying channel; the linearization based branch and bound algorithm \\
\hline
\cite{zhang2019task}& Computation offloading&
RSUs&
Tasks from vehicles &
Minimize the task delay;load balancing; SDN-based scheme\\
\hline
\cite{zhou2019energy}& Workload offloading&
UAVs&
Tasks from vehicles &
Maximize the energy efficiency; a low-complexity distributed method based on ADMM\\
\hline
\cite{xu2019multi}& Computation offloading&
RSUs&
Tasks from vehicles &
Multi-objective: reduce energy consumption and time delay while keep load balancing; non-dominated sorting genetic algorithm\\
\hline
\cite{luo2021selflearning}& Computation offloading&
RSUs&
Tasks from vehicles &
Distributed offloading decision-making game; self-learning based
distributed computation offloading\\
\hline
\cite{huang2019social}& Task offloading and container placement&
PVs&
Tasks from MUs &
Maximize the social welfare; convex optimization methods\\
\hline
\cite{huang2018parked}& Service provisioning&
PVs and ESs&
Tasks from MUs &
Maximize the cost of users; an interactive protocol; security and privacy constraints; Stackelberg game approach\\
\hline
\cite{zhang2019optimalPV}& Task offloading&
PVs and ESs&
Tasks from MUs and vehicles&
Maximize the overall coat; Stackelberg game approach\\
\hline
\cite{feng2017ave}& 
Task offloading&
Vehicles&
Tasks from vehicles&
Maximize the system utility; vehicle-to vehicle communication; ant colony optimization\\
\hline
\end{tabular}
\label{tabl works on CAVs}
\end{table*}

We summarize the studies on UAVs mentioned above in Table~\ref{tabl works on CAVs}.

\subsection{Video Service}
The video generated by smart devices has promoted the development of various applications, such as traffic control, autonomous driving, public surveillance and security, and AR/VR. Due to the limited storage and computing capabilities of smart devices, it may be inefficient to process the computation-intensive and bandwidth-hungry videos locally. 
Scheduling video service to the edge to process is a feasible method to meet the low-latency requirement.

In~\cite{hung2018videoedge}, VideoEdge was proposed to optimize the placement of computer vision components, where two challenges were addressed including exponentially large search space caused by multiple resource providers and merging conflicts. Yi \textit{et al.} in~\cite{yi2017lavea} presented LAVEA, a video analytics edge computing platform. They formulated the task selection and prioritized for offloading as an optimization problem. LAVEA can provide low-latency computation offloading service based on serverless architecture. 
For the AR applications in video services, Ali \textit{et al.} in~\cite{al2017energy} proposed a resource allocation scheme, which involved both communication and computing resources. They leveraged the inherently collaborative nature of AR applications and solved the energy expenditure minimization problem with low-latency constraint by the successive convex approximation algorithm. 
Further, Liu \textit{et al.} in~\cite{liu2019code} considered the reliability of AR task offloading problem, where the components of an AR task was modeled as a directed acyclic graph with dependencies. To minimize the failure probability of AR service, an integer PSO-based algorithm was proposed.

\begin{table*}[!htb]
\newcommand{\tabincell}[5]{\begin{tabular}{@{}#1@{}}#2\end{tabular}}
\centering
\caption{Comparison of Papers Focusing on Video Service.  Acronyms used in this Table: edge server (ES), particle swarm optimization (PSO), augmented reality (AR).}
\begin{tabular}{|m{1cm}<{\centering}|m{2.7cm}<{\centering}|m{2cm}<{\centering}|m{3cm}<{\centering}|m{5cm}|}
\hline
\makecell*[c]{\textbf{Paper}} &\makecell*[c]{\textbf{Things}}
&\makecell*[c]{\textbf{Edge}}
&\makecell*[c]{\textbf{What's to be scheduled}}
&\makecell*[c]{\textbf{Key points}}
\\
\hline
\cite{hung2018videoedge}& 
IoT Cameras
& Private clusters and public clouds&
Components of computer visions &
Maximize the average query accuracy; trade-off between multiple resources and accuracy; the constraints: large search space and merging conflicts\\
\hline
\cite{yi2017lavea}& 
Smartphones, security/dash cameras &
Container-based ESs&
Components of videos&
Minimize response time; inter-edge collaboration\\
\hline
\cite{al2017energy}& 
Smartphones&
ESs&
Components of a AR application&
Minimize the energy expenditure and latency; component-based model of an AR application; successive convex approximation algorithm; \\
\hline
\cite{liu2019code}& 
MDs&
ESs&
AR Tasks&
Minimize the failure probability; the reliability and latency requirement; the dependency of sub-tasks; PSO-based algorithm\\
\hline
\end{tabular}
\label{tabl works on video service}
\end{table*}

We summarize the studies on video services mentioned above in Table~\ref{tabl works on video service}.

\subsection{Smart City}
In 2016, Alibaba put forward the concept of ``smart city", where multiple urban data are used to manage the city better. To manage and process  the smart city data characterized by diversity and heterogeneity and involved the privacy and security of residents, some studies focus on designing edge collaborative processing systems~\cite{wang2018big,hou2018green,li2018delay}. Also, some works on the optimal placement of edge resources provide convenient and fast computing services for emerging applications in smart cities~\cite{choi2019scalable,xu2019trust,canali2019gasp,zheng2020task}.
For large-scale smart cities, the authors in~\cite{choi2019scalable} presented the logical edge network formed in a tree topology to place edge service in a resource-effective way. Based on the logical edge network, they also designed a service placement scheme meeting the service demands of IoT devices as well as the resource capacity of edge servers. To process the quantities of services produced by IoT devices in smart cities, Xu \textit{et al.} in \cite{xu2019trust} proposed TSP as a trust-oriented IoT service placement scheme to tackle the improvement of resource usage, load balance and energy consumption while protecting the privacy of IoT devices. Similarly, to deal with data streams generated from sensors deployed in smart cities, Canali \textit{et al.} in~\cite{canali2019gasp} also studied the service problem and proposed a scalable heuristic-based genetic algorithm.

\begin{table*}[!htb]
\newcommand{\tabincell}[5]{\begin{tabular}{@{}#1@{}}#2\end{tabular}}
\centering
\caption{Comparison of Papers Focusing on Smart ``Things".  Acronyms used in this Table: alternating direction method of multipliers (ADMM), deep Q-learning (DQN), job shop scheduling (JSP).}
\begin{tabular}{|m{1cm}<{\centering}|m{2.5cm}|m{2.4cm}|m{3cm}<{\centering}|m{5cm}|}
\hline
\makecell*[c]{\textbf{Paper}} &\makecell*[c]{\textbf{Domain}}
&\makecell*[c]{\textbf{Research issue}}
&\makecell*[c]{\textbf{What's to be scheduled}}
&\makecell*[c]{\textbf{Key points}}
\\
\hline

\cite{choi2019scalable}& 
Smart city&
Service placement&
Edge services&
Maximize the resource utilization; logical edge network \\
\hline

\cite{xu2019trust}& 
Smart city&
Service placement&
IoT services&
Optimize multiple performance metrics; the constraints: time and privacy; the strength Pareto evolutionary algorithm\\
\hline
\cite{alam2019edge}& 
Smart health&
Service provisioning&
Healthcare service and data&
Maximize the cost of healthcare system; a portfolio optimization approach; ADMM\\
\hline
\cite{chaudhry2017azspm}& 
Smart health&
Service provisioning&
Healthcare service&
A remote verification method; dynamic security composition; zero knowledge \\
\hline
\cite{lin2019smart}&
Smart manufacturing&
JSP&
Jobs generated by machine&
Maximize the job latency; DQN; job shop scheduling\\
\hline
\cite{sun2019ai}& 
Smart manufacturing&
Offloading&
Tasks generated by IoT devices&
Maximize the service accuracy; AI-enhanced offloading framework\\
\hline
\cite{zavalyshyn2018homepad}&
Smart home&
Data analysis&
Data generated by smart home devices&
Protect the privacy of users; a directed graph of elements; prolog rules; automatic verification\\
\hline
\cite{wang2017healthedge}& 
Smart home&
Task offloading&
Tasks generated by the healthcare system&
Minimize the task latency; health emergency and human behavior consideration\\
\hline
\end{tabular}
\label{tabl works on Smart "things"}
\end{table*}

\subsection{Smart Health}
The development of cloud computing, wireless broadband communication, BAN and wearable medical devices enhances mobile medical services and improves medical standards and medical conditions. However, as medical data grows exponentially, the cost of operating and maintaining the medical system is increasing. To alleviate this situation, deploying edge resources to process medical data at the edge has attracted much attention~\cite{alam2019edge,samie2016computation,manogaran2019wearable}.  Moreover, the establishment of edge-assisted medical systems can save costs for healthcare service providers~\cite{chaudhry2017azspm,ning2020mobile,nikoloudakis2019vulnerability}. Alam \textit{et al.}~\cite{alam2019edge} proposed an edge-of-things (EoT) computation framework for healthcare service provisioning, where an EoT is a bridge between service providers and healthcare consumers. The authors proposed a portfolio optimization approach for cost-effective service provisioning and used an ADMM method for healthcare data offloading. The security and privacy of healthcare data in smart health is very important. In~\cite{chaudhry2017azspm}, a security provisioning model named AZSPM, was proposed for medical devices in edge computing. AZSPM can build trust among medical devices with zero knowledge. For the wearable smart devices for physical monitoring, the work in~\cite{manogaran2019wearable} proposed an edge computing-based deep learning network system for physical monitoring by using multimedia technology with agile learning for real-time data processing, which improved the multiple performance metrics effectively.

\subsection{Smart Manufacturing}
Smart manufacturing refers to the realization of intelligent industrial operations through AI and big data technology. In smart manufacturing, the industrial devices 
need real-time control based on the generated data characterized with security and privacy. And the introduction of AI technology into the IIoT requires powerful computing capabilities to complete advanced fault prediction, demand forecasting and other big data processing tasks. Therefore, applying edge computing in smart manufacturing has become the direction of industry development, which can improve system performance, ensure data security and privacy, and reduce the cost of operation~\cite{chen2018edge,lin2019smart,sun2019ai,li2019hybrid}. Chen \textit{et al.}~\cite{chen2018edge} presented an edge computing architecture for IoT-based manufacturing, where edge computing acted as edge equipment, information fusion, network communication and cooperative mechanism with traditional computing. Job shop scheduling (JSP) problems are complex in smart manufacturing. In~\cite{lin2019smart}, Lin \textit{et al.} proposed an edge computing framework for smart manufacturing, which adjusted DQN to solve JSP problems. The work in~\cite{sun2019ai} designed an AI-enhanced offloading framework that combined the edge and cloud computing to maximize the service accuracy in IIoT. The authors introduced edge intelligence to smart manufacturing for the sake of many advantages it can bring, including personalization, responsiveness and privacy.

\subsection{Smart Home}
The development and enrichment of smart devices have made the system of smart homes reaches commercial maturity. Smart homes use lots of IoT devices (such as various sensors) to control and monitor the living environment in real-time. However, the ever-increasing number of smart devices, the multiple applications with low latency requirements, the big data generated by smart devices, and the extremely private home data, make it a tread to apply edge computing instead of cloud computing to smart homes. There are many works focusing on edge resource scheduling towards the smart home environment~\cite{cao2017edgeos_h,zavalyshyn2018homepad,zhai2020toward}. EdgeOSH, a home operating system, was proposed in~\cite{cao2017edgeos_h} to provide functions of the program interface and data management. 
In~\cite{zavalyshyn2018homepad}, HomePad was presented for home environments, and it allows IoT applications to execute at the edge. For users' privacy, HomePad was designed to enable users to determine how applications access and process sensitive data generated by smart devices. Besides, Wang \textit{et al.} in~\cite{wang2017healthedge} studied the resource management of the healthcare system in smart homes under the edge-cloud architecture, and presented a task scheduling scheme named HealthEdge, which can process different tasks based on priorities aiming to reduce the latency.

The studies on smart city, smart health, smart manufacturing and smart home are called the study on smart ``things" in our survey.
And we summarize the studies on smart "things" mentioned above in Table~\ref{tabl works on Smart "things"}. Notably, the application scenarios for smart "things" are deeply dependent on the development of IoT. We believe that the research on each application scenario will become more and more mature thanks to the explosive growth of edge computing in the field of IoT. 

\section{Challenges and Research Directions}
\label{sec challenges}
Despite the fact that the research on resource scheduling in edge computing has accumulated a lot of results, there are still many 
key issues that have not been well explored. This section discusses several open research challenges followed by future research directions.

\subsection{Model and Architecture}
\subsubsection{Computation and Communication Model}
To efficiently schedule edge resources to accomplish task processing, a computation model should be first established to reflect the relationship between task data size and the amount of computing capacity it requires.  In most existing works, it always utilizes a processing density (in CPU cycles/bit) to denote this kind of relationship; thus that the amount of computing capacity a task requires is equal to the product of task data size and processing density \cite{lin2019computation,miettinen2010energy}. Obviously, it is a linear representation. However, since different types of tasks have different processing densities, this kind of one-size-fits-all representation approach may not be suitable for various application tasks in edge computing. \textit{Therefore, more flexible computation models are worthy of further study}.
Besides, to better process application tasks, utilizing communication resources to offload part or all of the tasks to ENs is trending. During this process, the data transmission rate is a key concern for communication resource scheduling. Current representations of data transmission rate are mostly based on the Shannon-Hartley theorem, which tells a theoretical tightest upper bound on the data transmission rate over a communication channel of a specified bandwidth in the presence of noise. However, in the practical scenario of edge computing,  end-devices and ENs are always positioned in a complicated environment with extremely poor channel conditions, such as high mobility, shield, and interference \cite{li2020joint}. The actual data transmission rate can not achieve the theoretical value. \textit{Therefore, it is necessary to develop a more practical communication model based on field tests or considering different application scenarios. } 
\subsubsection{Computation Migration}
Since task processing always involves cooperation among multiple ENs or  end-devices, few studies focus on computation migration. Generally, to accomplish the computation migration, there are mainly six steps: migration environment sensing, task division, migration decision, task uploading, task execution, result return. Among them, task division and migration decision are the two most critical steps.  However, in most existing works that considered computation migration in resource scheduling, only the migration decision step is considered, and other steps are ignored \cite{chen2019dynamic,ma2017efficient}. Computation migration is more like a kind of concept of collaborative computing in current studies. \textit{Future research can focus more on the implement of computation migration considering the entire process.}  

\subsubsection{Task Partitioning and Integration}
Computation offloading has attracted much attention in resource scheduling in edge computing. A task can be divided into two parts, one part computed locally and the other part offloaded to ENs or other nodes for processing.  It is assumed that the offloaded part of a task is denoted by an offloaded ratio in most existing works\cite{yu2020joint,xiao2019task}. The resource scheduling process is to determine an optimal offloaded ratio and other optimization variables. Once the optimal offloaded ratio is obtained, this part of the task is directly offloaded \cite{zhou2018uav}. However, for a certain task, the divisible part may not be equal to the optimal offloaded part based on the optimization solution. \textit{Therefore, future research should step further on exploring the nature of tasks during task partitioning for computation offloading.} 
After the task is partitioned and processed by different nodes, it is necessary to integrate the dispersal results. Another concern may arise during this process: whether the integrated results are the same as those of none-partitioning processing? This concern leads to a future study on \textit{how to integrate the processing results from different nodes without losing the original information of the task.}  

\subsubsection{Green Energy}
To achieve energy saving and maintain longer battery life of IoT devices, it is a trend to utilize renewable green resources light and wind to strengthen energy support, which can significantly reduce carbon emissions and environmental pollution.  There are many studies on energy-harvesting or wireless-charging enabled edge computing \cite{bi2018computation,chen2018multi}. The introduction of extra energy supplement makes resource scheduling more complex since not only the energy consumption model during task transmission and task processing should be considered, but also the harvested energy.  Although marvelous solutions are proposed in existing works, most of them consider the extra energy can be harvested continuously \cite{feng2019computation,li2019dynamic}. However, in practice, the energy harvesting process may be unstable, which poses a significant challenge in designing an efficient resource scheduling strategy. \textit{Therefore, future research should focus more on the energy harvesting process.} 

\subsubsection{Heterogeneous Architecture}
The architecture of edge computing generally includes things layer, edge layer, and cloud layer. Most of the existing research on resource scheduling are under the thing-edge-cloud architecture. It is predicted that the integration of multidimensional networks such as space, air, and ground to form the space-air-ground integrated network (SAGIN) is the future trend to support the ever-increasing IoT applications \cite{hong2020space}. Under such a space-air-ground heterogeneous architecture, the SAGIN incorporated with edge computing can provide a myriad of services and applications, such as edge caching, computation offloading and cloud services  \cite{cheng2020comprehensive}. However, heterogeneous nodes ( end-devices, edge servers, CAVs, UAVs, and satellites) and the heterogeneous resources of those nodes make the resource management and scheduling complicated. Besides, heterogeneous nodes are subject to strong spatio-temporal constraints \cite{liu2020task}, which make the management and scheduling of heterogeneous resources more challenging. \textit{Therefore, it is necessary
to develop an efficient resource scheduling and management technology that can simultaneously orchestrate the heterogeneous nodes and resources in SAGIN.}
In this context, network slicing is a viable technique for efficient heterogeneous resource scheduling and management\cite{zhang2017software,zhang2018air}.

\subsection{Feasibility}
\subsubsection{Deployment}
There are relatively few studies on the deployment of ENs, including edge servers or IoT devices in resource scheduling. The geographical location of ENs has a great impact on resource scheduling.  Enlarging the service range of ENs can effectively improve edge resource utilization and effectively improve resource scheduling utility \cite{hong2019resource}. In many cases, the users are mobile, and ENs' deployment will be more complex. \textit{Therefore, future research can consider the deployment of ENs when designing resource scheduling mechanisms}.
\subsubsection{Management}
For the edge, scheduling computation tasks of users at the infrastructure is mostly limited to theoretical research. The technical issues on the implementation have not been well explored. Besides, the scalability of resource scheduling algorithms should be taken seriously.  With the rapid expansion of users' scale, the resource scheduling scheme is required to achieve flexible deployment and rapid configuration \cite{santos2019resource}.  Serverless computing has become a popular architectural alternative for building and running up-to-date applications and services \cite{suresh2019fnsched}. Serverless applications allow developers to focus on the code rather than on infrastructure configuration and management, which can speed up service provisioning and provide more efficient scaling \cite{jonas2019cloud}. The serverless computing architecture realizes the automatic scalability of services, pay-by-value, and automated high-availability management, which provides a powerful and convenience orchestration framework to schedule and manage edge resources. However, research on applying serverless architectures to edge computing is in its infancy, and many problems remain unsolved.
\textit{Therefore, more attention need to be paid to resource scheduling research based on the serverless edge architecture. }

\subsection{Security and privacy}
\subsubsection{System-level}
In the existing resource scheduling research, security and privacy issues have not been appreciated and fully explored.  In resource scheduling, the multi-layer architecture of edge computing makes the edge system vulnerable to hostile attacks \cite{he2019peace}. A system failure of an edge node or a failure caused by attacks may threaten the reliability and robustness of the entire edge system, thus making the resource scheduling meaningless.  \textit{Therefore, efforts are required to put into the fault tolerance research of edge systems in resource scheduling. Specifically, system robustness enhancement mechanism and intrusion detection strategy need to be developed.}
\subsubsection{Service-level}
In the existing research on computation offloading and service provisioning, the following issues are generally not considered: whether the offloaded edge node can be trusted, how to ensure that users can authorize the edge services, and how to protect the privacy of the data generated by the edge service. \textit{Therefore, designing authentication mechanisms for the users covered by a specific edge node is needed. Besides, the privacy module is also required for the edge data center to improve the trustworthiness of edge services.}
\subsubsection{Data-level}
In the process of resource scheduling, especially computation offloading, data collected by the edge or shared with IoT devices involve much private information.  In the existing research, the user data, the interaction data between ENs, and the computing data at the edge are unconditionally trusted and easily accessible \cite{duc2019machine}.  However, in real application scenarios such as smart home and smart health, these data involves privacy and even commercial secrets of users, and can be easily leaked during transmission and processing, causing huge losses \cite{ghobaei2019resource,markakis2017exegesis}. \textit{Therefore, more works are needed to focus on designing trust mechanisms 
and privacy preservation policies for the edge and users.}

\subsection{Dynamics}
In resource scheduling, users' mobility is a thorny challenge. In various application scenarios, users' mobile characteristics have not been well explored in current research, and most studies just conduct idealization and ignore this characteristic. The frequent mobility of users has a significant impact on task offloading and cache provisioning.  The offloading decision and cache decision at the current moment may not be applicable to users at the next moment, or even users have moved out of the service range of the edge node \cite{lin2019computation}. \textit{Therefore, incorporating the trajectory prediction of users into resource scheduling studies can effectively improve the users' QoS. Moreover, designing the mobility management policies to enable users to access ENs seamlessly can improve the service stability.}

\subsection{Joint Scheduling of Communication, Computing, Storage (CCS) Resources}
Task data should be received by processing nodes and cached in the data queue, waiting for processing to accomplish the offloaded tasks. The caching and queuing process is complicated and also very important for real-time task processing. However, in most existing works, the total task processing time is considered as the sum of local processing time, transmission time, and offloading processing time, ignoring the caching and queuing process. Besides, most studies on scheduling cache resources focus more on  caching popular content at the network edge to improve hit ratio and avoid duplicate transmissions of the same content, thus improving users’ QoE \cite{wang2017survey,wen2020joint}. A few works have been done to considered combining the joint allocation of communication and computing resources. \textit{Therefore, future work on joint scheduling of CCS resources should take the research further forward by considering the caching and queuing process.} 

\subsection{Evaluation}

\subsubsection{Workload}
The workload of users' requests has a non-negligible impact on resource scheduling. The requests from users are generally assumed to obey a specific distribution (e.g., Poisson distribution) in the current evaluation. Furthermore, the scheduled task's CPU, memory, and storage requirements are treated theoretically and idealistically without considering real system performance.  However, in the real environment, the peak situation of workload may put abnormal pressure on edge resources and even cause users' tasks to fail \cite{duc2019machine}. \textit{Therefore, resource provisioning based on workload prediction is an urgent problem for SPs. Also, for reliable service, a good load balancing strategy needs to be designed.}

\subsubsection{Test environment}
The performance evaluation of scheduling algorithms in current research is generally performed using simulation tools, including professional simulators for edge computing such as iFogSim \cite{gupta2017ifogsim}, EdgeCloudSim \cite{sonmez2018edgecloudsim}, and MyiFogSim \cite{lopes2017myifogsim}, and general simulation platforms like Matlab. Few studies evaluate their algorithms in real edge systems. \textit{Effort is required to focus on the feasibility of scheduling algorithms in real systems, e.g., designing testbeds or prototypes for evaluation.}

\section{Conclusion}
\label{sec conclusion}
In this survey, we conduct a systematic and comprehensive review of resource scheduling in edge computing. First, we lay the groundwork for the entire overview by elaborating on two fundamental questions of why resource scheduling is needed and what exactly resource scheduling refers to in edge computing. Second, we present the architecture and different collaborative manners for resource scheduling. Third, an in-depth overview of research issues and research techniques in resource scheduling is presented, which is the prominent effort of this survey. Regarding the key research issues, we first introduce a unified offloading model for edge computing. Then we summarize the current works from three research aspects including computation offloading, resource allocation, and resource provisioning. Regarding the key techniques, based on two operation modes, namely, centralized and distributed modes, the state-of-art works are investigated and explicitly categorized. Also, we summarize six performance indicators that frequently appear in the surveyed literature. Fourth, some typical application scenarios involved  in resource scheduling are introduced. Finally, for resource scheduling in edge computing to be investigated extensively and deeply, we shed light on the current research bottlenecks and challenges and look forward to more research investment in promising research directions.

\bibliographystyle{IEEEtran}
\bibliography{IEEEabrv,reference/survey}

\begin{thebibliography}{100}
\providecommand{\url}[1]{#1}
\csname url@samestyle\endcsname
\providecommand{\newblock}{\relax}
\providecommand{\bibinfo}[2]{#2}
\providecommand{\BIBentrySTDinterwordspacing}{\spaceskip=0pt\relax}
\providecommand{\BIBentryALTinterwordstretchfactor}{4}
\providecommand{\BIBentryALTinterwordspacing}{\spaceskip=\fontdimen2\font plus
\BIBentryALTinterwordstretchfactor\fontdimen3\font minus
  \fontdimen4\font\relax}
\providecommand{\BIBforeignlanguage}[2]{{%
\expandafter\ifx\csname l@#1\endcsname\relax
\typeout{** WARNING: IEEEtran.bst: No hyphenation pattern has been}%
\typeout{** loaded for the language `#1'. Using the pattern for}%
\typeout{** the default language instead.}%
\else
\language=\csname l@#1\endcsname
\fi
#2}}
\providecommand{\BIBdecl}{\relax}
\BIBdecl

\bibitem{hayes2008cloud}
B.~Hayes, ``Cloud computing,'' \emph{Communications of the ACM}, vol.~51,
  no.~7, pp. 9--11, 2008.

\bibitem{velte2009cloud}
T.~Velte, A.~Velte, and R.~Elsenpeter, \emph{Cloud computing, a practical
  approach}.\hskip 1em plus 0.5em minus 0.4em\relax McGraw-Hill, Inc., 2009.

\bibitem{xia2012internet}
F.~Xia, L.~T. Yang, L.~Wang, and A.~Vinel, ``Internet of things,''
  \emph{International journal of communication systems}, vol.~25, no.~9, p.
  1101, 2012.

\bibitem{CISCO2018}
Cisco, ``Edge-to-enterprise iot analytics for electric utilities solution
  overview,'' Website, 2018,
  \url{https://www.cisco.com/c/en/us/solutions/collateral/data-center-virtualization/big-data/solution-overview-c22-740248.html}.

\bibitem{hu2015mobile}
Y.~C. Hu, M.~Patel, D.~Sabella, N.~Sprecher, and V.~Young, ``Mobile edge
  computing—a key technology towards 5g,'' \emph{ETSI white paper}, vol.~11,
  no.~11, pp. 1--16, 2015.

\bibitem{shi2016edge}
W.~Shi, J.~Cao, Q.~Zhang, Y.~Li, and L.~Xu, ``Edge computing: Vision and
  challenges,'' \emph{IEEE internet of things journal}, vol.~3, no.~5, pp.
  637--646, 2016.

\bibitem{mach2017mobile}
P.~Mach and Z.~Becvar, ``Mobile edge computing: A survey on architecture and
  computation offloading,'' \emph{IEEE Communications Surveys \& Tutorials},
  vol.~19, no.~3, pp. 1628--1656, 2017.

\bibitem{tan2017virtual}
Z.~Tan, F.~R. Yu, X.~Li, H.~Ji, and V.~C. Leung, ``Virtual resource allocation
  for heterogeneous services in full duplex-enabled scns with mobile edge
  computing and caching,'' \emph{IEEE Transactions on Vehicular Technology},
  vol.~67, no.~2, pp. 1794--1808, 2017.

\bibitem{wang2018joint}
P.~Wang, C.~Yao, Z.~Zheng, G.~Sun, and L.~Song, ``Joint task assignment,
  transmission, and computing resource allocation in multilayer mobile edge
  computing systems,'' \emph{IEEE Internet of Things Journal}, vol.~6, no.~2,
  pp. 2872--2884, 2018.

\bibitem{lin2019computation}
L.~Lin, X.~Liao, H.~Jin, and P.~Li, ``Computation offloading toward edge
  computing,'' \emph{Proceedings of the IEEE}, vol. 107, no.~8, pp. 1584--1607,
  2019.

\bibitem{li2018end}
J.~Li, G.~Luo, N.~Cheng, Q.~Yuan, Z.~Wu, S.~Gao, and Z.~Liu, ``An end-to-end
  load balancer based on deep learning for vehicular network traffic control,''
  \emph{IEEE Internet of Things Journal}, vol.~6, no.~1, pp. 953--966, 2018.

\bibitem{wang2019edge}
S.~Wang, Y.~Zhao, J.~Xu, J.~Yuan, and C.-H. Hsu, ``Edge server placement in
  mobile edge computing,'' \emph{Journal of Parallel and Distributed
  Computing}, vol. 127, pp. 160--168, 2019.

\bibitem{chen2018efficient}
X.~Chen, W.~Li, S.~Lu, Z.~Zhou, and X.~Fu, ``Efficient resource allocation for
  on-demand mobile-edge cloud computing,'' \emph{IEEE Transactions on Vehicular
  Technology}, vol.~67, no.~9, pp. 8769--8780, 2018.

\bibitem{dinh2020online}
T.~Q. Dinh, B.~Liang, T.~Q. Quek, and H.~Shin, ``Online resource procurement
  and allocation in a hybrid edge-cloud computing system,'' \emph{IEEE
  Transactions on Wireless Communications}, vol.~19, no.~3, pp. 2137--2149,
  2020.

\bibitem{yoon2016low}
J.~Yoon, P.~Liu, and S.~Banerjee, ``Low-cost video transcoding at the wireless
  edge,'' in \emph{2016 IEEE/ACM Symposium on Edge Computing (SEC)}.\hskip 1em
  plus 0.5em minus 0.4em\relax IEEE, 2016, pp. 129--141.

\bibitem{xu2019edge}
X.~Xu, Y.~Xue, L.~Qi, Y.~Yuan, X.~Zhang, T.~Umer, and S.~Wan, ``An edge
  computing-enabled computation offloading method with privacy preservation for
  internet of connected vehicles,'' \emph{Future Generation Computer Systems},
  vol.~96, pp. 89--100, 2019.

\bibitem{zhou2018uav}
F.~Zhou, Y.~Wu, H.~Sun, and Z.~Chu, ``Uav-enabled mobile edge computing:
  Offloading optimization and trajectory design,'' in \emph{2018 IEEE
  International Conference on Communications (ICC)}.\hskip 1em plus 0.5em minus
  0.4em\relax IEEE, 2018, pp. 1--6.

\bibitem{li2018computing}
M.~Li, Q.~Wu, J.~Zhu, R.~Zheng, and M.~Zhang, ``A computing offloading game for
  mobile devices and edge cloud servers,'' \emph{Wireless Communications and
  Mobile Computing}, vol. 2018, 2018.

\bibitem{zhang2018jointresource}
J.~Zhang, X.~Hu, Z.~Ning, E.~C.-H. Ngai, L.~Zhou, J.~Wei, J.~Cheng, B.~Hu, and
  V.~C. Leung, ``Joint resource allocation for latency-sensitive services over
  mobile edge computing networks with caching,'' \emph{IEEE Internet of Things
  Journal}, vol.~6, no.~3, pp. 4283--4294, 2018.

\bibitem{dai2018joint}
Y.~Dai, D.~Xu, S.~Maharjan, and Y.~Zhang, ``Joint computation offloading and
  user association in multi-task mobile edge computing,'' \emph{IEEE
  Transactions on Vehicular Technology}, vol.~67, no.~12, pp. 12\,313--12\,325,
  2018.

\bibitem{wang2017online}
S.~Wang, M.~Zafer, and K.~K. Leung, ``Online placement of multi-component
  applications in edge computing environments,'' \emph{IEEE Access}, vol.~5,
  pp. 2514--2533, 2017.

\bibitem{guo2018computation}
H.~Guo, J.~Liu, and J.~Zhang, ``Computation offloading for multi-access mobile
  edge computing in ultra-dense networks,'' \emph{IEEE Communications
  Magazine}, vol.~56, no.~8, pp. 14--19, 2018.

\bibitem{meskar2018fair}
E.~Meskar and B.~Liang, ``Fair multi-resource allocation with external resource
  for mobile edge computing,'' in \emph{IEEE INFOCOM 2018-IEEE Conference on
  Computer Communications Workshops (INFOCOM WKSHPS)}.\hskip 1em plus 0.5em
  minus 0.4em\relax IEEE, 2018, pp. 184--189.

\bibitem{mao2017survey}
Y.~Mao, C.~You, J.~Zhang, K.~Huang, and K.~B. Letaief, ``A survey on mobile
  edge computing: The communication perspective,'' \emph{IEEE Communications
  Surveys \& Tutorials}, vol.~19, no.~4, pp. 2322--2358, 2017.

\bibitem{wang2017survey}
S.~Wang, X.~Zhang, Y.~Zhang, L.~Wang, J.~Yang, and W.~Wang, ``A survey on
  mobile edge networks: Convergence of computing, caching and communications,''
  \emph{Ieee Access}, vol.~5, pp. 6757--6779, 2017.

\bibitem{abbas2017mobile}
N.~Abbas, Y.~Zhang, A.~Taherkordi, and T.~Skeie, ``Mobile edge computing: A
  survey,'' \emph{IEEE Internet of Things Journal}, vol.~5, no.~1, pp.
  450--465, 2017.

\bibitem{peng2018survey}
K.~Peng, V.~Leung, X.~Xu, L.~Zheng, J.~Wang, and Q.~Huang, ``A survey on mobile
  edge computing: focusing on service adoption and provision,'' \emph{Wireless
  Communications and Mobile Computing}, vol. 2018, 2018.

\bibitem{tocze2018taxonomy}
K.~Tocz{\'e} and S.~Nadjm-Tehrani, ``A taxonomy for management and optimization
  of multiple resources in edge computing,'' \emph{Wireless Communications and
  Mobile Computing}, vol. 2018, 2018.

\bibitem{duc2019machine}
T.~L. Duc, R.~G. Leiva, P.~Casari, and P.-O. {\"O}stberg, ``Machine learning
  methods for reliable resource provisioning in edge-cloud computing: A
  survey,'' \emph{ACM Computing Surveys (CSUR)}, vol.~52, no.~5, pp. 1--39,
  2019.

\bibitem{hong2019resource}
C.-H. Hong and B.~Varghese, ``Resource management in fog/edge computing: a
  survey on architectures, infrastructure, and algorithms,'' \emph{ACM
  Computing Surveys (CSUR)}, vol.~52, no.~5, pp. 1--37, 2019.

\bibitem{ghobaei2019resource}
M.~Ghobaei-Arani, A.~Souri, and A.~A. Rahmanian, ``Resource management
  approaches in fog computing: A comprehensive review,'' \emph{Journal of Grid
  Computing}, pp. 1--42, 2019.

\bibitem{santos2019resource}
J.~Santos, T.~Wauters, B.~Volckaert, and F.~De~Turck, ``Resource provisioning
  in fog computing: From theory to practice,'' \emph{Sensors}, vol.~19, no.~10,
  p. 2238, 2019.

\bibitem{ren2019survey}
J.~Ren, D.~Zhang, S.~He, Y.~Zhang, and T.~Li, ``A survey on end-edge-cloud
  orchestrated network computing paradigms: Transparent computing, mobile edge
  computing, fog computing, and cloudlet,'' \emph{ACM Computing Surveys
  (CSUR)}, vol.~52, no.~6, pp. 1--36, 2019.

\bibitem{varghese2020survey}
B.~Varghese, N.~Wang, D.~Bermbach, C.-H. Hong, E.~de~Lara, W.~Shi, and
  C.~Stewart, ``A survey on edge benchmarking,'' \emph{arXiv preprint
  arXiv:2004.11725}, 2020.

\bibitem{samanta2019adaptive}
A.~Samanta and Z.~Chang, ``Adaptive service offloading for revenue maximization
  in mobile edge computing with delay-constraint,'' \emph{IEEE Internet of
  Things Journal}, vol.~6, no.~2, pp. 3864--3872, 2019.

\bibitem{huang2018parked}
X.~Huang, R.~Yu, J.~Liu, and L.~Shu, ``Parked vehicle edge computing:
  Exploiting opportunistic resources for distributed mobile applications,''
  \emph{IEEE Access}, vol.~6, pp. 66\,649--66\,663, 2018.

\bibitem{hou2016vehicular}
X.~Hou, Y.~Li, M.~Chen, D.~Wu, D.~Jin, and S.~Chen, ``Vehicular fog computing:
  A viewpoint of vehicles as the infrastructures,'' \emph{IEEE Transactions on
  Vehicular Technology}, vol.~65, no.~6, pp. 3860--3873, 2016.

\bibitem{abdelhamid2015vehicle}
S.~Abdelhamid, H.~S. Hassanein, and G.~Takahara, ``Vehicle as a resource
  (vaar),'' \emph{IEEE Network}, vol.~29, no.~1, pp. 12--17, 2015.

\bibitem{liu2019decentralized}
Y.~Liu, F.~R. Yu, X.~Li, H.~Ji, and V.~C. Leung, ``Decentralized resource
  allocation for video transcoding and delivery in blockchain-based system with
  mobile edge computing,'' \emph{IEEE Transactions on Vehicular Technology},
  vol.~68, no.~11, pp. 11\,169--11\,185, 2019.

\bibitem{li2020energy}
M.~Li, N.~Cheng, J.~Gao, Y.~Wang, L.~Zhao, and X.~Shen, ``Energy-efficient
  uav-assisted mobile edge computing: Resource allocation and trajectory
  optimization,'' \emph{IEEE Transactions on Vehicular Technology}, vol.~69,
  no.~3, pp. 3424--3438, 2020.

\bibitem{ren2019edge}
J.~Ren, Y.~He, G.~Huang, G.~Yu, Y.~Cai, and Z.~Zhang, ``An edge-computing based
  architecture for mobile augmented reality,'' \emph{IEEE Network}, vol.~33,
  no.~4, pp. 162--169, 2019.

\bibitem{taleb2017mobile}
T.~Taleb, S.~Dutta, A.~Ksentini, M.~Iqbal, and H.~Flinck, ``Mobile edge
  computing potential in making cities smarter,'' \emph{IEEE Communications
  Magazine}, vol.~55, no.~3, pp. 38--43, 2017.

\bibitem{lin2018task}
K.~Lin, S.~Pankaj, and D.~Wang, ``Task offloading and resource allocation for
  edge-of-things computing on smart healthcare systems,'' \emph{Computers \&
  Electrical Engineering}, vol.~72, pp. 348--360, 2018.

\bibitem{liao2019learning}
H.~Liao, Z.~Zhou, X.~Zhao, L.~Zhang, S.~Mumtaz, A.~Jolfaei, S.~H. Ahmed, and
  A.~K. Bashir, ``Learning-based context-aware resource allocation for
  edge-computing-empowered industrial iot,'' \emph{IEEE Internet of Things
  Journal}, vol.~7, no.~5, pp. 4260--4277, 2019.

\bibitem{sun2018double}
W.~Sun, J.~Liu, Y.~Yue, and H.~Zhang, ``Double auction-based resource
  allocation for mobile edge computing in industrial internet of things,''
  \emph{IEEE Transactions on Industrial Informatics}, vol.~14, no.~10, pp.
  4692--4701, 2018.

\bibitem{deng2019parallel}
Y.~Deng, Z.~Chen, X.~Yao, S.~Hassan, and A.~M. Ibrahim, ``Parallel offloading
  in green and sustainable mobile edge computing for delay-constrained iot
  system,'' \emph{IEEE Transactions on Vehicular Technology}, vol.~68, no.~12,
  pp. 12\,202--12\,214, 2019.

\bibitem{ali2019deep}
Z.~Ali, L.~Jiao, T.~Baker, G.~Abbas, Z.~H. Abbas, and S.~Khaf, ``A deep
  learning approach for energy efficient computational offloading in mobile
  edge computing,'' \emph{IEEE Access}, vol.~7, pp. 149\,623--149\,633, 2019.

\bibitem{wang2018energy}
C.~Wang, C.~Dong, J.~Qin, X.~Yang, and W.~Wen, ``Energy-efficient offloading
  policy for resource allocation in distributed mobile edge computing,'' in
  \emph{2018 IEEE Symposium on Computers and Communications (ISCC)}.\hskip 1em
  plus 0.5em minus 0.4em\relax IEEE, 2018, pp. 00\,366--00\,372.

\bibitem{liu2019deep}
Y.~Liu, H.~Yu, S.~Xie, and Y.~Zhang, ``Deep reinforcement learning for
  offloading and resource allocation in vehicle edge computing and networks,''
  \emph{IEEE Transactions on Vehicular Technology}, vol.~68, no.~11, pp.
  11\,158--11\,168, 2019.

\bibitem{yang2019energy}
Z.~Yang, C.~Pan, K.~Wang, and M.~Shikh-Bahaei, ``Energy efficient resource
  allocation in uav-enabled mobile edge computing networks,'' \emph{IEEE
  Transactions on Wireless Communications}, vol.~18, no.~9, pp. 4576--4589,
  2019.

\bibitem{chen2019efficient}
X.~Chen, Y.~Cai, Q.~Shi, M.~Zhao, B.~Champagne, and L.~Hanzo, ``Efficient
  resource allocation for relay-assisted computation offloading in mobile-edge
  computing,'' \emph{IEEE Internet of Things Journal}, vol.~7, no.~3, pp.
  2452--2468, 2019.

\bibitem{guo2018collaborative}
H.~Guo and J.~Liu, ``Collaborative computation offloading for multiaccess edge
  computing over fiber--wireless networks,'' \emph{IEEE Transactions on
  Vehicular Technology}, vol.~67, no.~5, pp. 4514--4526, 2018.

\bibitem{hong2019multi}
Z.~Hong, W.~Chen, H.~Huang, S.~Guo, and Z.~Zheng, ``Multi-hop cooperative
  computation offloading for industrial iot--edge--cloud computing
  environments,'' \emph{IEEE Transactions on Parallel and Distributed Systems},
  vol.~30, no.~12, pp. 2759--2774, 2019.

\bibitem{wang2019hetmec}
P.~Wang, Z.~Zheng, B.~Di, and L.~Song, ``Hetmec: Latency-optimal task
  assignment and resource allocation for heterogeneous multi-layer mobile edge
  computing,'' \emph{IEEE Transactions on Wireless Communications}, vol.~18,
  no.~10, pp. 4942--4956, 2019.

\bibitem{na2018frequency}
W.~Na, S.~Jang, Y.~Lee, L.~Park, N.-N. Dao, and S.~Cho, ``Frequency resource
  allocation and interference management in mobile edge computing for an
  internet of things system,'' \emph{IEEE Internet of Things Journal}, vol.~6,
  no.~3, pp. 4910--4920, 2018.

\bibitem{alameddine2019dynamic}
H.~A. Alameddine, S.~Sharafeddine, S.~Sebbah, S.~Ayoubi, and C.~Assi, ``Dynamic
  task offloading and scheduling for low-latency iot services in multi-access
  edge computing,'' \emph{IEEE Journal on Selected Areas in Communications},
  vol.~37, no.~3, pp. 668--682, 2019.

\bibitem{miao2020intelligent}
Y.~Miao, G.~Wu, M.~Li, A.~Ghoneim, M.~Al-Rakhami, and M.~S. Hossain,
  ``Intelligent task prediction and computation offloading based on mobile-edge
  cloud computing,'' \emph{Future Generation Computer Systems}, vol. 102, pp.
  925--931, 2020.

\bibitem{thai2019workload}
M.-T. Thai, Y.-D. Lin, Y.-C. Lai, and H.-T. Chien, ``Workload and capacity
  optimization for cloud-edge computing systems with vertical and horizontal
  offloading,'' \emph{IEEE Transactions on Network and Service Management},
  vol.~17, no.~1, pp. 227--238, 2019.

\bibitem{xu2017zenith}
J.~Xu, B.~Palanisamy, H.~Ludwig, and Q.~Wang, ``Zenith: Utility-aware resource
  allocation for edge computing,'' in \emph{2017 IEEE international conference
  on edge computing (EDGE)}.\hskip 1em plus 0.5em minus 0.4em\relax IEEE, 2017,
  pp. 47--54.

\bibitem{zhang2019dmra}
C.~Zhang, H.~Du, Q.~Ye, C.~Liu, and H.~Yuan, ``Dmra: A decentralized resource
  allocation scheme for multi-sp mobile edge computing,'' in \emph{2019 IEEE
  39th International Conference on Distributed Computing Systems
  (ICDCS)}.\hskip 1em plus 0.5em minus 0.4em\relax IEEE, 2019, pp. 390--398.

\bibitem{NTT2020eawp}
NTT, ``Edge accelerated web platform,'' Website, 2020,
  \url{https://www.ntt.co.jp/news2014/1401e/140123a.html}.

\bibitem{forecast2019cisco}
G.~M. D.~T. Forecast, ``Cisco visual networking index: global mobile data
  traffic forecast update, 2017--2022,'' \emph{Update}, vol. 2017, p. 2022,
  2019.

\bibitem{erman2011over}
J.~Erman, A.~Gerber, K.~Ramadrishnan, S.~Sen, and O.~Spatscheck, ``Over the top
  video: the gorilla in cellular networks,'' in \emph{Proceedings of the 2011
  ACM SIGCOMM conference on Internet measurement conference}, 2011, pp.
  127--136.

\bibitem{miettinen2010energy}
A.~P. Miettinen and J.~K. Nurminen, ``Energy efficiency of mobile clients in
  cloud computing.'' \emph{HotCloud}, vol.~10, no.~4, pp. 1--7, 2010.

\bibitem{amdahl1967validity}
G.~M. Amdahl, ``Validity of the single processor approach to achieving large
  scale computing capabilities,'' in \emph{Proceedings of the April 18-20,
  1967, spring joint computer conference}, 1967, pp. 483--485.

\bibitem{wang2016mobile}
Y.~Wang, M.~Sheng, X.~Wang, L.~Wang, and J.~Li, ``Mobile-edge computing:
  Partial computation offloading using dynamic voltage scaling,'' \emph{IEEE
  Transactions on Communications}, vol.~64, no.~10, pp. 4268--4282, 2016.

\bibitem{wang2017computation}
C.~Wang, C.~Liang, F.~R. Yu, Q.~Chen, and L.~Tang, ``Computation offloading and
  resource allocation in wireless cellular networks with mobile edge
  computing,'' \emph{IEEE Transactions on Wireless Communications}, vol.~16,
  no.~8, pp. 4924--4938, 2017.

\bibitem{mao2017stochastic}
Y.~Mao, J.~Zhang, S.~Song, and K.~B. Letaief, ``Stochastic joint radio and
  computational resource management for multi-user mobile-edge computing
  systems,'' \emph{IEEE Transactions on Wireless Communications}, vol.~16,
  no.~9, pp. 5994--6009, 2017.

\bibitem{du2018computation}
J.~Du, L.~Zhao, J.~Feng, and X.~Chu, ``Computation offloading and resource
  allocation in mixed fog/cloud computing systems with min-max fairness
  guarantee,'' \emph{IEEE Transactions on Communications}, vol.~66, no.~4, pp.
  1594--1608, 2018.

\bibitem{kim2017dual}
Y.~Kim, J.~Kwak, and S.~Chong, ``Dual-side optimization for cost-delay tradeoff
  in mobile edge computing,'' \emph{IEEE Transactions on Vehicular Technology},
  vol.~67, no.~2, pp. 1765--1781, 2017.

\bibitem{ding2019code}
Y.~Ding, C.~Liu, X.~Zhou, Z.~Liu, and Z.~Tang, ``A code-oriented partitioning
  computation offloading strategy for multiple users and multiple mobile edge
  computing servers,'' \emph{IEEE Transactions on Industrial Informatics},
  vol.~16, no.~7, pp. 4800--4810, 2019.

\bibitem{feng2019computation}
J.~Feng, Q.~Pei, F.~R. Yu, X.~Chu, and B.~Shang, ``Computation offloading and
  resource allocation for wireless powered mobile edge computing with latency
  constraint,'' \emph{IEEE Wireless Communications Letters}, vol.~8, no.~5, pp.
  1320--1323, 2019.

\bibitem{gu2018context}
B.~Gu, Z.~Zhou, S.~Mumtaz, V.~Frascolla, and A.~K. Bashir, ``Context-aware task
  offloading for multi-access edge computing: matching with externalities,'' in
  \emph{2018 IEEE Global Communications Conference (GLOBECOM)}.\hskip 1em plus
  0.5em minus 0.4em\relax IEEE, 2018, pp. 1--6.

\bibitem{liu2019energy}
F.~Liu, Z.~Huang, and L.~Wang, ``Energy-efficient collaborative task
  computation offloading in cloud-assisted edge computing for iot sensors,''
  \emph{Sensors}, vol.~19, no.~5, p. 1105, 2019.

\bibitem{liu2020distributed}
B.~Liu, Y.~Cao, Y.~Zhang, and T.~Jiang, ``A distributed framework for task
  offloading in edge computing networks of arbitrary topology,'' \emph{IEEE
  Transactions on Wireless Communications}, vol.~19, no.~4, pp. 2855--2867,
  2020.

\bibitem{nguyen2019computation}
T.~T. Nguyen, L.~Le, and Q.~Le-Trung, ``Computation offloading in mimo based
  mobile edge computing systems under perfect and imperfect csi estimation,''
  \emph{IEEE Transactions on Services Computing}, 2019.

\bibitem{yang2018distributed}
L.~Yang, H.~Zhang, X.~Li, H.~Ji, and V.~C. Leung, ``A distributed computation
  offloading strategy in small-cell networks integrated with mobile edge
  computing,'' \emph{IEEE/ACM Transactions on Networking}, vol.~26, no.~6, pp.
  2762--2773, 2018.

\bibitem{yang2018mobile}
L.~Yang, H.~Zhang, M.~Li, J.~Guo, and H.~Ji, ``Mobile edge computing empowered
  energy efficient task offloading in 5g,'' \emph{IEEE Transactions on
  Vehicular Technology}, vol.~67, no.~7, pp. 6398--6409, 2018.

\bibitem{chen2018multi}
W.~Chen, D.~Wang, and K.~Li, ``Multi-user multi-task computation offloading in
  green mobile edge cloud computing,'' \emph{IEEE Transactions on Services
  Computing}, vol.~12, no.~5, pp. 726--738, 2018.

\bibitem{kiran2019joint}
N.~Kiran, C.~Pan, S.~Wang, and C.~Yin, ``Joint resource allocation and
  computation offloading in mobile edge computing for sdn based wireless
  networks,'' \emph{Journal of Communications and Networks}, vol.~22, no.~1,
  pp. 1--11, 2019.

\bibitem{liu2018offloading}
J.~Liu and Q.~Zhang, ``Offloading schemes in mobile edge computing for
  ultra-reliable low latency communications,'' \emph{Ieee Access}, vol.~6, pp.
  12\,825--12\,837, 2018.

\bibitem{ning2018cooperative}
Z.~Ning, P.~Dong, X.~Kong, and F.~Xia, ``A cooperative partial computation
  offloading scheme for mobile edge computing enabled internet of things,''
  \emph{IEEE Internet of Things Journal}, vol.~6, no.~3, pp. 4804--4814, 2018.

\bibitem{pan2018energy}
Y.~Pan, M.~Chen, Z.~Yang, N.~Huang, and M.~Shikh-Bahaei, ``Energy-efficient
  noma-based mobile edge computing offloading,'' \emph{IEEE Communications
  Letters}, vol.~23, no.~2, pp. 310--313, 2018.

\bibitem{saleem2018performance}
U.~Saleem, Y.~Liu, S.~Jangsher, and Y.~Li, ``Performance guaranteed partial
  offloading for mobile edge computing,'' in \emph{2018 IEEE Global
  Communications Conference (GLOBECOM)}.\hskip 1em plus 0.5em minus 0.4em\relax
  IEEE, 2018, pp. 1--6.

\bibitem{shu2019multi}
C.~Shu, Z.~Zhao, Y.~Han, G.~Min, and H.~Duan, ``Multi-user offloading for edge
  computing networks: A dependency-aware and latency-optimal approach,''
  \emph{IEEE Internet of Things Journal}, vol.~7, no.~3, pp. 1678--1689, 2019.

\bibitem{xu2019joint}
X.~Xu, C.~He, Z.~Xu, L.~Qi, S.~Wan, and M.~Z.~A. Bhuiyan, ``Joint optimization
  of offloading utility and privacy for edge computing enabled iot,''
  \emph{IEEE Internet of Things Journal}, vol.~7, no.~4, pp. 2622--2629, 2019.

\bibitem{chen2018computation}
L.~Chen, S.~Zhou, and J.~Xu, ``Computation peer offloading for
  energy-constrained mobile edge computing in small-cell networks,''
  \emph{IEEE/ACM Transactions on Networking}, vol.~26, no.~4, pp. 1619--1632,
  2018.

\bibitem{guo2018anefficient}
F.~Guo, H.~Zhang, H.~Ji, X.~Li, and V.~C. Leung, ``An efficient computation
  offloading management scheme in the densely deployed small cell networks with
  mobile edge computing,'' \emph{IEEE/ACM Transactions on Networking}, vol.~26,
  no.~6, pp. 2651--2664, 2018.

\bibitem{jovsilo2020computation}
S.~Jo{\v{s}}ilo and G.~D{\'a}n, ``Computation offloading scheduling for
  periodic tasks in mobile edge computing,'' \emph{IEEE/ACM Transactions on
  Networking}, vol.~28, no.~2, pp. 667--680, 2020.

\bibitem{zhang2017optimal}
K.~Zhang, Y.~Mao, S.~Leng, S.~Maharjan, and Y.~Zhang, ``Optimal delay
  constrained offloading for vehicular edge computing networks,'' in \emph{2017
  IEEE International Conference on Communications (ICC)}.\hskip 1em plus 0.5em
  minus 0.4em\relax IEEE, 2017, pp. 1--6.

\bibitem{zhao2019computation}
J.~Zhao, Q.~Li, Y.~Gong, and K.~Zhang, ``Computation offloading and resource
  allocation for cloud assisted mobile edge computing in vehicular networks,''
  \emph{IEEE Transactions on Vehicular Technology}, vol.~68, no.~8, pp.
  7944--7956, 2019.

\bibitem{chen2015early}
Z.~Chen, L.~Jiang, W.~Hu, K.~Ha, B.~Amos, P.~Pillai, A.~Hauptmann, and
  M.~Satyanarayanan, ``Early implementation experience with wearable cognitive
  assistance applications,'' in \emph{Proceedings of the 2015 workshop on
  Wearable Systems and Applications}, 2015, pp. 33--38.

\bibitem{lin2007enhancing}
Y.~Lin, B.~Kemme, M.~Patino-Martinez, and R.~Jimenez-Peris, ``Enhancing edge
  computing with database replication,'' in \emph{2007 26th IEEE International
  Symposium on Reliable Distributed Systems (SRDS 2007)}.\hskip 1em plus 0.5em
  minus 0.4em\relax IEEE, 2007, pp. 45--54.

\bibitem{gao2003application}
L.~Gao, M.~Dahlin, A.~Nayate, J.~Zheng, and A.~Iyengar, ``Application specific
  data replication for edge services,'' in \emph{Proceedings of the 12th
  international conference on World Wide Web}, 2003, pp. 449--460.

\bibitem{luo2020edgevcd}
Q.~Luo, C.~Li, T.~H. Luan, and W.~Shi, ``Edgevcd: Intelligent
  algorithm-inspired content distribution in vehicular edge computing
  network,'' \emph{IEEE Internet of Things Journal}, vol.~7, no.~6, pp.
  5562--5579, 2020.

\bibitem{amento2016focusstack}
B.~Amento, B.~Balasubramanian, R.~J. Hall, K.~Joshi, G.~Jung, and K.~H. Purdy,
  ``Focusstack: Orchestrating edge clouds using location-based focus of
  attention,'' in \emph{2016 IEEE/ACM Symposium on Edge Computing (SEC)}.\hskip
  1em plus 0.5em minus 0.4em\relax IEEE, 2016, pp. 179--191.

\bibitem{liu2016paradrop}
P.~Liu, D.~Willis, and S.~Banerjee, ``Paradrop: Enabling lightweight
  multi-tenancy at the network’s extreme edge,'' in \emph{2016 IEEE/ACM
  Symposium on Edge Computing (SEC)}.\hskip 1em plus 0.5em minus 0.4em\relax
  IEEE, 2016, pp. 1--13.

\bibitem{chen2019dynamic}
M.~Chen, W.~Li, G.~Fortino, Y.~Hao, L.~Hu, and I.~Humar, ``A dynamic service
  migration mechanism in edge cognitive computing,'' \emph{ACM Transactions on
  Internet Technology (TOIT)}, vol.~19, no.~2, pp. 1--15, 2019.

\bibitem{ma2017efficient}
L.~Ma, S.~Yi, and Q.~Li, ``Efficient service handoff across edge servers via
  docker container migration,'' in \emph{Proceedings of the Second ACM/IEEE
  Symposium on Edge Computing}, 2017, pp. 1--13.

\bibitem{xiao2017qoe}
Y.~Xiao and M.~Krunz, ``Qoe and power efficiency tradeoff for fog computing
  networks with fog node cooperation,'' in \emph{IEEE INFOCOM 2017-IEEE
  Conference on Computer Communications}.\hskip 1em plus 0.5em minus
  0.4em\relax IEEE, 2017, pp. 1--9.

\bibitem{luo2020collaborative}
Q.~Luo, C.~Li, T.~H. Luan, and W.~Shi, ``Collaborative data scheduling for
  vehicular edge computing via deep reinforcement learning,'' \emph{IEEE
  Internet of Things Journal}, vol.~7, no.~10, pp. 9637--9650, 2020.

\bibitem{hu2019learning}
M.~Hu, L.~Zhuang, D.~Wu, Y.~Zhou, X.~Chen, and L.~Xiao, ``Learning driven
  computation offloading for asymmetrically informed edge computing,''
  \emph{IEEE Transactions on Parallel and Distributed Systems}, vol.~30, no.~8,
  pp. 1802--1815, 2019.

\bibitem{li2019energy}
S.~Li, D.~Zhai, P.~Du, and T.~Han, ``Energy-efficient task offloading, load
  balancing, and resource allocation in mobile edge computing enabled iot
  networks,'' \emph{Science China Information Sciences}, vol.~62, no.~2, p.
  29307, 2019.

\bibitem{liu2019dynamic}
C.-F. Liu, M.~Bennis, M.~Debbah, and H.~V. Poor, ``Dynamic task offloading and
  resource allocation for ultra-reliable low-latency edge computing,''
  \emph{IEEE Transactions on Communications}, vol.~67, no.~6, pp. 4132--4150,
  2019.

\bibitem{mazouzi2019dm2}
H.~Mazouzi, N.~Achir, and K.~Boussetta, ``Dm2-ecop: An efficient computation
  offloading policy for multi-user multi-cloudlet mobile edge computing
  environment,'' \emph{ACM Transactions on Internet Technology (TOIT)},
  vol.~19, no.~2, pp. 1--24, 2019.

\bibitem{yang2018multivessel}
T.~Yang, H.~Feng, C.~Yang, Y.~Wang, J.~Dong, and M.~Xia, ``Multivessel
  computation offloading in maritime mobile edge computing network,''
  \emph{IEEE Internet of Things Journal}, vol.~6, no.~3, pp. 4063--4073, 2018.

\bibitem{yu2020joint}
Z.~Yu, Y.~Gong, S.~Gong, and Y.~Guo, ``Joint task offloading and resource
  allocation in uav-enabled mobile edge computing,'' \emph{IEEE Internet of
  Things Journal}, vol.~7, no.~4, pp. 3147--3159, 2020.

\bibitem{xiao2019task}
K.~Xiao, Z.~Gao, C.~Yao, Q.~Wang, Z.~Mo, and Y.~Yang, ``Task offloading and
  resources allocation based on fairness in edge computing,'' in \emph{2019
  IEEE Wireless Communications and Networking Conference (WCNC)}.\hskip 1em
  plus 0.5em minus 0.4em\relax IEEE, 2019, pp. 1--6.

\bibitem{luo2021minimizing}
Q.~Luo, C.~Li, T.~H. Luan, and W.~Shi, ``Minimizing the delay and cost of
  computation offloading for vehicular edge computing,'' \emph{IEEE
  Transactions on Services Computing}, pp. 1--1, 2021, early access, doi:
  {10.1109/TSC.2021.3064579}.

\bibitem{liu2019reliability}
J.~Liu and Q.~Zhang, ``Reliability and latency aware code-partitioning
  offloading in mobile edge computing,'' in \emph{2019 IEEE Wireless
  Communications and Networking Conference (WCNC)}.\hskip 1em plus 0.5em minus
  0.4em\relax IEEE, 2019, pp. 1--7.

\bibitem{bi2018computation}
S.~Bi and Y.~J. Zhang, ``Computation rate maximization for wireless powered
  mobile-edge computing with binary computation offloading,'' \emph{IEEE
  Transactions on Wireless Communications}, vol.~17, no.~6, pp. 4177--4190,
  2018.

\bibitem{chen2018task}
M.~Chen and Y.~Hao, ``Task offloading for mobile edge computing in software
  defined ultra-dense network,'' \emph{IEEE Journal on Selected Areas in
  Communications}, vol.~36, no.~3, pp. 587--597, 2018.

\bibitem{elgendy2019resource}
I.~A. Elgendy, W.~Zhang, Y.-C. Tian, and K.~Li, ``Resource allocation and
  computation offloading with data security for mobile edge computing,''
  \emph{Future Generation Computer Systems}, vol. 100, pp. 531--541, 2019.

\bibitem{gu2018optimal}
Q.~Gu, G.~Wang, J.~Liu, R.~Fan, D.~Fan, and Z.~Zhong, ``Optimal offloading with
  non-orthogonal multiple access in mobile edge computing,'' in \emph{2018 IEEE
  Global Communications Conference (GLOBECOM)}.\hskip 1em plus 0.5em minus
  0.4em\relax IEEE, 2018, pp. 1--5.

\bibitem{bahreini2019energy}
T.~Bahreini, H.~Badri, and D.~Grosu, ``Energy-aware capacity provisioning and
  resource allocation in edge computing systems,'' in \emph{International
  Conference on Edge Computing}.\hskip 1em plus 0.5em minus 0.4em\relax
  Springer, 2019, pp. 31--45.

\bibitem{kiani2019hierarchical}
A.~Kiani, N.~Ansari, and A.~Khreishah, ``Hierarchical capacity provisioning for
  fog computing,'' \emph{IEEE/ACM Transactions on Networking}, vol.~27, no.~3,
  pp. 962--971, 2019.

\bibitem{khalili2019joint}
A.~Khalili, S.~Zarandi, and M.~Rasti, ``Joint resource allocation and
  offloading decision in mobile edge computing,'' \emph{IEEE Communications
  Letters}, vol.~23, no.~4, pp. 684--687, 2019.

\bibitem{kuang2019partial}
Z.~Kuang, L.~Li, J.~Gao, L.~Zhao, and A.~Liu, ``Partial offloading scheduling
  and power allocation for mobile edge computing systems,'' \emph{IEEE Internet
  of Things Journal}, vol.~6, no.~4, pp. 6774--6785, 2019.

\bibitem{li2018energyaware}
L.~Li, X.~Zhang, K.~Liu, F.~Jiang, and J.~Peng, ``An energy-aware task
  offloading mechanism in multiuser mobile-edge cloud computing,'' \emph{Mobile
  Information Systems}, vol. 2018, 2018.

\bibitem{lyu2018selective}
X.~Lyu, H.~Tian, L.~Jiang, A.~Vinel, S.~Maharjan, S.~Gjessing, and Y.~Zhang,
  ``Selective offloading in mobile edge computing for the green internet of
  things,'' \emph{IEEE Network}, vol.~32, no.~1, pp. 54--60, 2018.

\bibitem{xu2019computation}
X.~Xu, Q.~Liu, Y.~Luo, K.~Peng, X.~Zhang, S.~Meng, and L.~Qi, ``A computation
  offloading method over big data for iot-enabled cloud-edge computing,''
  \emph{Future Generation Computer Systems}, vol.~95, pp. 522--533, 2019.

\bibitem{yu2018computation}
S.~Yu, R.~Langar, X.~Fu, L.~Wang, and Z.~Han, ``Computation offloading with
  data caching enhancement for mobile edge computing,'' \emph{IEEE Transactions
  on Vehicular Technology}, vol.~67, no.~11, pp. 11\,098--11\,112, 2018.

\bibitem{xu2018joint}
J.~Xu, L.~Chen, and P.~Zhou, ``Joint service caching and task offloading for
  mobile edge computing in dense networks,'' in \emph{IEEE INFOCOM 2018-IEEE
  Conference on Computer Communications}.\hskip 1em plus 0.5em minus
  0.4em\relax IEEE, 2018, pp. 207--215.

\bibitem{nikoloudakis2018edge}
Y.~Nikoloudakis, E.~Markakis, G.~Alexiou, S.~Bourazani, G.~Mastorakis,
  E.~Pallis, I.~Politis, C.~Skianis, and C.~Mavromoustakis, ``Edge caching
  architecture for media delivery over p2p networks,'' in \emph{2018 IEEE 23rd
  International Workshop on Computer Aided Modeling and Design of Communication
  Links and Networks (CAMAD)}.\hskip 1em plus 0.5em minus 0.4em\relax IEEE,
  2018, pp. 1--5.

\bibitem{guo2018joint}
F.~Guo, L.~Ma, H.~Zhang, H.~Ji, and X.~Li, ``Joint load management and resource
  allocation in the energy harvesting powered small cell networks with mobile
  edge computing,'' in \emph{IEEE INFOCOM 2018-IEEE Conference on Computer
  Communications Workshops (INFOCOM WKSHPS)}.\hskip 1em plus 0.5em minus
  0.4em\relax IEEE, 2018, pp. 299--304.

\bibitem{guo2019adaptive}
F.~Guo, F.~R. Yu, H.~Zhang, H.~Ji, M.~Liu, and V.~C. Leung, ``Adaptive resource
  allocation in future wireless networks with blockchain and mobile edge
  computing,'' \emph{IEEE Transactions on Wireless Communications}, vol.~19,
  no.~3, pp. 1689--1703, 2019.

\bibitem{qian2019noma}
L.~P. Qian, B.~Shi, Y.~Wu, B.~Sun, and D.~H. Tsang, ``Noma-enabled mobile edge
  computing for internet of things via joint communication and computation
  resource allocations,'' \emph{IEEE Internet of Things Journal}, vol.~7,
  no.~1, pp. 718--733, 2019.

\bibitem{wang2018dynamic}
F.~Wang and X.~Zhang, ``Dynamic interface-selection and resource allocation
  over heterogeneous mobile edge-computing wireless networks with energy
  harvesting,'' in \emph{IEEE INFOCOM 2018-IEEE Conference on Computer
  Communications Workshops (INFOCOM WKSHPS)}.\hskip 1em plus 0.5em minus
  0.4em\relax IEEE, 2018, pp. 190--195.

\bibitem{wang2019effective}
Y.~Wang, X.~Tao, Y.~T. Hou, and P.~Zhang, ``Effective capacity-based resource
  allocation in mobile edge computing with two-stage tandem queues,''
  \emph{IEEE Transactions on Communications}, vol.~67, no.~9, pp. 6221--6233,
  2019.

\bibitem{xing2019joint}
H.~Xing, L.~Liu, J.~Xu, and A.~Nallanathan, ``Joint task assignment and
  resource allocation for d2d-enabled mobile-edge computing,'' \emph{IEEE
  Transactions on Communications}, vol.~67, no.~6, pp. 4193--4207, 2019.

\bibitem{yang2019efficientresource}
Z.~Yang, C.~Pan, J.~Hou, and M.~Shikh-Bahaei, ``Efficient resource allocation
  for mobile-edge computing networks with noma: Completion time and energy
  minimization,'' \emph{IEEE Transactions on Communications}, vol.~67, no.~11,
  pp. 7771--7784, 2019.

\bibitem{zhao2019context}
P.~Zhao, H.~Tian, K.-C. Chen, S.~Fan, and G.~Nie, ``Context-aware tdd
  configuration and resource allocation for mobile edge computing,'' \emph{IEEE
  Transactions on Communications}, vol.~68, no.~2, pp. 1118--1131, 2019.

\bibitem{zhao2020mobile}
C.~Zhao, Y.~Cai, A.~Liu, M.~Zhao, and L.~Hanzo, ``Mobile edge computing meets
  mmwave communications: Joint beamforming and resource allocation for system
  delay minimization,'' \emph{IEEE Transactions on Wireless Communications},
  vol.~19, no.~4, pp. 2382--2396, 2020.

\bibitem{elgendy2020efficient}
I.~A. Elgendy, W.-Z. Zhang, Y.~Zeng, H.~He, Y.-C. Tian, and Y.~Yang,
  ``Efficient and secure multi-user multi-task computation offloading for
  mobile-edge computing in mobile iot networks,'' \emph{IEEE Transactions on
  Network and Service Management}, vol.~17, no.~4, pp. 2410--2422, 2020.

\bibitem{wang2019edgeai}
X.~Wang, Y.~Han, C.~Wang, Q.~Zhao, X.~Chen, and M.~Chen, ``In-edge ai:
  Intelligentizing mobile edge computing, caching and communication by
  federated learning,'' \emph{IEEE Network}, vol.~33, no.~5, pp. 156--165,
  2019.

\bibitem{liang2017energy}
C.~Liang, Y.~He, F.~R. Yu, and N.~Zhao, ``Energy-efficient resource allocation
  in software-defined mobile networks with mobile edge computing and caching,''
  in \emph{2017 IEEE Conference on Computer Communications Workshops (INFOCOM
  WKSHPS)}.\hskip 1em plus 0.5em minus 0.4em\relax IEEE, 2017, pp. 121--126.

\bibitem{zhou2017resource}
Y.~Zhou, F.~R. Yu, J.~Chen, and Y.~Kuo, ``Resource allocation for
  information-centric virtualized heterogeneous networks with in-network
  caching and mobile edge computing,'' \emph{IEEE Transactions on Vehicular
  Technology}, vol.~66, no.~12, pp. 11\,339--11\,351, 2017.

\bibitem{cui2017energy}
Y.~Cui, W.~He, C.~Ni, C.~Guo, and Z.~Liu, ``Energy-efficient resource
  allocation for cache-assisted mobile edge computing,'' in \emph{2017 IEEE
  42nd Conference on Local Computer Networks (LCN)}.\hskip 1em plus 0.5em minus
  0.4em\relax IEEE, 2017, pp. 640--648.

\bibitem{hao2018energy}
Y.~Hao, M.~Chen, L.~Hu, M.~S. Hossain, and A.~Ghoneim, ``Energy efficient task
  caching and offloading for mobile edge computing,'' \emph{IEEE Access},
  vol.~6, pp. 11\,365--11\,373, 2018.

\bibitem{yang2019cloudlet}
S.~Yang, F.~Li, M.~Shen, X.~Chen, X.~Fu, and Y.~Wang, ``Cloudlet placement and
  task allocation in mobile edge computing,'' \emph{IEEE Internet of Things
  Journal}, vol.~6, no.~3, pp. 5853--5863, 2019.

\bibitem{breitbach2019context}
M.~Breitbach, D.~Sch{\"a}fer, J.~Edinger, and C.~Becker, ``Context-aware data
  and task placement in edge computing environments,'' in \emph{2019 IEEE
  International Conference on Pervasive Computing and Communications
  (PerCom)}.\hskip 1em plus 0.5em minus 0.4em\relax IEEE, 2019, pp. 1--10.

\bibitem{fan2017deadline}
J.~Fan, X.~Wei, T.~Wang, T.~Lan, and S.~Subramaniam, ``Deadline-aware task
  scheduling in a tiered iot infrastructure,'' in \emph{GLOBECOM 2017-2017 IEEE
  Global Communications Conference}.\hskip 1em plus 0.5em minus 0.4em\relax
  IEEE, 2017, pp. 1--7.

\bibitem{cao2018performance}
Z.~Cao, H.~Zhang, and B.~Liu, ``Performance and stability of application
  placement in mobile edge computing system,'' in \emph{2018 IEEE 37th
  International Performance Computing and Communications Conference
  (IPCCC)}.\hskip 1em plus 0.5em minus 0.4em\relax IEEE, 2018, pp. 1--8.

\bibitem{mahmud2019quality}
R.~Mahmud, S.~N. Srirama, K.~Ramamohanarao, and R.~Buyya, ``Quality of
  experience (qoe)-aware placement of applications in fog computing
  environments,'' \emph{Journal of Parallel and Distributed Computing}, vol.
  132, pp. 190--203, 2019.

\bibitem{mahmud2020profit}
------, ``Profit-aware application placement for integrated fog--cloud
  computing environments,'' \emph{Journal of Parallel and Distributed
  Computing}, vol. 135, pp. 177--190, 2020.

\bibitem{fan2019cost}
Q.~Fan and N.~Ansari, ``On cost aware cloudlet placement for mobile edge
  computing,'' \emph{IEEE/CAA Journal of Automatica Sinica}, vol.~6, no.~4, pp.
  926--937, 2019.

\bibitem{santoyo2020network}
A.~Santoyo-Gonz{\'a}lez and C.~Cervell{\'o}-Pastor, ``Network-aware placement
  optimization for edge computing infrastructure under 5g,'' \emph{IEEE
  access}, vol.~8, pp. 56\,015--56\,028, 2020.

\bibitem{jin2018cooperative}
J.~Jin, Y.~Li, and J.~Luo, ``Cooperative storage by exploiting graph-based data
  placement algorithm for edge computing environment,'' \emph{Concurrency and
  Computation: Practice and Experience}, vol.~30, no.~20, p. e4914, 2018.

\bibitem{chiti2019virtual}
F.~Chiti, R.~Fantacci, F.~Paganelli, and B.~Picano, ``Virtual functions
  placement with time constraints in fog computing: A matching theory
  perspective,'' \emph{IEEE Transactions on Network and Service Management},
  vol.~16, no.~3, pp. 980--989, 2019.

\bibitem{choi2019scalable}
J.~Choi and S.~Ahn, ``Scalable service placement in the fog computing
  environment for the iot-based smart city,'' \emph{Journal of Information
  Processing Systems}, vol.~15, no.~2, pp. 440--448, 2019.

\bibitem{maia2019optimized}
A.~M. Maia, Y.~Ghamri-Doudane, D.~Vieira, and M.~F. de~Castro, ``Optimized
  placement of scalable iot services in edge computing,'' in \emph{2019
  IFIP/IEEE Symposium on Integrated Network and Service Management (IM)}.\hskip
  1em plus 0.5em minus 0.4em\relax IEEE, 2019, pp. 189--197.

\bibitem{suresh2019fnsched}
A.~Suresh and A.~Gandhi, ``Fnsched: An efficient scheduler for serverless
  functions,'' in \emph{Proceedings of the 5th International Workshop on
  Serverless Computing}, 2019, pp. 19--24.

\bibitem{aske2018supporting}
A.~Aske and X.~Zhao, ``Supporting multi-provider serverless computing on the
  edge,'' in \emph{Proceedings of the 47th International Conference on Parallel
  Processing Companion}, 2018, pp. 1--6.

\bibitem{li2018energy}
Y.~Li and S.~Wang, ``An energy-aware edge server placement algorithm in mobile
  edge computing,'' in \emph{2018 IEEE International Conference on Edge
  Computing (EDGE)}.\hskip 1em plus 0.5em minus 0.4em\relax IEEE, 2018, pp.
  66--73.

\bibitem{meng2019joint}
J.~Meng, C.~Zeng, H.~Tan, Z.~Li, B.~Li, and X.-Y. Li, ``Joint heterogeneous
  server placement and application configuration in edge computing,'' in
  \emph{2019 IEEE 25th International Conference on Parallel and Distributed
  Systems (ICPADS)}.\hskip 1em plus 0.5em minus 0.4em\relax IEEE, 2019, pp.
  488--497.

\bibitem{xiao2018heuristic}
K.~Xiao, Z.~Gao, Q.~Wang, and Y.~Yang, ``A heuristic algorithm based on
  resource requirements forecasting for server placement in edge computing,''
  in \emph{2018 IEEE/ACM Symposium on Edge Computing (SEC)}.\hskip 1em plus
  0.5em minus 0.4em\relax IEEE, 2018, pp. 354--355.

\bibitem{li2019joint}
C.~Li, J.~Bai, and J.~Tang, ``Joint optimization of data placement and
  scheduling for improving user experience in edge computing,'' \emph{Journal
  of Parallel and Distributed Computing}, vol. 125, pp. 93--105, 2019.

\bibitem{lin2019time}
B.~Lin, F.~Zhu, J.~Zhang, J.~Chen, X.~Chen, N.~N. Xiong, and J.~L. Mauri, ``A
  time-driven data placement strategy for a scientific workflow combining edge
  computing and cloud computing,'' \emph{IEEE Transactions on Industrial
  Informatics}, vol.~15, no.~7, pp. 4254--4265, 2019.

\bibitem{tang2019new}
Y.~Tang, H.~Wang, K.~Guo, T.~Luo, and T.~Chi, ``A new replica placement
  mechanism for mobile media streaming in edge computing,'' \emph{Concurrency
  and Computation: Practice and Experience}, p. e5361, 2019.

\bibitem{chen2019effective}
Z.~Chen, J.~Hu, G.~Min, and X.~Chen, ``Effective data placement for scientific
  workflows in mobile edge computing using genetic particle swarm
  optimization,'' \emph{Concurrency and Computation: Practice and Experience},
  p. e5413, 2019.

\bibitem{huang2019latency}
T.~Huang, W.~Lin, Y.~Li, L.~He, and S.~Peng, ``A latency-aware multiple data
  replicas placement strategy for fog computing,'' \emph{Journal of Signal
  Processing Systems}, vol.~91, no.~10, pp. 1191--1204, 2019.

\bibitem{gao2019winning}
B.~Gao, Z.~Zhou, F.~Liu, and F.~Xu, ``Winning at the starting line: Joint
  network selection and service placement for mobile edge computing,'' in
  \emph{IEEE INFOCOM 2019-IEEE Conference on Computer Communications}.\hskip
  1em plus 0.5em minus 0.4em\relax IEEE, 2019, pp. 1459--1467.

\bibitem{ouyang2018follow}
T.~Ouyang, Z.~Zhou, and X.~Chen, ``Follow me at the edge: Mobility-aware
  dynamic service placement for mobile edge computing,'' \emph{IEEE Journal on
  Selected Areas in Communications}, vol.~36, no.~10, pp. 2333--2345, 2018.

\bibitem{ouyang2019adaptive}
T.~Ouyang, R.~Li, X.~Chen, Z.~Zhou, and X.~Tang, ``Adaptive user-managed
  service placement for mobile edge computing: An online learning approach,''
  in \emph{IEEE INFOCOM 2019-IEEE Conference on Computer Communications}.\hskip
  1em plus 0.5em minus 0.4em\relax IEEE, 2019, pp. 1468--1476.

\bibitem{poularakis2019joint}
K.~Poularakis, J.~Llorca, A.~M. Tulino, I.~Taylor, and L.~Tassiulas, ``Joint
  service placement and request routing in multi-cell mobile edge computing
  networks,'' in \emph{IEEE INFOCOM 2019-IEEE Conference on Computer
  Communications}.\hskip 1em plus 0.5em minus 0.4em\relax IEEE, 2019, pp.
  10--18.

\bibitem{yousefpour2019fogplan}
A.~Yousefpour, A.~Patil, G.~Ishigaki, I.~Kim, X.~Wang, H.~C. Cankaya, Q.~Zhang,
  W.~Xie, and J.~P. Jue, ``Fogplan: a lightweight qos-aware dynamic fog service
  provisioning framework,'' \emph{IEEE Internet of Things Journal}, vol.~6,
  no.~3, pp. 5080--5096, 2019.

\bibitem{chen2019collaborative}
L.~Chen, C.~Shen, P.~Zhou, and J.~Xu, ``Collaborative service placement for
  edge computing in dense small cell networks,'' \emph{IEEE Transactions on
  Mobile Computing}, vol.~20, no.~2, pp. 377--390, 2021.

\bibitem{goudarzi2020application}
M.~Goudarzi, H.~Wu, M.~S. Palaniswami, and R.~Buyya, ``An application placement
  technique for concurrent iot applications in edge and fog computing
  environments,'' \emph{IEEE Transactions on Mobile Computing}, 2020.

\bibitem{mutichiro2019usage}
B.~Mutichiro, H.~Yang, and Y.~Kim, ``Usage aware vnf placement for improved qos
  in edge computing,'' in \emph{2019 International Conference on Information
  and Communication Technology Convergence (ICTC)}.\hskip 1em plus 0.5em minus
  0.4em\relax IEEE, 2019, pp. 808--812.

\bibitem{wu2019efficient}
H.~Wu, W.~J. Knottenbelt, and K.~Wolter, ``An efficient application
  partitioning algorithm in mobile environments,'' \emph{IEEE Transactions on
  Parallel and Distributed Systems}, vol.~30, no.~7, pp. 1464--1480, 2019.

\bibitem{neto2018uloof}
J.~L.~D. Neto, S.-Y. Yu, D.~F. Macedo, J.~M.~S. Nogueira, R.~Langar, and
  S.~Secci, ``Uloof: A user level online offloading framework for mobile edge
  computing,'' \emph{IEEE Transactions on Mobile Computing}, vol.~17, no.~11,
  pp. 2660--2674, 2018.

\bibitem{wang2020joint}
C.~Wang, S.~Zhang, Z.~Qian, M.~Xiao, J.~Wu, B.~Ye, and S.~Lu, ``Joint server
  assignment and resource management for edge-based mar system,''
  \emph{IEEE/ACM Transactions on Networking}, vol.~28, no.~5, pp. 2378--2391,
  2020.

\bibitem{zhao2021offloading}
G.~Zhao, H.~Xu, Y.~Zhao, C.~Qiao, and L.~Huang, ``Offloading tasks with
  dependency and service caching in mobile edge computing,'' \emph{IEEE
  Transactions on Parallel and Distributed Systems}, vol.~32, no.~11, pp.
  2777--2792, 2021.

\bibitem{li2018energyawareedge}
Y.~Li and S.~Wang, ``An energy-aware edge server placement algorithm in mobile
  edge computing,'' in \emph{2018 IEEE International Conference on Edge Compung
  (EDGE)}.\hskip 1em plus 0.5em minus 0.4em\relax IEEE, 2018, pp. 66--73.

\bibitem{plachy2016dynamic}
J.~Plachy, Z.~Becvar, and E.~C. Strinati, ``Dynamic resource allocation
  exploiting mobility prediction in mobile edge computing,'' in \emph{2016 IEEE
  27th Annual International Symposium on Personal, Indoor, and Mobile Radio
  Communications (PIMRC)}.\hskip 1em plus 0.5em minus 0.4em\relax IEEE, 2016,
  pp. 1--6.

\bibitem{OpenStack2019}
``Build the future of open infrastructure,'' Website, 2019,
  \url{https://www.openstack.org}.

\bibitem{Kubernetes2020}
``Production-grade container orchestration—kubernetes,'' Website, 2020,
  \url{https://kubernetes.io}.

\bibitem{OpenEdge2019}
``Openedge support and learning,'' Website, 2019,
  \url{https://www.progress.com/support/openedge}.

\bibitem{tao2019survey}
Z.~Tao, Q.~Xia, Z.~Hao, C.~Li, L.~Ma, S.~Yi, and Q.~Li, ``A survey of virtual
  machine management in edge computing,'' \emph{Proceedings of the IEEE}, vol.
  107, no.~8, pp. 1482--1499, 2019.

\bibitem{morabito2017virtualization}
R.~Morabito, ``Virtualization on internet of things edge devices with container
  technologies: a performance evaluation,'' \emph{IEEE Access}, vol.~5, pp.
  8835--8850, 2017.

\bibitem{zhang2021joint}
J.~Zhang, X.~Zhou, T.~Ge, X.~Wang, and T.~Hwang, ``Joint task scheduling and
  containerizing for efficient edge computing,'' \emph{IEEE Transactions on
  Parallel and Distributed Systems}, vol.~32, no.~8, pp. 2086--2100, 2021.

\bibitem{aslanpour2021serverless}
M.~S. Aslanpour, A.~N. Toosi, C.~Cicconetti, B.~Javadi, P.~Sbarski, D.~Taibi,
  M.~Assuncao, S.~S. Gill, R.~Gaire, and S.~Dustdar, ``Serverless edge
  computing: vision and challenges,'' in \emph{2021 Australasian Computer
  Science Week Multiconference}, 2021, pp. 1--10.

\bibitem{he2019peace}
X.~He, R.~Jin, and H.~Dai, ``Peace: Privacy-preserving and cost-efficient task
  offloading for mobile-edge computing,'' \emph{IEEE Transactions on Wireless
  Communications}, vol.~19, no.~3, pp. 1814--1824, 2019.

\bibitem{lyu2017optimal}
X.~Lyu, W.~Ni, H.~Tian, R.~P. Liu, X.~Wang, G.~B. Giannakis, and A.~Paulraj,
  ``Optimal schedule of mobile edge computing for internet of things using
  partial information,'' \emph{IEEE Journal on Selected Areas in
  Communications}, vol.~35, no.~11, pp. 2606--2615, 2017.

\bibitem{mao2016dynamic}
Y.~Mao, J.~Zhang, and K.~B. Letaief, ``Dynamic computation offloading for
  mobile-edge computing with energy harvesting devices,'' \emph{IEEE Journal on
  Selected Areas in Communications}, vol.~34, no.~12, pp. 3590--3605, 2016.

\bibitem{zhang2020dynamic}
Q.~Zhang, L.~Gui, F.~Hou, J.~Chen, S.~Zhu, and F.~Tian, ``Dynamic task
  offloading and resource allocation for mobile-edge computing in dense cloud
  ran,'' \emph{IEEE Internet of Things Journal}, vol.~7, no.~4, pp. 3282--3299,
  2020.

\bibitem{li2019dynamic}
C.~Li, J.~Tang, and Y.~Luo, ``Dynamic multi-user computation offloading for
  wireless powered mobile edge computing,'' \emph{Journal of Network and
  Computer Applications}, vol. 131, pp. 1--15, 2019.

\bibitem{kherraf2019optimized}
N.~Kherraf, H.~A. Alameddine, S.~Sharafeddine, C.~M. Assi, and A.~Ghrayeb,
  ``Optimized provisioning of edge computing resources with heterogeneous
  workload in iot networks,'' \emph{IEEE Transactions on Network and Service
  Management}, vol.~16, no.~2, pp. 459--474, 2019.

\bibitem{saleem2020latency}
U.~Saleem, Y.~Liu, S.~Jangsher, X.~Tao, and Y.~Li, ``Latency minimization for
  d2d-enabled partial computation offloading in mobile edge computing,''
  \emph{IEEE Transactions on Vehicular Technology}, vol.~69, no.~4, pp.
  4472--4486, 2020.

\bibitem{wang2020optimal}
F.~Wang, J.~Xu, and S.~Cui, ``Optimal energy allocation and task offloading
  policy for wireless powered mobile edge computing systems,'' \emph{IEEE
  Transactions on Wireless Communications}, vol.~19, no.~4, pp. 2443--2459,
  2020.

\bibitem{liu2019uav}
Y.~Liu, K.~Xiong, Q.~Ni, P.~Fan, and K.~B. Letaief, ``Uav-assisted wireless
  powered cooperative mobile edge computing: Joint offloading, cpu control, and
  trajectory optimization,'' \emph{IEEE Internet of Things Journal}, vol.~7,
  no.~4, pp. 2777--2790, 2019.

\bibitem{yang2019joint}
X.~Yang, Z.~Fei, J.~Zheng, N.~Zhang, and A.~Anpalagan, ``Joint multi-user
  computation offloading and data caching for hybrid mobile cloud/edge
  computing,'' \emph{IEEE Transactions on Vehicular Technology}, vol.~68,
  no.~11, pp. 11\,018--11\,030, 2019.

\bibitem{do2015proximal}
C.~T. Do, N.~H. Tran, C.~Pham, M.~G.~R. Alam, J.~H. Son, and C.~S. Hong, ``A
  proximal algorithm for joint resource allocation and minimizing carbon
  footprint in geo-distributed fog computing,'' in \emph{2015 International
  Conference on Information Networking (ICOIN)}.\hskip 1em plus 0.5em minus
  0.4em\relax IEEE, 2015, pp. 324--329.

\bibitem{zhou2017virtual}
Y.~Zhou, F.~R. Yu, J.~Chen, and Y.~Kuo, ``Virtual resource allocation for
  information-centric heterogeneous networks with mobile edge computing,'' in
  \emph{2017 IEEE Conference on Computer Communications Workshops (INFOCOM
  WKSHPS)}.\hskip 1em plus 0.5em minus 0.4em\relax IEEE, 2017, pp. 235--240.

\bibitem{badri2019energy}
H.~Badri, T.~Bahreini, D.~Grosu, and K.~Yang, ``Energy-aware application
  placement in mobile edge computing: a stochastic optimization approach,''
  \emph{IEEE Transactions on Parallel and Distributed Systems}, vol.~31, no.~4,
  pp. 909--922, 2019.

\bibitem{meng2019closed}
X.~Meng, W.~Wang, Y.~Wang, V.~K. Lau, and Z.~Zhang, ``Closed-form delay-optimal
  computation offloading in mobile edge computing systems,'' \emph{IEEE
  Transactions on Wireless Communications}, vol.~18, no.~10, pp. 4653--4667,
  2019.

\bibitem{guo2020user}
Y.~Guo, S.~Wang, A.~Zhou, J.~Xu, J.~Yuan, and C.-H. Hsu, ``User
  allocation-aware edge cloud placement in mobile edge computing,''
  \emph{Software: Practice and Experience}, vol.~50, no.~5, pp. 489--502, 2020.

\bibitem{lu2019cost}
S.~Lu, J.~Wu, Y.~Duan, N.~Wang, and J.~Fang, ``Cost-efficient resource
  provision for multiple mobile users in fog computing,'' in \emph{2019 IEEE
  25th International Conference on Parallel and Distributed Systems
  (ICPADS)}.\hskip 1em plus 0.5em minus 0.4em\relax IEEE, 2019, pp. 422--429.

\bibitem{pasteris2019service}
S.~Pasteris, S.~Wang, M.~Herbster, and T.~He, ``Service placement with provable
  guarantees in heterogeneous edge computing systems,'' in \emph{IEEE INFOCOM
  2019-IEEE Conference on Computer Communications}.\hskip 1em plus 0.5em minus
  0.4em\relax IEEE, 2019, pp. 514--522.

\bibitem{luo2018optimal}
Q.~Luo, C.~Li, T.~H. Luan, and Y.~Wen, ``Optimal utility of vehicles in lte-v
  scenario: An immune clone-based spectrum allocation approach,'' \emph{IEEE
  Transactions on Intelligent Transportation Systems}, vol.~20, no.~5, pp.
  1942--1953, 2019.

\bibitem{li2019virtual}
D.~Li, P.~Hong, K.~Xue, and J.~Pei, ``Virtual network function placement and
  resource optimization in nfv and edge computing enabled networks,''
  \emph{Computer Networks}, vol. 152, pp. 12--24, 2019.

\bibitem{zhang2020energy}
W.~Zhang, Z.~Zhang, S.~Zeadally, H.-C. Chao, and V.~C. Leung,
  ``Energy-efficient workload allocation and computation resource configuration
  in distributed cloud/edge computing systems with stochastic workloads,''
  \emph{IEEE Journal on Selected Areas in Communications}, vol.~38, no.~6, pp.
  1118--1132, 2020.

\bibitem{canali2019gasp}
C.~Canali and R.~Lancellotti, ``Gasp: Genetic algorithms for service placement
  in fog computing systems,'' \emph{Algorithms}, vol.~12, no.~10, p. 201, 2019.

\bibitem{peng2019energy}
K.~Peng, M.~Zhu, Y.~Zhang, L.~Liu, J.~Zhang, V.~C. Leung, and L.~Zheng, ``An
  energy-and cost-aware computation offloading method for workflow applications
  in mobile edge computing,'' \emph{EURASIP Journal on Wireless Communications
  and Networking}, vol. 2019, no.~1, p. 207, 2019.

\bibitem{xu2020dynamic}
X.~Xu, H.~Cao, Q.~Geng, X.~Liu, F.~Dai, and C.~Wang, ``Dynamic resource
  provisioning for workflow scheduling under uncertainty in edge computing
  environment,'' \emph{Concurrency and Computation: Practice and Experience},
  p. e5674, 2020.

\bibitem{hu2019dynamic}
S.~Hu and G.~Li, ``Dynamic request scheduling optimization in mobile edge
  computing for iot applications,'' \emph{IEEE Internet of Things Journal},
  vol.~7, no.~2, pp. 1426--1437, 2019.

\bibitem{xu2019energy}
X.~Xu, Y.~Li, T.~Huang, Y.~Xue, K.~Peng, L.~Qi, and W.~Dou, ``An energy-aware
  computation offloading method for smart edge computing in wireless
  metropolitan area networks,'' \emph{Journal of Network and Computer
  Applications}, vol. 133, pp. 75--85, 2019.

\bibitem{guo2018efficient2}
F.~Guo, H.~Zhang, H.~Ji, X.~Li, and V.~C. Leung, ``An efficient computation
  offloading management scheme in the densely deployed small cell networks with
  mobile edge computing,'' \emph{IEEE/ACM Transactions on Networking}, vol.~26,
  no.~6, pp. 2651--2664, 2018.

\bibitem{mseddi2019joint}
A.~Mseddi, W.~Jaafar, H.~Elbiaze, and W.~Ajib, ``Joint container placement and
  task provisioning in dynamic fog computing,'' \emph{IEEE Internet of Things
  Journal}, vol.~6, no.~6, pp. 10\,028--10\,040, 2019.

\bibitem{wu2020efficient}
Y.~Wu, J.~Wu, L.~Chen, J.~Yan, and Y.~Luo, ``Efficient task scheduling for
  servers with dynamic states in vehicular edge computing,'' \emph{Computer
  Communications}, vol. 150, pp. 245--253, 2020.

\bibitem{huang2019bilevel}
P.-Q. Huang, Y.~Wang, K.~Wang, and Z.-Z. Liu, ``A bilevel optimization approach
  for joint offloading decision and resource allocation in cooperative mobile
  edge computing,'' \emph{IEEE transactions on cybernetics}, vol.~50, no.~10,
  pp. 4228--4241, 2019.

\bibitem{qiu2019online}
X.~Qiu, L.~Liu, W.~Chen, Z.~Hong, and Z.~Zheng, ``Online deep reinforcement
  learning for computation offloading in blockchain-empowered mobile edge
  computing,'' \emph{IEEE Transactions on Vehicular Technology}, vol.~68,
  no.~8, pp. 8050--8062, 2019.

\bibitem{ning2019deep}
Z.~Ning, P.~Dong, X.~Wang, J.~J. Rodrigues, and F.~Xia, ``Deep reinforcement
  learning for vehicular edge computing: An intelligent offloading system,''
  \emph{ACM Transactions on Intelligent Systems and Technology (TIST)},
  vol.~10, no.~6, pp. 1--24, 2019.

\bibitem{lu2020optimization}
H.~Lu, C.~Gu, F.~Luo, W.~Ding, and X.~Liu, ``Optimization of lightweight task
  offloading strategy for mobile edge computing based on deep reinforcement
  learning,'' \emph{Future Generation Computer Systems}, vol. 102, pp.
  847--861, 2020.

\bibitem{shen2019computation}
S.~Shen, Y.~Han, X.~Wang, and Y.~Wang, ``Computation offloading with multiple
  agents in edge-computing--supported iot,'' \emph{ACM Transactions on Sensor
  Networks (TOSN)}, vol.~16, no.~1, pp. 1--27, 2019.

\bibitem{liu2020resource}
X.~Liu, J.~Yu, J.~Wang, and Y.~Gao, ``Resource allocation with edge computing
  in iot networks via machine learning,'' \emph{IEEE Internet of Things
  Journal}, vol.~7, no.~4, pp. 3415--3426, 2020.

\bibitem{wang2019computation}
J.~Wang, J.~Hu, G.~Min, W.~Zhan, Q.~Ni, and N.~Georgalas, ``Computation
  offloading in multi-access edge computing using a deep sequential model based
  on reinforcement learning,'' \emph{IEEE Communications Magazine}, vol.~57,
  no.~5, pp. 64--69, 2019.

\bibitem{zhang2019deep}
K.~Zhang, Y.~Zhu, S.~Leng, Y.~He, S.~Maharjan, and Y.~Zhang, ``Deep learning
  empowered task offloading for mobile edge computing in urban informatics,''
  \emph{IEEE Internet of Things Journal}, vol.~6, no.~5, pp. 7635--7647, 2019.

\bibitem{xiong2020resource}
X.~Xiong, K.~Zheng, L.~Lei, and L.~Hou, ``Resource allocation based on deep
  reinforcement learning in iot edge computing,'' \emph{IEEE Journal on
  Selected Areas in Communications}, vol.~38, no.~6, pp. 1133--1146, 2020.

\bibitem{zhai2020toward}
Y.~Zhai, T.~Bao, L.~Zhu, M.~Shen, X.~Du, and M.~Guizani, ``Toward
  reinforcement-learning-based service deployment of 5g mobile edge computing
  with request-aware scheduling,'' \emph{IEEE Wireless Communications},
  vol.~27, no.~1, pp. 84--91, 2020.

\bibitem{yu2017computation}
S.~Yu, X.~Wang, and R.~Langar, ``Computation offloading for mobile edge
  computing: A deep learning approach,'' in \emph{2017 IEEE 28th Annual
  International Symposium on Personal, Indoor, and Mobile Radio Communications
  (PIMRC)}.\hskip 1em plus 0.5em minus 0.4em\relax IEEE, 2017, pp. 1--6.

\bibitem{chen2020resource}
X.~Chen, F.~Zhu, Z.~Chen, G.~Min, X.~Zheng, and C.~Rong, ``Resource allocation
  for cloud-based software services using prediction-enabled feedback control
  with reinforcement learning,'' \emph{IEEE Transactions on Cloud Computing},
  2020.

\bibitem{lasaulce2011game}
S.~Lasaulce and H.~Tembine, \emph{Game theory and learning for wireless
  networks: fundamentals and applications}.\hskip 1em plus 0.5em minus
  0.4em\relax Academic Press, 2011.

\bibitem{li2019cooperative}
Q.~Li, J.~Zhao, and Y.~Gong, ``Cooperative computation offloading and resource
  allocation for mobile edge computing,'' in \emph{2019 IEEE International
  Conference on Communications Workshops (ICC Workshops)}.\hskip 1em plus 0.5em
  minus 0.4em\relax IEEE, 2019, pp. 1--6.

\bibitem{liu2017price}
M.~Liu and Y.~Liu, ``Price-based distributed offloading for mobile-edge
  computing with computation capacity constraints,'' \emph{IEEE Wireless
  Communications Letters}, vol.~7, no.~3, pp. 420--423, 2017.

\bibitem{ranadheera2018computation}
S.~Ranadheera, S.~Maghsudi, and E.~Hossain, ``Computation offloading and
  activation of mobile edge computing servers: A minority game,'' \emph{IEEE
  Wireless Communications Letters}, vol.~7, no.~5, pp. 688--691, 2018.

\bibitem{zhang2018joint}
J.~Zhang, W.~Xia, F.~Yan, and L.~Shen, ``Joint computation offloading and
  resource allocation optimization in heterogeneous networks with mobile edge
  computing,'' \emph{IEEE Access}, vol.~6, pp. 19\,324--19\,337, 2018.

\bibitem{asheralieva2019hierarchical}
A.~Asheralieva and D.~Niyato, ``Hierarchical game-theoretic and reinforcement
  learning framework for computational offloading in uav-enabled mobile edge
  computing networks with multiple service providers,'' \emph{IEEE Internet of
  Things Journal}, vol.~6, no.~5, pp. 8753--8769, 2019.

\bibitem{bai2020risk}
Y.~Bai, L.~Chen, L.~Song, and J.~Xu, ``Risk-aware edge computation offloading
  using bayesian stackelberg game,'' \emph{IEEE Transactions on Network and
  Service Management}, vol.~17, no.~2, pp. 1000--1012, 2020.

\bibitem{meng2019fault}
S.~Meng, Q.~Li, T.~Wu, W.~Huang, J.~Zhang, and W.~Li, ``A fault-tolerant
  dynamic scheduling method on hierarchical mobile edge cloud computing,''
  \emph{Computational Intelligence}, vol.~35, no.~3, pp. 577--598, 2019.

\bibitem{zhan2020deep}
Y.~Zhan, S.~Guo, P.~Li, and J.~Zhang, ``A deep reinforcement learning based
  offloading game in edge computing,'' \emph{IEEE Transactions on Computers},
  vol.~69, no.~6, pp. 883--893, 2020.

\bibitem{zhang2017data}
T.~Zhang, ``Data offloading in mobile edge computing: A coalition and pricing
  based approach,'' \emph{IEEE Access}, vol.~6, pp. 2760--2767, 2017.

\bibitem{pham2018decentralized}
Q.-V. Pham, T.~Leanh, N.~H. Tran, B.~J. Park, and C.~S. Hong, ``Decentralized
  computation offloading and resource allocation for mobile-edge computing: A
  matching game approach,'' \emph{IEEE Access}, vol.~6, pp. 75\,868--75\,885,
  2018.

\bibitem{gu2019task}
B.~Gu and Z.~Zhou, ``Task offloading in vehicular mobile edge computing: A
  matching-theoretic framework,'' \emph{IEEE Vehicular Technology Magazine},
  vol.~14, no.~3, pp. 100--106, 2019.

\bibitem{he2019truthful}
J.~He, D.~Zhang, Y.~Zhou, and Y.~Zhang, ``A truthful online mechanism for
  collaborative computation offloading in mobile edge computing,'' \emph{IEEE
  Transactions on Industrial Informatics}, vol.~16, no.~7, pp. 4832--4841,
  2019.

\bibitem{li2019online}
G.~Li and J.~Cai, ``An online incentive mechanism for collaborative task
  offloading in mobile edge computing,'' \emph{IEEE Transactions on Wireless
  Communications}, vol.~19, no.~1, pp. 624--636, 2019.

\bibitem{jiao2018social}
Y.~Jiao, P.~Wang, D.~Niyato, and Z.~Xiong, ``Social welfare maximization
  auction in edge computing resource allocation for mobile blockchain,'' in
  \emph{2018 IEEE international conference on communications (ICC)}.\hskip 1em
  plus 0.5em minus 0.4em\relax IEEE, 2018, pp. 1--6.

\bibitem{ren2019federated}
J.~Ren, H.~Wang, T.~Hou, S.~Zheng, and C.~Tang, ``Federated learning-based
  computation offloading optimization in edge computing-supported internet of
  things,'' \emph{IEEE Access}, vol.~7, pp. 69\,194--69\,201, 2019.

\bibitem{qian2019privacy}
Y.~Qian, L.~Hu, J.~Chen, X.~Guan, M.~M. Hassan, and A.~Alelaiwi,
  ``Privacy-aware service placement for mobile edge computing via federated
  learning,'' \emph{Information Sciences}, vol. 505, pp. 562--570, 2019.

\bibitem{xiong2018mobile}
Z.~Xiong, Y.~Zhang, D.~Niyato, P.~Wang, and Z.~Han, ``When mobile blockchain
  meets edge computing,'' \emph{IEEE Communications Magazine}, vol.~56, no.~8,
  pp. 33--39, 2018.

\bibitem{xu2019blockchain}
X.~Xu, Y.~Chen, X.~Zhang, Q.~Liu, X.~Liu, and L.~Qi, ``A blockchain-based
  computation offloading method for edge computing in 5g networks,''
  \emph{Software: Practice and Experience}, 2019.

\bibitem{xiao2020edgeabc}
K.~Xiao, Z.~Gao, W.~Shi, X.~Qiu, Y.~Yang, and L.~Rui, ``Edgeabc: An
  architecture for task offloading and resource allocation in the internet of
  things,'' \emph{Future Generation Computer Systems}, vol. 107, pp. 498--508,
  2020.

\bibitem{gu2015matching}
Y.~Gu, W.~Saad, M.~Bennis, M.~Debbah, and Z.~Han, ``Matching theory for future
  wireless networks: Fundamentals and applications,'' \emph{IEEE Communications
  Magazine}, vol.~53, no.~5, pp. 52--59, 2015.

\bibitem{zhou2018mobile}
Z.~Zhou, J.~Feng, B.~Gu, B.~Ai, S.~Mumtaz, J.~Rodriguez, and M.~Guizani, ``When
  mobile crowd sensing meets uav: Energy-efficient task assignment and route
  planning,'' \emph{IEEE Transactions on Communications}, vol.~66, no.~11, pp.
  5526--5538, 2018.

\bibitem{zhang2019near}
D.~Zhang, L.~Tan, J.~Ren, M.~K. Awad, S.~Zhang, Y.~Zhang, and P.-J. Wan,
  ``Near-optimal and truthful online auction for computation offloading in
  green edge-computing systems,'' \emph{IEEE Transactions on Mobile Computing},
  vol.~19, no.~4, pp. 880--893, 2019.

\bibitem{jin2015auction}
A.-L. Jin, W.~Song, P.~Wang, D.~Niyato, and P.~Ju, ``Auction mechanisms toward
  efficient resource sharing for cloudlets in mobile cloud computing,''
  \emph{IEEE Transactions on Services Computing}, vol.~9, no.~6, pp. 895--909,
  2015.

\bibitem{konevcny2015Federated}
J.~Kone{\v{c}}n{\`y}, B.~McMahan, and D.~Ramage, ``Federated optimization:
  Distributed optimization beyond the datacenter,'' \emph{arXiv preprint
  arXiv:1511.03575}, 2015.

\bibitem{luong2018optimal}
N.~C. Luong, Z.~Xiong, P.~Wang, and D.~Niyato, ``Optimal auction for edge
  computing resource management in mobile blockchain networks: A deep learning
  approach,'' in \emph{2018 IEEE International Conference on Communications
  (ICC)}.\hskip 1em plus 0.5em minus 0.4em\relax IEEE, 2018, pp. 1--6.

\bibitem{huang2021resource}
Y.~Huang, J.~Zhang, J.~Duan, B.~Xiao, F.~Ye, and Y.~Yang, ``Resource allocation
  and consensus of blockchains in pervasive edge computing environments,''
  \emph{IEEE Transactions on Mobile Computing}, 2021.

\bibitem{bai2019energy}
T.~Bai, J.~Wang, Y.~Ren, and L.~Hanzo, ``Energy-efficient computation
  offloading for secure uav-edge-computing systems,'' \emph{IEEE Transactions
  on Vehicular Technology}, vol.~68, no.~6, pp. 6074--6087, 2019.

\bibitem{dai2018jointoffloading}
Y.~Dai, D.~Xu, S.~Maharjan, and Y.~Zhang, ``Joint offloading and resource
  allocation in vehicular edge computing and networks,'' in \emph{2018 IEEE
  Global Communications Conference (GLOBECOM)}.\hskip 1em plus 0.5em minus
  0.4em\relax IEEE, 2018, pp. 1--7.

\bibitem{wang2020agent}
R.~Wang, Y.~Cao, A.~Noor, T.~A. Alamoudi, and R.~Nour, ``Agent-enabled task
  offloading in uav-aided mobile edge computing,'' \emph{Computer
  Communications}, vol. 149, pp. 324--331, 2020.

\bibitem{alam2019edge}
M.~G.~R. Alam, M.~S. Munir, M.~Z. Uddin, M.~S. Alam, T.~N. Dang, and C.~S.
  Hong, ``Edge-of-things computing framework for cost-effective provisioning of
  healthcare data,'' \emph{Journal of Parallel and Distributed Computing}, vol.
  123, pp. 54--60, 2019.

\bibitem{dinh2018learning}
T.~Q. Dinh, Q.~D. La, T.~Q. Quek, and H.~Shin, ``Learning for computation
  offloading in mobile edge computing,'' \emph{IEEE Transactions on
  Communications}, vol.~66, no.~12, pp. 6353--6367, 2018.

\bibitem{li2020joint}
S.~Li, S.~Lin, L.~Cai, W.~Li, and G.~Zhu, ``Joint resource allocation and
  computation offloading with time-varying fading channel in vehicular edge
  computing,'' \emph{IEEE Transactions on Vehicular Technology}, vol.~69,
  no.~3, pp. 3384--3398, 2020.

\bibitem{chen2020stackelberg}
Y.~Chen, Z.~Li, B.~Yang, K.~Nai, and K.~Li, ``A stackelberg game approach to
  multiple resources allocation and pricing in mobile edge computing,''
  \emph{Future Generation Computer Systems}, vol. 108, pp. 273--287, 2020.

\bibitem{shi2019share}
W.~Shi, J.~Zhang, and R.~Zhang, ``Share-based edge computing paradigm with
  mobile-to-wired offloading computing,'' \emph{IEEE Communications Letters},
  vol.~23, no.~11, pp. 1953--1957, 2019.

\bibitem{ullah2020task}
I.~Ullah and H.~Y. Youn, ``Task classification and scheduling based on k-means
  clustering for edge computing,'' \emph{Wireless Personal Communications}, pp.
  1--14, 2020.

\bibitem{zheng2020task}
X.~Zheng, M.~Li, and J.~Guo, ``Task scheduling using edge computing system in
  smart city,'' \emph{International Journal of Communication Systems}, p.
  e4422, 2020.

\bibitem{cao2018mobile}
X.~Cao, J.~Xu, and R.~Zhang, ``Mobile edge computing for cellular-connected
  uav: Computation offloading and trajectory optimization,'' in \emph{2018 IEEE
  19th International Workshop on Signal Processing Advances in Wireless
  Communications (SPAWC)}.\hskip 1em plus 0.5em minus 0.4em\relax IEEE, 2018,
  pp. 1--5.

\bibitem{liu2019minimization}
J.~Liu, L.~Li, F.~Yang, X.~Liu, X.~Li, X.~Tang, and Z.~Han, ``Minimization of
  offloading delay for two-tier uav with mobile edge computing,'' in \emph{2019
  15th International Wireless Communications \& Mobile Computing Conference
  (IWCMC)}.\hskip 1em plus 0.5em minus 0.4em\relax IEEE, 2019, pp. 1534--1538.

\bibitem{wang2019joint}
Y.~Wang, Z.-Y. Ru, K.~Wang, and P.-Q. Huang, ``Joint deployment and task
  scheduling optimization for large-scale mobile users in multi-uav-enabled
  mobile edge computing,'' \emph{IEEE transactions on cybernetics}, vol.~50,
  no.~9, pp. 3984--3997, 2019.

\bibitem{zhang2019computation}
J.~Zhang, L.~Zhou, F.~Zhou, B.-C. Seet, H.~Zhang, Z.~Cai, and J.~Wei,
  ``Computation-efficient offloading and trajectory scheduling for multi-uav
  assisted mobile edge computing,'' \emph{IEEE Transactions on Vehicular
  Technology}, vol.~69, no.~2, pp. 2114--2125, 2019.

\bibitem{hu2019uav}
X.~Hu, K.-K. Wong, K.~Yang, and Z.~Zheng, ``Uav-assisted relaying and edge
  computing: Scheduling and trajectory optimization,'' \emph{IEEE Transactions
  on Wireless Communications}, vol.~18, no.~10, pp. 4738--4752, 2019.

\bibitem{peng2021multi}
H.~Peng and X.~Shen, ``Multi-agent reinforcement learning based resource
  management in mec-and uav-assisted vehicular networks,'' \emph{IEEE Journal
  on Selected Areas in Communications}, vol.~39, no.~1, pp. 131--141, 2021.

\bibitem{zhang2019task}
J.~Zhang, H.~Guo, J.~Liu, and Y.~Zhang, ``Task offloading in vehicular edge
  computing networks: A load-balancing solution,'' \emph{IEEE Transactions on
  Vehicular Technology}, vol.~69, no.~2, pp. 2092--2104, 2019.

\bibitem{zhou2019energy}
Z.~Zhou, J.~Feng, Z.~Chang, and X.~Shen, ``Energy-efficient edge computing
  service provisioning for vehicular networks: A consensus admm approach,''
  \emph{IEEE Transactions on Vehicular Technology}, vol.~68, no.~5, pp.
  5087--5099, 2019.

\bibitem{wang2019computationvehicular}
J.~Wang, D.~Feng, S.~Zhang, J.~Tang, and T.~Q. Quek, ``Computation offloading
  for mobile edge computing enabled vehicular networks,'' \emph{IEEE Access},
  vol.~7, pp. 62\,624--62\,632, 2019.

\bibitem{xu2019multi}
X.~Xu, R.~Gu, F.~Dai, L.~Qi, and S.~Wan, ``Multi-objective computation
  offloading for internet of vehicles in cloud-edge computing,'' \emph{Wireless
  Networks}, vol.~26, no.~3, pp. 1611--1629, 2019.

\bibitem{luo2021selflearning}
Q.~Luo, C.~Li, T.~H. Luan, W.~Shi, and W.~Weigang, ``Self-learning based
  computation offloading for internet of vehicles: Model and algorithm,''
  \emph{IEEE Transactions on Wireless Communications}, pp. 1--1, 2021, early
  access, doi: {10.1109/TWC.2021.3071248}.

\bibitem{huang2019social}
X.~Huang, P.~Li, and R.~Yu, ``Social welfare maximization in container-based
  task scheduling for parked vehicle edge computing,'' \emph{IEEE
  Communications Letters}, vol.~23, no.~8, pp. 1347--1351, 2019.

\bibitem{zhang2019optimalPV}
J.~Zhang, X.~Huang, and R.~Yu, ``Optimal task assignment with delay constraint
  for parked vehicle assisted edge computing: A stackelberg game approach,''
  \emph{IEEE Communications Letters}, vol.~24, no.~3, pp. 598--602, 2019.

\bibitem{feng2017ave}
J.~Feng, Z.~Liu, C.~Wu, and Y.~Ji, ``Ave: Autonomous vehicular edge computing
  framework with aco-based scheduling,'' \emph{IEEE Transactions on Vehicular
  Technology}, vol.~66, no.~12, pp. 10\,660--10\,675, 2017.

\bibitem{hung2018videoedge}
C.-C. Hung, G.~Ananthanarayanan, P.~Bodik, L.~Golubchik, M.~Yu, P.~Bahl, and
  M.~Philipose, ``Videoedge: Processing camera streams using hierarchical
  clusters,'' in \emph{2018 IEEE/ACM Symposium on Edge Computing (SEC)}.\hskip
  1em plus 0.5em minus 0.4em\relax IEEE, 2018, pp. 115--131.

\bibitem{yi2017lavea}
S.~Yi, Z.~Hao, Q.~Zhang, Q.~Zhang, W.~Shi, and Q.~Li, ``Lavea: Latency-aware
  video analytics on edge computing platform,'' in \emph{Proceedings of the
  Second ACM/IEEE Symposium on Edge Computing}, 2017, pp. 1--13.

\bibitem{al2017energy}
A.~Al-Shuwaili and O.~Simeone, ``Energy-efficient resource allocation for
  mobile edge computing-based augmented reality applications,'' \emph{IEEE
  Wireless Communications Letters}, vol.~6, no.~3, pp. 398--401, 2017.

\bibitem{liu2019code}
J.~Liu and Q.~Zhang, ``Code-partitioning offloading schemes in mobile edge
  computing for augmented reality,'' \emph{IEEE Access}, vol.~7, pp.
  11\,222--11\,236, 2019.

\bibitem{wang2018big}
T.~Wang, M.~Z.~A. Bhuiyan, G.~Wang, M.~A. Rahman, J.~Wu, and J.~Cao, ``Big data
  reduction for a smart city’s critical infrastructural health monitoring,''
  \emph{IEEE Communications Magazine}, vol.~56, no.~3, pp. 128--133, 2018.

\bibitem{hou2018green}
W.~Hou, Z.~Ning, and L.~Guo, ``Green survivable collaborative edge computing in
  smart cities,'' \emph{IEEE Transactions on Industrial informatics}, vol.~14,
  no.~4, pp. 1594--1605, 2018.

\bibitem{li2018delay}
M.~Li, P.~Si, and Y.~Zhang, ``Delay-tolerant data traffic to software-defined
  vehicular networks with mobile edge computing in smart city,'' \emph{IEEE
  Transactions on Vehicular Technology}, vol.~67, no.~10, pp. 9073--9086, 2018.

\bibitem{xu2019trust}
X.~Xu, X.~Liu, Z.~Xu, F.~Dai, X.~Zhang, and L.~Qi, ``Trust-oriented iot service
  placement for smart cities in edge computing,'' \emph{IEEE Internet of Things
  Journal}, vol.~7, no.~5, pp. 4084--4091, 2019.

\bibitem{chaudhry2017azspm}
J.~Chaudhry, K.~Saleem, R.~Islam, A.~Selamat, M.~Ahmad, and C.~Valli, ``Azspm:
  Autonomic zero-knowledge security provisioning model for medical control
  systems in fog computing environments,'' in \emph{2017 IEEE 42nd Conference
  on Local Computer Networks Workshops (LCN Workshops)}.\hskip 1em plus 0.5em
  minus 0.4em\relax IEEE, 2017, pp. 121--127.

\bibitem{lin2019smart}
C.-C. Lin, D.-J. Deng, Y.-L. Chih, and H.-T. Chiu, ``Smart manufacturing
  scheduling with edge computing using multiclass deep q network,'' \emph{IEEE
  Transactions on Industrial Informatics}, vol.~15, no.~7, pp. 4276--4284,
  2019.

\bibitem{sun2019ai}
W.~Sun, J.~Liu, and Y.~Yue, ``Ai-enhanced offloading in edge computing: When
  machine learning meets industrial iot,'' \emph{IEEE Network}, vol.~33, no.~5,
  pp. 68--74, 2019.

\bibitem{zavalyshyn2018homepad}
I.~Zavalyshyn, N.~O. Duarte, and N.~Santos, ``Homepad: A privacy-aware smart
  hub for home environments,'' in \emph{2018 IEEE/ACM Symposium on Edge
  Computing (SEC)}.\hskip 1em plus 0.5em minus 0.4em\relax IEEE, 2018, pp.
  58--73.

\bibitem{wang2017healthedge}
H.~Wang, J.~Gong, Y.~Zhuang, H.~Shen, and J.~Lach, ``Healthedge: Task
  scheduling for edge computing with health emergency and human behavior
  consideration in smart homes,'' in \emph{2017 IEEE International Conference
  on Big Data (Big Data)}.\hskip 1em plus 0.5em minus 0.4em\relax IEEE, 2017,
  pp. 1213--1222.

\bibitem{samie2016computation}
F.~Samie, V.~Tsoutsouras, L.~Bauer, S.~Xydis, D.~Soudris, and J.~Henkel,
  ``Computation offloading and resource allocation for low-power iot edge
  devices,'' in \emph{2016 IEEE 3rd World Forum on Internet of Things
  (WF-IoT)}.\hskip 1em plus 0.5em minus 0.4em\relax IEEE, 2016, pp. 7--12.

\bibitem{manogaran2019wearable}
G.~Manogaran, P.~M. Shakeel, H.~Fouad, Y.~Nam, S.~Baskar, N.~Chilamkurti, and
  R.~Sundarasekar, ``Wearable iot smart-log patch: An edge computing-based
  bayesian deep learning network system for multi access physical monitoring
  system,'' \emph{Sensors}, vol.~19, no.~13, p. 3030, 2019.

\bibitem{ning2020mobile}
Z.~Ning, P.~Dong, X.~Wang, X.~Hu, L.~Guo, B.~Hu, Y.~Guo, T.~Qiu, and R.~Kwok,
  ``Mobile edge computing enabled 5g health monitoring for internet of medical
  things: A decentralized game theoretic approach,'' \emph{IEEE Journal on
  Selected Areas in Communications}, vol.~39, no.~2, pp. 463 -- 478, 2021.

\bibitem{nikoloudakis2019vulnerability}
Y.~Nikoloudakis, E.~Pallis, G.~Mastorakis, C.~X. Mavromoustakis, C.~Skianis,
  and E.~K. Markakis, ``Vulnerability assessment as a service for fog-centric
  ict ecosystems: A healthcare use case,'' \emph{Peer-to-Peer Networking and
  Applications}, vol.~12, no.~5, pp. 1216--1224, 2019.

\bibitem{chen2018edge}
B.~Chen, J.~Wan, A.~Celesti, D.~Li, H.~Abbas, and Q.~Zhang, ``Edge computing in
  iot-based manufacturing,'' \emph{IEEE Communications Magazine}, vol.~56,
  no.~9, pp. 103--109, 2018.

\bibitem{li2019hybrid}
X.~Li, J.~Wan, H.-N. Dai, M.~Imran, M.~Xia, and A.~Celesti, ``A hybrid
  computing solution and resource scheduling strategy for edge computing in
  smart manufacturing,'' \emph{IEEE Transactions on Industrial Informatics},
  vol.~15, no.~7, pp. 4225--4234, 2019.

\bibitem{cao2017edgeos_h}
J.~Cao, L.~Xu, R.~Abdallah, and W.~Shi, ``Edgeos\_h: a home operating system
  for internet of everything,'' in \emph{2017 IEEE 37th international
  conference on distributed computing systems (ICDCS)}.\hskip 1em plus 0.5em
  minus 0.4em\relax IEEE, 2017, pp. 1756--1764.

\bibitem{hong2020space}
T.~Hong, W.~Zhao, R.~Liu, and M.~Kadoch, ``Space-air-ground iot network and
  related key technologies,'' \emph{IEEE Wireless Communications}, vol.~27,
  no.~2, pp. 96--104, 2020.

\bibitem{cheng2020comprehensive}
N.~Cheng, W.~Quan, W.~Shi, H.~Wu, Q.~Ye, H.~Zhou, W.~Zhuang, X.~S. Shen, and
  B.~Bai, ``A comprehensive simulation platform for space-air-ground integrated
  network,'' \emph{IEEE Wireless Communications}, vol.~27, no.~1, pp. 178--185,
  2020.

\bibitem{liu2020task}
J.~Liu, X.~Du, J.~Cui, M.~Pan, and D.~Wei, ``Task-oriented intelligent
  networking architecture for the space--air--ground--aqua integrated
  network,'' \emph{IEEE Internet of Things Journal}, vol.~7, no.~6, pp.
  5345--5358, 2020.

\bibitem{zhang2017software}
N.~Zhang, S.~Zhang, P.~Yang, O.~Alhussein, W.~Zhuang, and X.~S. Shen,
  ``Software defined space-air-ground integrated vehicular networks: Challenges
  and solutions,'' \emph{IEEE Communications Magazine}, vol.~55, no.~7, pp.
  101--109, 2017.

\bibitem{zhang2018air}
S.~Zhang, W.~Quan, J.~Li, W.~Shi, P.~Yang, and X.~Shen, ``Air-ground integrated
  vehicular network slicing with content pushing and caching,'' \emph{IEEE
  Journal on Selected Areas in Communications}, vol.~36, no.~9, pp. 2114--2127,
  2018.

\bibitem{jonas2019cloud}
E.~Jonas, J.~Schleier-Smith, V.~Sreekanti, C.-C. Tsai, A.~Khandelwal, Q.~Pu,
  V.~Shankar, J.~Carreira, K.~Krauth, N.~Yadwadkar \emph{et~al.}, ``Cloud
  programming simplified: A berkeley view on serverless computing,''
  \emph{arXiv preprint arXiv:1902.03383}, 2019.

\bibitem{markakis2017exegesis}
E.~K. Markakis, K.~Karras, N.~Zotos, A.~Sideris, T.~Moysiadis, A.~Corsaro,
  G.~Alexiou, C.~Skianis, G.~Mastorakis, C.~X. Mavromoustakis \emph{et~al.},
  ``Exegesis: Extreme edge resource harvesting for a virtualized fog
  environment,'' \emph{IEEE Communications Magazine}, vol.~55, no.~7, pp.
  173--179, 2017.

\bibitem{wen2020joint}
W.~Wen, Y.~Cui, T.~Q. Quek, F.-C. Zheng, and S.~Jin, ``Joint optimal software
  caching, computation offloading and communications resource allocation for
  mobile edge computing,'' \emph{IEEE Transactions on Vehicular Technology},
  vol.~69, no.~7, pp. 7879 -- 7894, 2020.

\bibitem{gupta2017ifogsim}
H.~Gupta, A.~Vahid~Dastjerdi, S.~K. Ghosh, and R.~Buyya, ``ifogsim: A toolkit
  for modeling and simulation of resource management techniques in the internet
  of things, edge and fog computing environments,'' \emph{Software: Practice
  and Experience}, vol.~47, no.~9, pp. 1275--1296, 2017.

\bibitem{sonmez2018edgecloudsim}
C.~Sonmez, A.~Ozgovde, and C.~Ersoy, ``Edgecloudsim: An environment for
  performance evaluation of edge computing systems,'' \emph{Transactions on
  Emerging Telecommunications Technologies}, vol.~29, no.~11, p. e3493, 2018.

\bibitem{lopes2017myifogsim}
M.~M. Lopes, W.~A. Higashino, M.~A. Capretz, and L.~F. Bittencourt,
  ``Myifogsim: A simulator for virtual machine migration in fog computing,'' in
  \emph{Companion Proceedings of the10th International Conference on Utility
  and Cloud Computing}, 2017, pp. 47--52.

\end{thebibliography}
\begin{IEEEbiography}[{\includegraphics[width=1in,height=1.25in,clip,keepaspectratio]{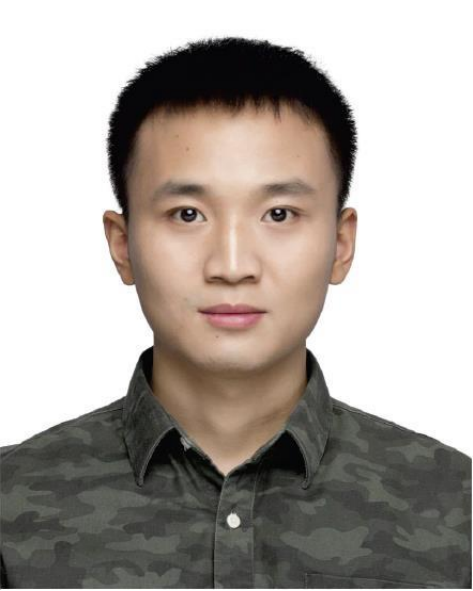}}]{Quyuan Luo}
received the Ph.D. degree in communication and information system from Xidian University, Xi'an, China, in 2020. He had been a visiting scholar with computer science, Wayne State University, USA from 2019 to 2020. He is currently an assistant professor with the School of Information Science and Technology, Southwest Jiaotong University. His current research interests include intelligent transportation systems, content distribution, edge computing and resource allocation in vehicular networks.
\end{IEEEbiography}

\begin{IEEEbiography}[{\includegraphics[width=1in,height=1.25in,clip,keepaspectratio]{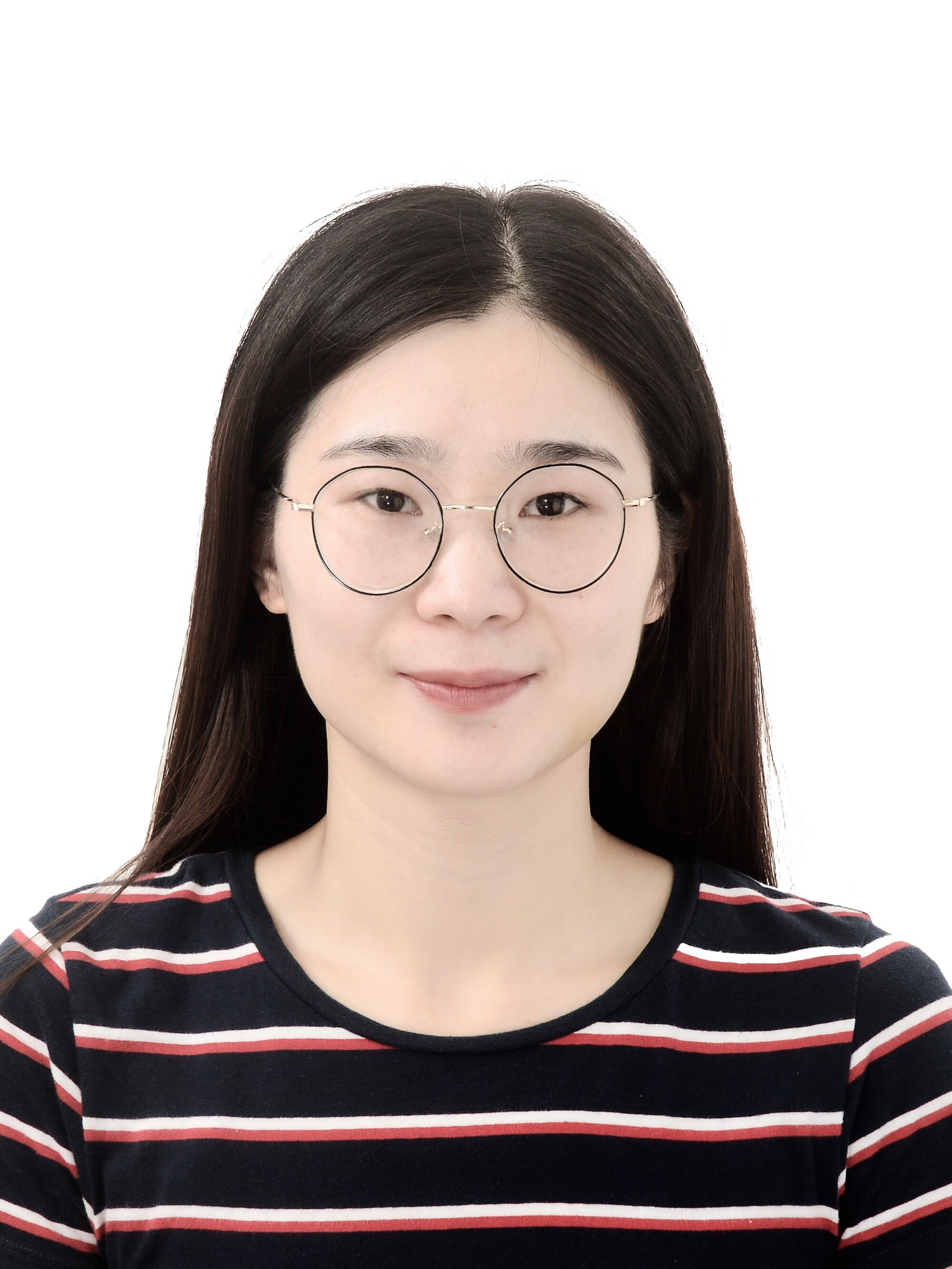}}]{Shihong Hu}
received the bachelor's degree in communication engineering from Jiangnan University in 2016. She is a PhD. candidate of the school of Artificial Intelligence and Computer, Jiangnan University. She had been a Visiting Scholar in Prof. Weisong Shi's MIST Lab for research on resource scheduling in edge computing project, Wayne State University, USA, from 2019 to 2020. Her research interests include wireless sensor networks and edge computing.
\end{IEEEbiography}

\begin{IEEEbiography}[{\includegraphics[width=1in,height=1.25in,clip,keepaspectratio]{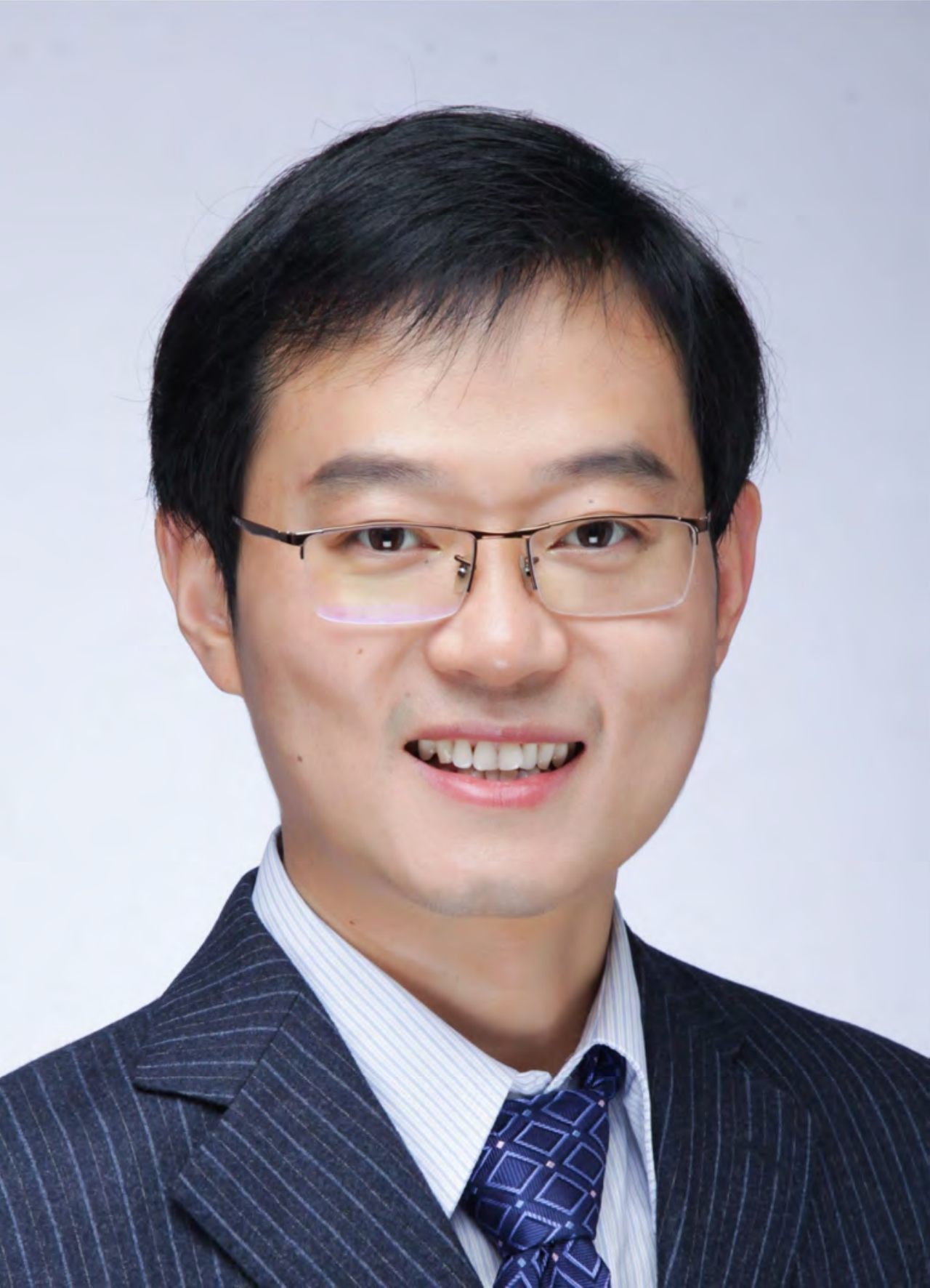}}]{Changle Li}
(M'09-SM'16) received the Ph.D. degree in communication and information system from Xidian University, Xi'an, China, in 2005. He conducted his postdoctoral research in Canada and the National Institute of information and Communications Technology, Japan, respectively. He had been a Visiting Scholar with the University of Technology Sydney and is currently a Professor with the State Key Laboratory of Integrated Services Networks, Xidian University. His research interests include intelligent transportation systems, vehicular networks, mobile ad hoc networks, and wireless sensor networks.
\end{IEEEbiography}

\begin{IEEEbiography}[{\includegraphics[width=1in,height=1.25in,clip,keepaspectratio]{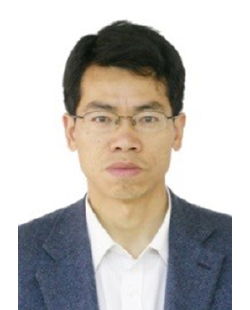}}]{Guanghui Li}
received the Ph.D. degree from the Institute of Computing Technology, Chinese Academy of Sciences, Beijing, China, in 2005. He is currently a Professor with the Department of Computer Science, Jiangnan University, Wuxi, China. He has published over 70 papers in journal or conferences. His research interests include wireless sensor networks, fault tolerant computing, and nondestructive testing and evaluation. His research was supported by the National Foundation of China, Zhejiang, Jiangsu Provincial Science and Technology Foundation, and other governmental and industrial agencies.
\end{IEEEbiography}

\begin{IEEEbiography}[{\includegraphics[width=1in,height=1.25in,clip,keepaspectratio]{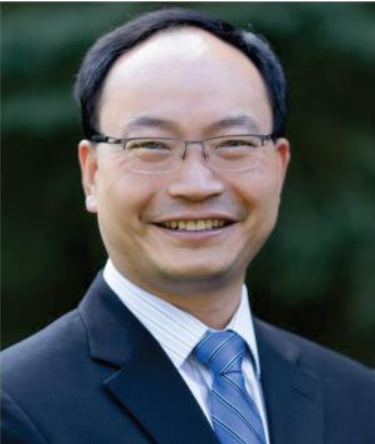}}]{Weisong Shi}
received the B.S. degree fromXidian University, Xi’an, China, in 1995, and thePh.D. degree from the Chinese Academy of Sci-ences, in 2000, both in computer engineering.Weisong Shi is a Charles H. Gershenson Distin-guished Faculty Fellow and a Professor of ComputerScience with Wayne State University, USA, wherehe directs the Mobile and Internet SysTems Labora-tory (MIST) and Connected and Autonomous dRiv-ing Laboratory (CAR), investigating performance,reliability, power- and energy-efficiency, trust andprivacy issues of networked computer systems, and applications. He is one ofthe world leaders in the edge computing research community and publishedthe first book on edge computing. His paper entitled “Edge Computing: Visionand Challenges” has been cited more than 1700 times. In 2018, Dr. Shiled the development of IEEE Course on Edge Computing. In 2019, Dr. Shiserved as the lead guest editor for the edge computing special issue on theprestigious Proceedings of the IEEE journal. He is the Founding SteeringCommittee Chair of the ACM/IEEE Symposium on Edge Computing (SEC)and the IEEE/ACM Connected Health: Applications, Systems and Engineering(CHASE). He is an IEEE Fellow and an ACM Distinguished Scientist.
\end{IEEEbiography}

\end{document}